\documentclass{ar-1col}

\usepackage[comma]{natbib}
\usepackage{url}
\usepackage{amsmath, amssymb}
\usepackage{enumitem}
\usepackage{hyperref}

\jname{Xxxx. Xxx. Xxx. Xxx.}
\jvol{AA}
\jyear{YYYY}
\doi{10.1146/((please add article doi))}

\newcommand\maspm{\ensuremath{\text{mas~yr}^{-1}}}
\newcommand\muas{\ensuremath{\mu\text{as}}}
\newcommand\muaspm{\ensuremath{\mu\text{as~yr}^{-1}}}
\newcommand\kms{\ensuremath{\text{km}~\text{s}^{-1}}}
\newcommand\mura{\ensuremath{\mu_{\alpha*}}}
\newcommand\mudec{\ensuremath{\mu_\delta}}
\newcommand\vect[1]{\ensuremath{{\boldsymbol{#1}}}}
\newcommand\gaia{\textit{Gaia}}
\newcommand\gdr[1]{\gaia~DR#1}
\newcommand\edr[1]{\gaia~EDR#1}
\newcommand\msun{\ensuremath{M_\odot}}
\newcommand\bpminrp{\ensuremath{(G_\mathrm{BP}-G_\mathrm{RP})}}
\newcommand\grvs{\ensuremath{G_\mathrm{RVS}}}
\newcommand\gbp{\ensuremath{G_\mathrm{BP}}}
\newcommand\grp{\ensuremath{G_\mathrm{RP}}}

\begin{document}

\markboth{Anthony G.A.~Brown}{Microarcsecond Astrometry}

\title{Microarcsecond Astrometry: Science Highlights from {\gaia}}

\author{Anthony G.A.~Brown$\null$
\affil{$\null$Leiden Observatory, Leiden University, Niels Bohrweg 2, 2333 CA Leiden, The Netherlands; email:
brown@strw.leidenuniv.nl}}

\begin{abstract}
    Access to microarcsecond astrometry is now routine in the radio, infrared, and optical domains. In particular the
    publication of the second data release from the {\gaia} mission made it possible for every astronomer to work with
    easily accessible, high-precision astrometry for $1.7$ billion sources to 21st magnitude over the full sky.
    \begin{itemize}[rightmargin=80pt]
        \item {\gaia} provides splendid astrometry but at the limits of the data small systematic errors are present. A
            good understanding of the Hipparcos/{\gaia} astrometry concept, and of the data collection and processing,
            provides insights into the origins of the systematic errors and how to mitigate their effects.
        \item A selected set of results from {\gaia} highlight the breadth of exciting science and unexpected results,
            from the solar system to the distant universe, to creative uses of the data.
        \item \gdr{2} provides for the first time a dense sampling of Galactic phase space with high precision
            astrometry, photometry, and radial velocities, allowing to uncover subtle features in phase space and the
            observational HR diagram. 
        \item In the coming decade, we can look forward to more accurate and richer {\gaia} data releases, and new
            photometric and spectroscopic surveys coming online that will provide essential complementary data. 
        \item The longer term promises exciting new opportunities for microarcsecond astrometry and beyond, including
            the plans for an infrared version of {\gaia} which would offer the dense sampling of phase space deep into
            the Milky Way's nuclear regions.
    \end{itemize}
\end{abstract}

\maketitle

\tableofcontents

\section{INTRODUCTION}
\label{sec:intro}

Advances in astrometric techniques and instrumentation over the past two decades have brought us to the point where the
measurement uncertainties have become so small as to make it convenient to express these in units of microarcseconds
(where $1$~\muas\ corresponds to $5$~picoradians). In this review ``microarcsecond astrometry'' refers to instruments and
surveys that routinely reach astrometric measurement uncertainties at the tens of {\muas} level. These include the radio
VLBI technique, the GRAVITY instrument in the infrared, and the Hubble Space Telescope in the optical in its spatial
scanning mode. However, above all it is the {\gaia} mission and its second data release which truly opened up the
microarcsecond era in the optical domain, revolutionizing all fields of astronomy.

This review has two main objectives, to explain ``how {\gaia} works'' and to summarize a selection of science highlights
from the first two {\gaia} data releases. The aim of reviewing the Hipparcos/{\gaia} concept of making absolute
astrometric measurements is to provide the astronomer using the {\gaia} data with a basic understanding of how the data
is collected and processed. The main drivers of the precision of the measurements are explained and particular attention
is paid to the sources of systematic errors in the {\gaia} astrometry and how these can be mitigated. The resulting
enhanced understanding of the {\gaia} catalogue data will improve the scientific interpretation thereof. 

Microarcsecond astrometry opens up an enormous number of exciting science areas. At radio wavelengths, the VLBI
technique allows for establishing geometric distances to distant star forming regions which can be used to trace the
Milky Way's spiral arms. In the infrared domain, exquisite studies of the dynamics of stars orbiting the Milky Way's
central black hole are possible with the GRAVITY instrument, leading to an incredibly precise determination of the
distance to the Galactic centre as well as tests of general relativity. At optical wavelengths, microarcsecond
astrometry enables a recalibration of the distance scale of the universe through geometric distances to standard candles
(HST, \gaia), and the {\gaia} data alone provide a fantastic showcase of the power of the highly accurate fundamental
astronomical data, positions, parallaxes, and proper motions. The first and second {\gaia} data releases, \gdr{1} and
\gdr{2}, have revolutionized the studies of the structure, dynamics, and formation history of the Milky Way \citep[see
for example the review by][]{Helmi2020}, but have made possible much more beyond this core goal of the {\gaia} mission.

Precise star positions enable the study of the shapes and possibly atmospheres of Kuiper Belt Objects through stellar
occultations. The accurate {\gaia} parallaxes for large numbers of stars reveal subtle and as yet unexplained features
in the observational Hertzsprung-Russell diagram, and allows us to peer deep into the interiors of white dwarfs. The
dense sampling with parallaxes and proper motions (and {\gaia} radial velocities) of the phase space around the sun
uncovered the ``phase spiral'' and the exciting story of the interaction between the Sagittarius dwarf galaxy and the
Milky Way. The {\gaia} proper motion measurements in distant stellar systems allow the detailed mapping of globular
cluster and satellite galaxy orbits, the uncovering of the many new streams, and provides us with the equivalent of an
integral field unit measurement of the tangential motion fields in the large and small Magellanic clouds. Finally,
{\gaia} reaches all the way to distant universe, providing discoveries of new lensed quasar systems and insights into
AGN accretion disk and jets. The other main objective of this review is thus to summarize these and other science
highlights from {\gaia}. Only a highly selected number of science topics will be discussed, intended primarily to
illustrate the breath of science that can be addressed with {\gaia} data, hopefully inspiring further creative uses of
the data. Many important topics are missing, which is entirely the choice of the author, and the absence of a particular
topic does not in any way reflect on its relevance.

This review starts in Section~\ref{sec:hist} with a brief historical overview of 20th century astrometry, focused on
motivating the need for astrometry from space. The modern context for the {\gaia} mission is also provided, discussing
several instruments and techniques that are highly complementary to {\gaia} and can outperform it in astrometric
precision. Section~\ref{sec:hipgaia} explains in some detail how the Hipparcos/{\gaia} concept of astrometry works.
Section~\ref{sec:scanningsat} explains the measurement concept, which is followed in Section~\ref{sec:precision} by a
discussion of the drivers of the astrometric precision of {\gaia} and how these can be used to make mission parameter
trade-offs. Section~\ref{sec:dataprocessing} discusses the astrometric data processing for {\gaia} and is focused on
providing the elements needed to understand the origin of systematic errors in {\gaia} astrometry, which are discussed
in Section~\ref{sec:systematics}. The main aim of Section~\ref{sec:hipgaia} is to provide the user of the {\gaia} data
with enough understanding of the mission and its measurement concepts to profit from this knowledge when interpreting the
{\gaia} data. For the interested reader many entry points to the Hipparcos/{\gaia} literature are provided in which much
more details can be found. Section~\ref{sec:highlights} presents selected science highlights, mostly from \gdr{2}, and
the topics are roughly ordered by ``distance'', from the solar system to the quasars. In addition some of the more
unexpected uses of the {\gaia} data are highlighted. Section~\ref{sec:highlights} closes with a few lessons learned so
far from the scientific exploitation of the {\gaia} data. Section~\ref{sec:nearfuture} presents the prospects for the
coming decade in which many more results from {\gaia} will appear, increasing in accuracy and richness, alongside the
large spectroscopic and photometric surveys soon starting their operations. In Section~\ref{sec:future} future
directions for astrometry are examined, including the essential task of maintaining the dense and highly accurate
optical astrometric reference frame provided by {\gaia}. Section~\ref{sec:conclusions} summarizes the conclusions and
highlights a number of important issues to address in the future.

\section{SPLITTING THE ARCSECOND}
\label{sec:hist}

Optical astrometric programmes from the 20th century are summarized, emphasizing the motivations for space
astrometry\footnote{The section title refers to the reply by C.~Wynne, when asked about the design of the Hipparcos
    telescope, ``I’m not saying that it won’t [work] but I do know that seconds of arc don’t split into milliseconds of
    arc very easily!'' \citep[quoted in][]{Perryman2010}. This was paraphrased by L.~Lindegren at the IAU Symposium 248
\textit{A Giant Step: From Milli- to Micro-Arcsecond Astrometry} to remind the participants that ``Milliarcseconds do
not split into microarcseconds very easily''.}. The focus is on the state of affairs prior to the appearance of the
Hipparcos Catalogue in 1997. Subsequent developments in ground based optical astrometry are not discussed. An extensive
overview of the historical developments is provided by \cite{Perryman2012}. Modern radio and optical/IR interferometric
instruments, which can outperform {\gaia} in terms of astrometric precision, are briefly described, stressing the
powerful combination of different microarcsecond astrometry techniques.

\subsection{Optical astrometry in the 20th century; the atmospheric barrier}
\label{sec:20th}

Prior to the Hipparcos mission \citep{HipparcosCat}, the discipline of astrometry was divided into three fields
\citep{Perryman2012}: parallax programmes to establish the distances to stars; large-scale surveys to collect positions
and proper motions in support of Galactic structure studies; and the construction of astrometric reference frames
through the measurement of precise absolute positions of a limited number of stars spread over the whole sky.

\begin{marginnote}
    \entry{Astrometry}{The branch of astronomy concerned with the accurate measurement of celestial positions of
    astronomical sources.}
    \entry{Parallax}{Apparent annual motion of a source on the sky as an observer orbits around the solar system
    barycentre.}
    \entry{Proper motion}{Displacement of a source on the sky due to its motion with respect to the solar system
    barycentre.}
    \entry{Celestial reference system}{Set of prescriptions and conventions together with the modelling required to
    define, at any time, a triad of axes.}
    \entry{Celestial reference frame}{Practical realization of a celestial reference system defined by fiducial
    directions in agreement with the reference system's concepts.}
\end{marginnote}

The highest relative astrometric measurement accuracies were needed for parallax programmes. These relied on measuring
the parallactic motion of a target star through differential position measurements with respect to a presumably very
distant reference star. The reviews by \cite{Vasilevskis1966}, \cite{VanDeKamp1971}, and \cite{VanAltena1983} illustrate
the major efforts that went into improving the long-focus astrometry technique, for which the principles were
established at the start of the 20th century \citep{Perryman2012}. Prior to the publication of the Hipparcos Catalogue,
the state of the art was ``The general catalogue of trigonometric parallaxes'' \citep{gcts1995}, which listed parallaxes
for just over $8000$ stars.  The quoted uncertainties were smaller than 2 milliarcseconds for 10 percent of the sample
(the mode of the uncertainty distribution was located at $\sim10$~mas).  Apart from the large uncertainties (for present
day standards), the catalogue also suffered from the inhomogeneity of the observations and data underlying the parallax
results.

\begin{textbox}[t]
    \section{Relative vs.\ absolute astrometry}
    The terms ``relative'' and ``absolute'' astrometry are regularly contrasted. The former is often also referred to as
    ``narrow-angle'' astrometry and the latter as ``global'' or ``wide-angle'' astrometry. In relative astrometry the
    position of the target source is determined relative to a nearby (within less than $\sim1^\circ$) reference source.
    This allows for the elimination of many sources of error and offers the highest astrometric \emph{precision}. For
    applications where the absolute position of the sources is not relevant, such as exoplanet searches, this is the
    method of choice. It is also employed in all ground-based parallax programmes. The major drawback is that the
    reference source may have a non-zero parallax, which necessitates estimating corrections to the parallax of the
    target in order to put it on an absolute scale. This involves the uncertain modelling of the distribution of
    parallaxes of the reference source population and thus leads to lower \emph{accuracy}. Choosing a distant
    extragalactic reference tends to severely restrict the sky area accessible to parallax measurements.

    Absolute astrometry involves the measurement of directions to sources widely separated on the sky in order to
    establish their positions with respect to a fixed reference frame, instead of each other. The term ``absolute''
    indicates that the positions and proper motions are given with respect to a quasi-inertial coordinate system with a
    known reference plane and pole (the $x$-$y$ plane and the direction of the $z$-axis, loosely speaking). A
    quasi-inertial system should not rotate. This is essential for the dynamical interpretation of the proper motions
    which assumes the absence of any centrifugal or Coriolis forces that would appear in a rotating frame. Absolute
    astrometry requires a known and stable reference platform from which the observations are carried out. In practice
    one solves for the source positions and the parameters of the observing platform at the same time (in a ``global''
    solution). 

    In the case of Hipparcos and {\gaia} the use of two telescopes with viewing directions separated by a large angle
    ($\sim90^\circ$) establishes a network of angles measured between sources widely separated on the sky. Parallactic
    motions of different sources can then be disentangled which allows the measurement of absolute parallaxes, without
    reference to distant extragalactic sources (Section~\ref{sec:framesplxs}).
\end{textbox}

All-sky surveys took off at the end of the 19th century. This was driven by the advent of photography that enabled the
positions of many stars to be measured simultaneously, at the expense of abandoning the highest positional accuracies
that were achievable with parallax programmes. Repeating such position measurements over time enabled the derivation of
proper motions. The plate material from these surveys was eventually digitized, which allowed the construction of
astrometric catalogues listing positions for up to hundreds of millions of sources and proper motions for tens of
millions. Examples include the series of catalogues from the US Naval Observatory and the Guide Star Catalogue produced
to support Hubble Space Telescope operations. For an extensive overview of astrometric sky surveys from the 20th
century refer to \cite{Perryman2009}.

The third astrometric task concerns the construction of an all-sky network of stars for which accurate absolute
positions and proper motions are known. Such networks can be used to construct reference frames which serve as anchors
for all other astrometric measurements. Throughout the 20th century the instrument of choice for reference frame
observations remained the meridian circle, which collects position measurements of stars by precisely timing their
transit across an observatory's local meridian. The consequence was that only a sparse network of reference stars could
be established at relatively bright magnitudes. The enormous efforts to construct the optical astrometric reference
frame prior to Hipparcos culminated with the fifth edition of the ``Fundamental Katalog'' \citep[FK5][]{Fricke1988},
which listed the positions and proper motions for 1535 stars, with the majority at $V$ magnitudes between $4$ and $11$.

Astrometric surveys and reference frame programmes required collecting observations with different telescopes spread
over different sites around the globe. Efforts were made to use the same telescope/instrument designs or the same
measurement technique. Nevertheless astrometric catalogues based on inhomogeneous observations (from different
telescope/instrument combinations at different observing sites) inevitably led to severe systematic limitations to the
accuracies of the surveys and the reference frame. These were manifest in the so-called ``zonal'' or ``regional'' errors
which led to, among others, systematic proper motion biases which differ from one region on the sky to another. Local
systematic errors in the positions in the Guide Star Catalogue version 1.1 (referenced to the FK5) were of the order of
1 arcsecond \citep{Perryman2009}, while the distortions in the FK5 reference frame reached 100 mas or more
\citep[e.g.][]{Schwan2002}. In addition \cite{Lindegren1980} showed that traditional (long-focus) methods of
narrow-angle differential astrometry were also limited by the effects of the Earth's atmosphere, predicting that at best
parallaxes with precisions of order 1~mas could be collected at the rate of 100 per year.

Thus optical astrometry was in danger of stalling at positional accuracies at the arcsecond level for wide field
surveys, corresponding to proper motion accuracies of at best $10$--$20$~\maspm\ for the time baselines covered (and
beset with the above mentioned regional errors), and parallaxes limited to milliarcsecond level precision for modest
numbers of stars (suffering from the accuracy limitations inherent to relative parallax measurements). The best existing
reference frame was sparse and covered only stars of fairly bright magnitudes. The need for denser and fainter optical
reference frames was stressed by \cite{Monet1988}, in particular to enable accurate pointing of the Hubble Space
Telescope to ensure its high imaging resolution could be used together with other high angular resolution instruments,
such as the Very Large Array. This remains an important issue as discussed in Section~\ref{sec:crfmaintenance}. 

\subsection{Removing the barrier: astrometry from space}
\label{sec:spaceastrometry}

The unsatisfactory state of affairs for optical astrometry was recognized already in the 1960s when ideas started to be
developed for overcoming the limitation of the Earth as an observing platform, by going to space \citep{Perryman2011}. In
1967 P.~Lacroute presented the proposal for what eventually became the Hipparcos mission \citep{iau1967}. The proposed
concept solved several problems in one go \citep{Perryman2012}:
\begin{itemize}
    \item The move to space would ensure that the measurements were not hampered by the effects of the Earth's
        atmosphere and that the instruments would work in a thermally stable, gravity-free environment, thus eliminating
        two major causes of systematic errors in ground based astrometric surveys.
    \item A single instrument could be used to observe the entire sky, ensuring the homogeneity of the survey and further
        eliminating causes for zonal errors.
    \item The use of two telescopes with viewing directions separated by a wide angle of order $90$ degrees, and with
        the images projected onto a common focal plane, enabled the construction of a rigid reference frame spanning the
        entire sky and the measurement of absolute as opposed to relative parallaxes.
\end{itemize}
It is this concept that provided both the Hipparcos and {\gaia} missions with the following key capabilities, rolling the
three primary tasks of astrometry into one survey:
\begin{itemize}
    \item The efficient collection of high accuracy absolute parallaxes for large numbers of stars over a wide range of
        magnitudes.
    \item A homogeneous and highly accurate survey of positions and proper motions for the same stars, free from zonal
        errors.
    \item The establishment of a dense, accurate, and rigid network of reference positions on the celestial sphere, free
        from regional errors. In the case of {\gaia} for the first time the optical celestial reference frame is
        realized directly through observations of distant QSOs \citep{gcrf2}.
\end{itemize}
A key element of Hipparcos and {\gaia} astrometry is the precise determination of source image locations in the data
stream. This process also produces image fluxes, but accurate determination of either image parameter requires
the knowledge of source colours in order to account for instrument chromatic effects. In practice a precise,
simultaneous, and homogeneous multi-colour photometric survey is required to complement the astrometry. Thus Hipparcos
and {\gaia} provide astrophysical information for all observed sources, in the case of {\gaia} also through the medium
resolution spectra gathered by its Radial Velocity Spectrograph \citep{gaiamission, Cropper2018}. 

\begin{textbox}[t]
    \section{Astrometry with the Hubble Space Telescope}
    Differential astrometry from space has been carried out with the Hubble Space Telescope since the 1990s. Its fine
    guidance sensors have been used to measure parallaxes to sub-milliarcsecond uncertainty levels \citep{Benedict2017},
    while the cameras have been used to measure proper motions through imaging campaigns spread over several years.
    These programmes enabled studies of the internal kinematics of globular clusters \citep[e.g.][]{Bellini2015} and the
    Magellanic Clouds \citep{VanderMarel2014}, and the tangential motions of dwarf galaxies
    \citep[e.g.][]{Kallivayalil2006} and M31 \citep{Sohn2012}. The publication of \gdr{1} allowed for the anchoring of
    the HST astrometry to a dense and much more accurate net of reference sources, facilitating the measurement of
    absolute proper motions and, for example, the 3D internal motions in the Sculptor dwarf galaxy \citep{Massari2018}.
    Based on a new astrometric technique that employs spatial scanning with the HST \citep{Riess2014, Casertano2016},
    \cite{Riess2018} presented parallax measurements of Cepheid variables at $30$--$50$~\muas\ precision. HST thus
    remains very complementary to {\gaia}, especially for bright star parallax work and for proper motion studies of the
    most crowded regions in globular clusters, the Milky Way bulge, and the Magellanic Clouds.
\end{textbox}

\subsection{Splitting the milliarcsecond with interferometry}
\label{sec:interferometry}

The limitations to ground-based optical astrometry can also be overcome through interferometry, which first came to
fruition in the radio domain \citep{Counselman1976}. Today, \muas\ astrometric precision is possible from Earth with
radio or infrared interferometry. This technique combines the electromagnetic signal received by telescopes separated by
a long baseline in order to achieve the necessary sensitivity to the exact direction to a source. \cite{ReidHonma2014}
reviewed microarcsecond astrometry with Very Long Baseline Interferometry, while \cite{JohnstonDeVegt1999} discussed the
application of VLBI to the construction of reference frames. In the infrared domain the GRAVITY instrument coupled to
ESO's Very Large Telescope Interferometer is capable of achieving microarcsecond astrometry \citep{Gravity2017}.

\subsubsection{Narrow angle astrometry}
\label{sec:narrowangle}

With VLBI single measurement positional precisions down to $\sim10$~\muas\ can be achieved through relative measurements
over narrow angles ($\sim1^\circ$) on the sky. \cite{ReidHonma2014} outline the principles of the technique. One
observes the target and reference source located close together on the sky at nearly the same time. For both sources the
delay between the arrival time of the signal at one of the telescopes is measured. Differencing the two measurements
allows one to eliminate sources of error due to delays from the troposphere, ionosphere, the antenna location
uncertainties, and instrumental delays, as all these terms are very similar over small angles on the sky. Errors due to
source structure can be handled by examining the source images and calculating the expected phase shifts. The errors due
to thermal noise can be ignored for sufficiently high signal to noise ratio measurements.

The GRAVITY instrument combines the light from ESO's four VLT unit telescopes and can also employ the auxiliary
telescopes on the Paranal site. A technique similar to VLBI is used to achieve high precision narrow angle astrometry,
in this case over fields of view of $2$--$4$ arcseconds. This again allows for the cancellation of several sources of
error by differencing the target and reference source measurements. The operation at infrared wavelengths demands a much
more complex instrument to control the phase differences (delays) between target and reference. GRAVITY features a
variety of innovations which allow reaching microarcsecond astrometry \citep{Gravity2017}. For broad band observations
the GRAVITY instrument has been demonstrated to achieve single measurement positional precisions in the
$30$--$100$~\muas\ range \citep{Gravity2019}. Over a sufficiently narrow wavelength range, the spectro-differential
astrometry technique allows reaching relative positional precisions of a few \muas\ by tracking the phase differences
between continuum and line emission from a source \citep{Gravity2017}.

The interferometric instruments achieve astrometric precisions over narrow angles that surpass the performance of
{\gaia}, and with a suitably distant reference source (such as QSOs) the relative astrometry can be placed on an
absolute scale. {\gaia} and the interferometry instruments are highly complementary. {\gaia} provides access to
astrometry for vast numbers of sources over the entire sky, but its astrometry is limited, or not available, in obscured
regions in or near the Galactic plane, and in crowded areas such as the centres of globular clusters.
\cite{ReidHonma2014} summarize the science applications of VLBI astrometry which complement {\gaia}, such as: the
accurate determination of distances and motions of maser sources in star forming regions, which allow us to trace the
spiral arms of the Milky Way; the access to astrometry for asymptotic giant branch giant stars through masers in their
envelopes \citep[where {\gaia} astrometry is affected by the photocentre displacements due to the large convective
atmospheres,][]{Chiavassa2018}; the access to astrometry of pulsars; and the possibility to study megamasers in other
galaxies to make direct estimates of the Hubble constant. Astrometry from GRAVITY has been used in combination with
older adaptive optics data, radial velocities, and the VLBI proper motion measurements of Sgr A*, to very accurately
model the orbit of the S2 star around the massive black hole and derive a distance to the Galactic centre with only
$0.3$\% uncertainty \citep{Gravity2019}. This fixes a Galactic structural parameter which can be used in Milky Way
studies with {\gaia} data. These synergies highlight the powerful combination of microarcsecond astrometry techniques at
the disposal of astronomers in the 21st century.

\subsubsection{Reference frames}
\label{sec:refframes}

On January 1 1998 the International Astronomical Union adopted the International Celestial Reference System (ICRS) as a
celestial reference system based on directions to a set of extragalactic sources. The ICRS represents a quasi-inertial
reference system to replace the older systems (e.g., FK5) in which source coordinates were referred to a system
primarily based on the dynamics of the solar system \citep{Feissel1998}\footnote{There are two important consequences of
this change, often not realized by many of us: there is no epoch associated with the ICRS, and changes of source
coordinates between different epochs can be calculated from the proper motions alone \citep[where for rigorous epoch
propagation the parallax and radial velocity are also needed, e.g.][]{Butkevich2014}.}. The practical materialisation of
the ICRS is the Internal Celestial Reference Frame (ICRF), which consists of a list of celestial positions and their
uncertainties for a set of extragalactic sources. For details on reference systems refer to \cite{JohnstonDeVegt1999}.

The ICRF was first set up in the radio domain through VLBI astrometry of carefully selected extragalactic radio sources
\citep{Ma1998}, observed over decades. Although the basic observables are the same as for narrow angle VLBI astrometry
(delays in the signal as received by different antennae), a global solution must be made where the source positions are
solved for along with the positions and velocities of the observing stations, accounting for the Earth's deformations and
its orientation at the time of observation. In addition the effects of the ionosphere and troposphere are calibrated or
modelled out (where the troposphere is one of the main factors limiting the astrometric accuracy). For details refer to
\cite{Ma1998} and the chapter by Fomalont in \cite{VanAltena2013}. The most recent version of the reference frame is the
ICRF3\footnote{\url{http://hpiers.obspm.fr/icrs-pc/newwww/icrf/index.php}} which consists of 303 ``defining'' sources
and an additional 4285 sources which are being observed regularly and may enter the defining set at some future time.
The median position uncertainty for the ICRF3 (as given by the semi-major axis of the uncertainty ellipse) is $0.23$~mas
for the full set of sources and $50$~\muas\ for the defining sources observed in the 8.4~GHz band ($80$~\muas\ for
ICRF2). Hence also reference frames have now firmly entered the microarcsecond era.

Upon the introduction of the ICRF the IAU adopted the Hipparcos Catalogue as the optical materialisation of the ICRS
\citep{Feissel1998}. The optical frame was aligned to the radio frame (considered as the primary reference) through
several intermediate steps, given that no extragalactic sources (except 3C 273) were observed by Hipparcos
\citep{Lindegren1995}. {\gaia} observes millions of quasars, which for the first time enables the realization of
an optical reference frame at sub-milliarcsecond precision (median $0.4$~mas), built solely on direct observations of
extragalactic sources \citep{gcrf2}. The {\gaia}-CRF2 is based on the positions of some $550\,000$ quasars and has a
substantial overlap with the radio ICRF, which allowed for the alignment of the optical reference frame to a prototype
of ICRF3 at the $20$--$30$~\muas\ level. The {\gaia} DR2 catalogue together with the {\gaia}-CRF2 thus represents a vast
improvement for optical reference frames, providing mas-level positions to magnitude 21 for a dense network of sources
all over the sky. This allows the astrometric anchoring of ongoing ground-based optical/IR surveys and will be essential
for the operation of future extremely large telescopes.

\section{GLOBAL ASTROMETRY WITH THE HIPPARCOS/GAIA CONCEPT}
\label{sec:hipgaia}

The concept for performing global astrometry with {\gaia} follows the same principles that were used for the Hipparcos
mission \citep{Lindegren2005}:
\begin{enumerate}
    \item Collect observations simultaneously from two fields of view separated by a large angle.
    \item Scan roughly along a great circle passing through both fields of view.
    \item Make mainly one-dimensional measurements along the scanning direction.
    \item The ``basic angle'' between the two fields must be known and extremely stable.
    \item Repeat the measurements as many times as required to reach the desired astrometric accuracy, and with scans in
        varying orientations in order to cover the whole sky.
\end{enumerate}
Below I summarize the motivations for this concept, the drivers of the astrometric precision, the basic elements of the
astrometric data processing for {\gaia}, and the sources of systematic errors and how these might be mitigated. The
focus is on explaining the concepts at a high level, but with sufficient detail such that anyone using the {\gaia} data
has a basic understanding of how the astrometry is derived from the individual measurements, and what the main
limitations are that should be considered when making investigations with {\gaia} catalogue data. \cite{gaiamission}
provide a detailed overview of the {\gaia} spacecraft, mission, and scientific instruments, and a summary of the entire
DPAC data processing chain, including photometry, radial velocities, and higher level data products.

\begin{marginnote}
    \entry{DPAC}{{\gaia} Data Processing and Analysis Consortium, tasked with turning the raw {\gaia} telemetry into the
    data releases.}
\end{marginnote}

\begin{figure}[t]
    \centering
    \includegraphics[width=0.9\textwidth]{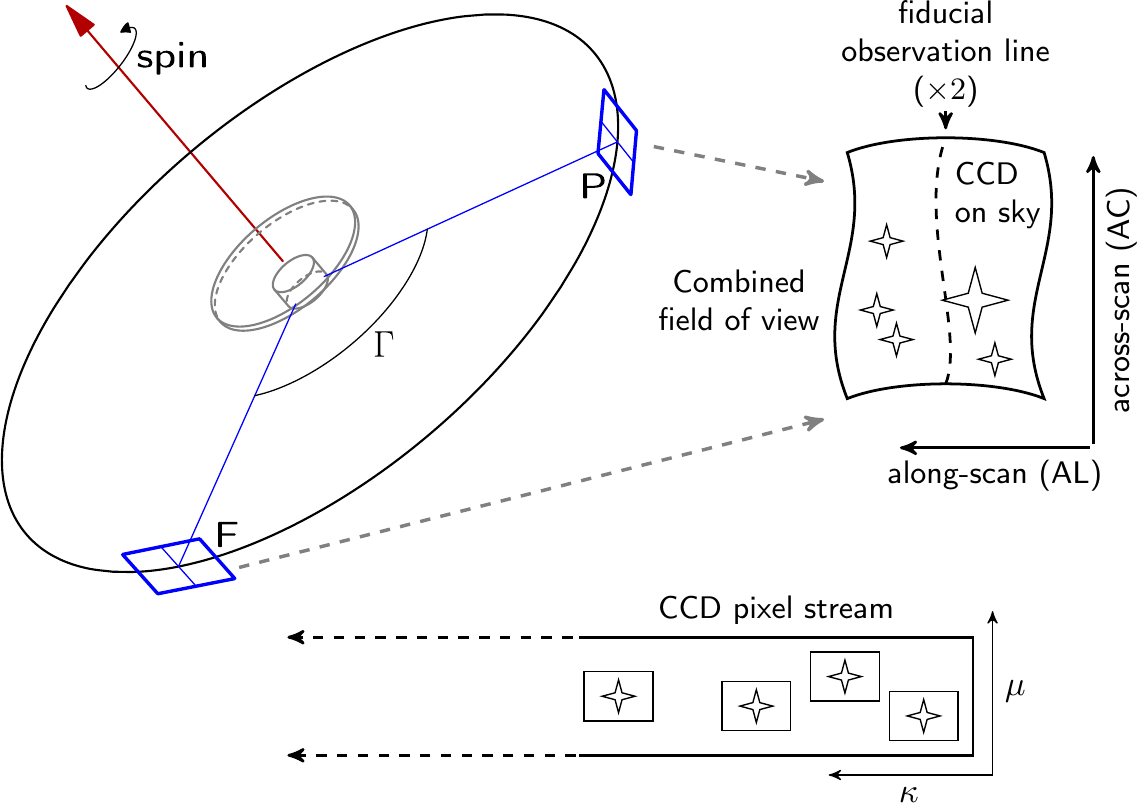}
    \caption{Astrometric observations with {\gaia}. At top left {\gaia} is schematically depicted, with the spacecraft spin
        axis pointing away from the sunshield, and the lines of sight of the two telescopes at right angles to the spin
        axis, separated by the basic angle $\Gamma$. The projection on the sky of the focal plane through both fields of
        view is indicated with P and F. The top right illustrates schematically the collection of the basic observables
        $t_\mathrm{obs}$ (the observing time, the moment a source crosses the fiducial observation line) and
        $\mu_\mathrm{obs}$ (the AC location of the source). The bottom part of the figure shows the data stream in the
        form of continuously read out CCD pixels. Only the pixels in small windows around the sources are sent to the
        ground. The observables are derived from the source locations $\kappa$ (AL) and $\mu$ (AC) in this pixel
        stream.}
    \label{fig:gaiaobs}
\end{figure}

\begin{textbox}[t]
    \section{Gaia in brief\label{sec:gaiabox}}
    {\gaia} is the ESA space astrometry mission, launched in December 2013, collecting accurate positions, parallaxes,
    and proper motions for all sources to magnitude $20.7$ in its white-light photometric band $G$ (covering the range
    $330$--$1050$~nm). Multi-colour photometry is obtained for all stars and radial velocities are collected for stars
    brighter than $G\approx17$. {\gaia}'s photometric instrument consists of two low-resolution fused-silica prisms
    dispersing the light entering the field of view. One disperser --- called BP for Blue Photometer --- operates in
    the wavelength range $330$--$680$ nm; the other --- called RP for Red Photometer --- covers $640$--$1050$ nm. From
    the integrated flux in the prism spectra two broad band blue and red magnitudes, {\gbp} and {\grp}, are defined.
    Radial velocities are measured with the Radial Velocity Spectrograph (RVS) which collects spectra over the
    wavelength range $847$--$874$~nm at a resolution of $\sim11\,000$. The apparent brightness of sources as measured by
    the integrated flux over this wavelength range is referred to as \grvs.

    {\gaia} carries two telescopes with fields of view separated by $106.5^\circ$, of which the light is combined onto a
    single focal plane. Every six hours {\gaia} spins around the axis perpendicular the lines of sight of the
    telescopes. Sources observed by {\gaia} thus drift across the focal plane, and Time-Delayed Integration (TDI, or
    drift-scanning) is used to accumulate photo-electrons into a sharp image as the sources travel across a CCD.
    The wide dynamic range of {\gaia} is achieved by progressively reducing the integration time for bright sources
    ($G<13$) through gates in the CCD detectors, which hold back and discard the photoelectrons accumulated before the
    gate \citep[cf.][]{gaiamission}. To fit the data collected by {\gaia} into the telemetry budget, only the pixels
    immediately around a source are read out and transmitted (cf.\ Figure~\ref{fig:gaiaobs}). For bright sources
    the full 2D window is transmitted, while for fainter sources the window pixels are summed in the direction
    perpendicular to the scanning direction, leading to one-dimensional image profiles being transmitted to Earth.
\end{textbox}

\subsection{Astrometry with {\gaia}}
\label{sec:scanningsat}

Figure~\ref{fig:gaiaobs} illustrates how {\gaia} collects astrometric observables. The {\gaia} spacecraft consists of a
three meter tall cylindrical structure, housing the service module and science instruments, which is kept in the shade
by a ten meter diameter sunshield \citep[see Fig.~1 in][]{gaiamission}. This structure is schematically shown in
Figure~\ref{fig:gaiaobs}, top left, together with the spacecraft spin axis and the lines of sight of the two telescopes,
at right angles to the spin axis and separated by the basic angle $\Gamma$. During one spacecraft revolution the fields
of view of the telescopes scan along the indicated great circle. The light from the telescopes is imaged onto the same
focal plane, covered with CCD detectors, and the projection on the sky of the focal plane through both fields of view is
indicated with P for ``preceding'' and F for ``following'' in the sense of the spin direction of the spacecraft. The
top right of Figure~\ref{fig:gaiaobs} schematically shows the projection on the sky of a CCD detector in the focal
plane. The distortions indicate symbolically that details of the optical projection from telescope mirrors to the
focal plane, in combination with the CCD location and orientation, as well as the properties of the CCD pixel grid, lead
to a projected pixel grid that is not necessarily rectangular on the sky. This will be discussed further in the section
on astrometric data processing. As the spacecraft spins the stars will drift from left to right across the field of
view, in the so-called along-scan (AL) direction\footnote{Note that stars drift along the focal plane in a direction
opposite to the scanning direction dictated by the spacecraft spin vector. The stars drift from left to right along
the CCD but the AL coordinate (equivalent to time) increases from right to left}. The perpendicular direction is
referred to as across-scan (AC). The basic observables are the times $t_\mathrm{obs}$ at which stars cross a fiducial
observing line on the detector and the across-scan locations of the stars $\mu_\mathrm{obs}$ (both for each field of
view). The observables are derived from the CCD image samples. To ensure the vast stream of pixel
data fits in the telemetry budget, only the pixels immediately around a source image are transmitted, the so-called
window (bottom part of Figure~\ref{fig:gaiaobs}). In most cases the pixels in the window are summed on-board in the AC
direction, leading to one dimensional profiles as source ``images'' \citep[see][for the details on the focal plane
read-out scheme]{gaiamission}. The observation times and across-scan locations are derived from the positions of the
sources in these windows, together with information that allows placing the windows in the overall data stream.

\begin{marginnote}
    \entry{AL}{Along-scan. Direction along the great circle scanned by {\gaia}'s telescopes during one spacecraft
    revolution}
    \entry{AC}{Across-scan: Direction perpendicular to the along-scan direction.}
    \entry{$t_\mathrm{obs}$}{Observation time of a source, derived from the centroid of the source image.}
    \entry{$\mu_\mathrm{obs}$}{AC location of a source at the time of observation $t_\mathrm{obs}$.}
\end{marginnote}

Put simplistically, the observation times together with a knowledge of the spacecraft attitude (its orientation and spin
phase) allow us to reconstruct the instantaneous celestial positions of the observed sources. Repeated measurement of
the source positions then leads to determination of the parallax and proper motions. The actual estimation of the
astrometric parameters is much more complicated as will be discussed in Section~\ref{sec:dataprocessing}.

The observing concept outlined above is based on the following considerations \citep{LindegrenBastian2010,Lindegren2005}
for a space mission that implements in a single survey the traditional astrometric programmes (parallaxes, surveys,
reference frames, Section \ref{sec:20th}).

\subsubsection{Wide-angle measurements for reference frames and absolute parallaxes}
\label{sec:framesplxs} 

The advantages of moving to space pointed out in Section~\ref{sec:spaceastrometry} enable the bridging of large ($\sim1$
radian) angles when measuring positions of sources with respect to each other. This is the only way to ensure that
for any two sources separated by an angle $\rho$, the uncertainty on $\rho$ is independent of its value
\citep{LindegrenBastian2010}. The astrometric catalogue is thus free of the zonal errors which result from the
accumulation of errors when combining relative position measurements made in different parts of the sky over small
angles. The presence of zonal errors makes for a less rigid astrometric reference frame and leads to systematic
proper motion errors correlated over large scales, which could, for example, introduce erroneous interpretations of
proper motions in terms of Milky Way dynamics.

\begin{figure}[t]
    \centering
    \includegraphics[width=0.7\textwidth]{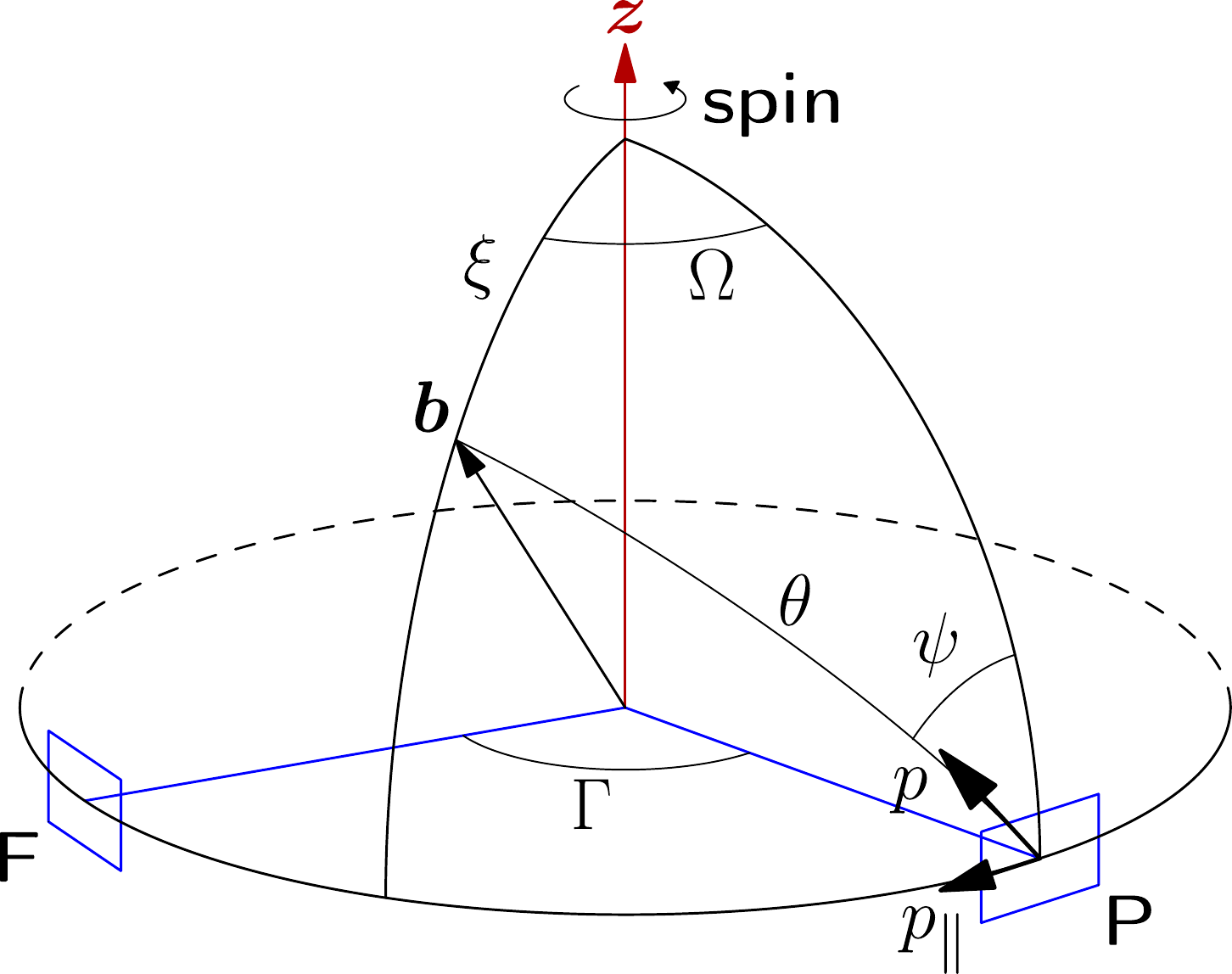}
    \caption{The parallax shift of a source as seen by {\gaia}. The spacecraft schematic is omitted here (cf.\
        Figure~\ref{fig:gaiaobs}). The source is seen in the preceding field of view which is at an azimuth $\Omega$
        with respect to the meridian through the spacecraft spin axis (\vect{z}) and the direction to the solar system
        barycentre (\vect{b}). The angle between the source direction and \vect{b} is $\theta$, while $\xi$ is the fixed
        angle between the spin axis and \vect{b} (the ``solar aspect angle''). The parallax shift of the source is
        toward \vect{b} and proportional to $\sin\theta$. The vectors \vect{p} and $\vect{p}_\parallel$ indicate
        respectively the parallactic shift of the source and its projection on the along scan direction.
    \textit{Credits: Adapted from \cite{LindegrenBastian2010}, right panel of their Fig.~4}.}
    \label{fig:alplxshift}
\end{figure}

Less intuitively, wide-angle astrometry enables the measurement of absolute parallaxes. This is schematically
illustrated in Fig.~4 in \cite{Lindegren2005} for a traditional parallax measurement, based on the observed angles
between a target and a reference source, as having the reference source at 90 degrees from the target. For the specific
case of {\gaia} the way parallax affects the measured positions (observing times) of sources is illustrated in
Figure~\ref{fig:alplxshift}. The parallax shift \vect{p} for any source (the difference between its direction as seen
from the solar system barycentre and from the observer) is directed toward the solar system barycentre and is given as
$p=\Vert\vect{p}\Vert=\varpi R\sin\theta$, where $\varpi$ is the parallax of the source and $R$ the distance from the
observer to the solar system barycentre in au. What matters in the {\gaia} measurements is the parallax shift
$p_\parallel$ in the along-scan direction, which is $p_\parallel = p\sin\psi = \varpi R\sin\theta\sin\psi$ (see
Figure~\ref{fig:alplxshift}). From the law of sines in spherical trigonometry it follows that
$\sin\theta\sin\psi=\sin\xi\sin\Omega$, hence $p_\parallel = \varpi R\sin\xi\sin\Omega$.

\begin{marginnote}
    \entry{$\Omega$}{Azimuth of a source around the {\gaia} spin axis as defined in Figure~\ref{fig:alplxshift}.}
    \entry{$\xi$}{Solar aspect angle. Fixed angle between {\gaia}'s spin axis and the direction to the sun.}
\end{marginnote}

Now consider two sources, one located in the preceding and one in the following field of view, with parallaxes
$\varpi_\mathrm{P}$ and $\varpi_\mathrm{F}$ The large basic angle leads to large differences in the parallax factor
($\sin\xi\sin\Omega$). In particular if the source in the preceding field of view is observed at $\Omega=0$, it will
have zero parallax shift along scan, while for the other source the shift is $\varpi_\mathrm{F}R\sin\xi\sin\Gamma$. The
situation is reversed if the source in the following field of view is at $\Omega=0$ \citep[see Fig.~2
in][]{gaiamission}. It is this property of wide-angle astrometric measurements that allows us to disentangle the
parallactic motions for different sources, meaning that the measurements are sensitive to each of $\varpi_\mathrm{P}$
and $\varpi_\mathrm{F}$. For narrow-field astrometry the parallax factors for all the stars in the field of view are
nearly the same (as $\theta$ is nearly the same) and one is only sensitive to $\varpi_\mathrm{P}-\varpi_\mathrm{F}$,
which makes additive corrections to the measured parallaxes necessary.

\subsubsection{One-dimensional measurements}
\label{sec:1dobs} 

Ideally one would measure the actual angular separation between sources on the sky. However, when measuring sources
simultaneously in two fields of view separated by a large angle (of order 90 degrees) it is sufficient to measure only
the distance between the sources projected on the great circle passing through the two fields of view. As long as the
fields of view themselves are relatively small (of order 1 degree) the difference between the projected and actual
angular separation can be ignored \citep{Lindegren2005}.  Alternatively, this means that the measurement precision
across-scan can be much lower (by a factor 100) than along-scan. This explains a number of aspects of the {\gaia}
design: the smaller across-scan size of the {\gaia} telescope mirrors, the larger AC size of the CCD pixels, and the CCD
read-out method (drift-scanning or time-delayed integration, TDI). The one-dimensional measurements also lead to
simplifications in the calibrations (Section \ref{sec:dataprocessing}) and allow for significant savings in the amount
of data to be sent down by numerically binning the CCD images in the AC direction. This also lowers the relative
contribution of the read-out noise to the counts in the image samples. For the full design details refer to
\cite{gaiamission}.

\subsubsection{Scanning law}
\label{sec:scanlaw} 

To see the parallactic and proper motions of the sources observed by {\gaia} it is necessary to measure them multiple
times, with the scans intersecting at large angles in order to build up a rigid 2D network of angles between sources.
The required continuous re-orientation of the spacecraft should be carried out as smoothly as possible and the full sky
should be covered in a sufficiently short amount of time to allow for repeated scanning of the entire sky. The most
efficient way found so far is to use the so-called uniform revolving scanning as implemented for both Hipparcos and
{\gaia} \citep{LindegrenBastian2010}. The principle is that the spacecraft spins around the axis perpendicular to the
two lines of sight, at a fixed rate ($60''$~s$^{-1}$ for {\gaia}, i.e.\ a 6-h spin period), with the telescopes thus
scanning a great circle during one revolution. The spin-axis itself is made to precess around the direction to the sun,
with a period of 63 days for {\gaia}, maintaining a fixed angle $\xi$ (Figure~\ref{fig:alplxshift}) between the
spin-axis and the spacecraft-Sun direction. As the Sun moves along the ecliptic (as seen from {\gaia}) the spin-axis
makes a slow looping motion around the direction to the sun. In combination with the spinning motion the full sky can be
covered in about 3--4 months \citep[see Figs.~6 and 7 in][]{gaiamission}.

The parallax effect as seen by {\gaia} scales with $\sin\xi$ which would suggest $\xi=90^\circ$ as the best choice. However
in this case the Sun would shine into the {\gaia} telescopes during every spacecraft revolution. In practice $\xi$ should
be substantially less than 90 degrees, which follows from considerations on the size needed for the sunshield and the
energy that can be collected by solar arrays mounted on the illuminated side of the sunshield (maximum for $\xi=0$). For
{\gaia} the solar aspect angle is $\xi=45^\circ$. For more details on the {\gaia} scanning law see \cite{gaiamission} and
\cite{LindegrenBastian2010}.

\subsubsection{Basic angle}
\label{sec:ba}

In order to conduct wide-angle astrometry the value of the basic angle should be large (of order 1 radian) and in
principle the best value would be 90 degrees. However, while this is a good choice when a global astrometric solution is
made (Section~\ref{sec:dataprocessing}), in practice simple fractions $m/n$ of 360 degrees should be avoided. For
Hipparcos this was motivated by the fact that the data processing proceeded in three steps \citep[in order to keep the
problem computationally tractable,][]{Kovalevsky1992, Lindegren1992}, where in the first step the star positions were
reduced to coordinates along a great circle. If the relevant system of equations is examined it results that values of
the basic angle that are simple fractions of 360 degrees lead to much higher variances on the derived positions
\citep{LindegrenBastian2010}. While for {\gaia} enough computational power is available to avoid the data processing in
steps, the so-called ``First Look'' astrometric solution still solves for star positions along a great circle in order
to get quick (daily) and detailed insights into the health of the {\gaia} instruments \citep{Fabricius2016, Jordan2005}.
Hence also for {\gaia} the basic angle value, $106.5^\circ$, was chosen to avoid simple fractions of $360^\circ$. It is
essential that the basic angle is extremely stable in order to avoid an overall zero-point error on the parallaxes
(Section~\ref{sec:systematics}). For {\gaia} stability at the few {\muas} level was required.

\subsection{Astrometric precision}
\label{sec:precision}

Before describing the data processing required to turn the {\gaia} measurements into an astrometric catalogue, I discuss
here the basic drivers of the astrometric precision. Issues relating to accuracy are discussed in Section
\ref{sec:systematics}. Ultimately the astrometric precision depends on how well a source image can be located in the
{\gaia} data stream. For the image location problem, \cite{Lindegren2005} shows that the astrometric (i.e.\ angular)
precision $\sigma$ scales as:
\begin{equation}
    \sigma \propto \frac{\lambda_\mathrm{eff}}{B\sqrt{N}}\,,
    \label{eq:precisionscaling}
\end{equation}
where $\lambda$ is the effective wavelength of the measurements, $N$ the number of photons collected, and $B$
the aperture size of the mirrors (or the baseline of an interferometric system). This formula can be used to make
basic design choices for a scanning astrometry mission. Typically a certain parallax accuracy is targeted, which
scales as (cf.\ Section~\ref{sec:framesplxs}):
\begin{equation}
    \sigma_\varpi \propto \frac{\lambda_\mathrm{eff}}{B\sqrt{N}\sin\xi}\,.
    \label{eq:plxprecisionscaling}
\end{equation}
The number of photons collected is proportional to: the average time spent observing a source, which is dictated by the
field of view size as a fraction of the full sky, multiplied by the overall mission length; the photon collection
efficiency of the system (driven by the telescope transmission and detector quantum efficiency); the photon flux
received from a source; and the aperture area. The solid angle of the field of view is determined by the focal plane
area and the telescope focal length. In the above formulae many complications are skipped over, such as the fact that
multiple astrometric parameters as well as calibration information must be extracted form the same observations.
Nevertheless the formulae allow making basic trade-offs in the design of a mission like {\gaia}. Cost drivers such as
the value of $\xi$, aperture size, and focal plane size, can be compensated for by choosing a longer mission length.
\cite{Lindegren2005} describes how the enormous gains made from Hipparcos to {\gaia} can be understood from the basic
precision scaling.

\subsection{{\gaia} Astrometric data processing}
\label{sec:dataprocessing}

Having reviewed the basic motivations for wide-angle astrometry from space and the basic design drivers for reaching a
target accuracy, we can now turn to the question of how the astrometric parameters of the sources observed by {\gaia} are
derived from the basic observables, $t_\mathrm{obs}$ and $\mu_\mathrm{obs}$ (Figure \ref{fig:gaiaobs}). The data
processing problem for {\gaia} is formulated as a minimization problem \citep[][Eq.~1]{Lindegren2012}:
\begin{equation}
    \min_{\vect{s},\vect{n}} \Vert\vect{f}^\mathrm{obs}-\vect{f}^\mathrm{calc}(\vect{s},\vect{n})\Vert_{\cal M}\,.
    \label{eq:minimization}
\end{equation}
The task is to minimize the difference between the vector of observables $\vect{f}^\mathrm{obs}$ and their predicted
values $\vect{f}^\mathrm{calc}$. The latter depend on the source parameters \vect{s} and a set of ``nuisance
parameters'' \vect{n}, which have to be estimated along with the source parameters but are not themselves of interest.
The norm in the equation above is calculated for a metric $\cal M$, where in practice for {\gaia} a weighted least
squares solution is chosen taking the necessary precautions to make the solution robust. 

\begin{figure}[t]
    \centering
    \begin{minipage}[b]{0.48\textwidth}
        \includegraphics[width=\textwidth]{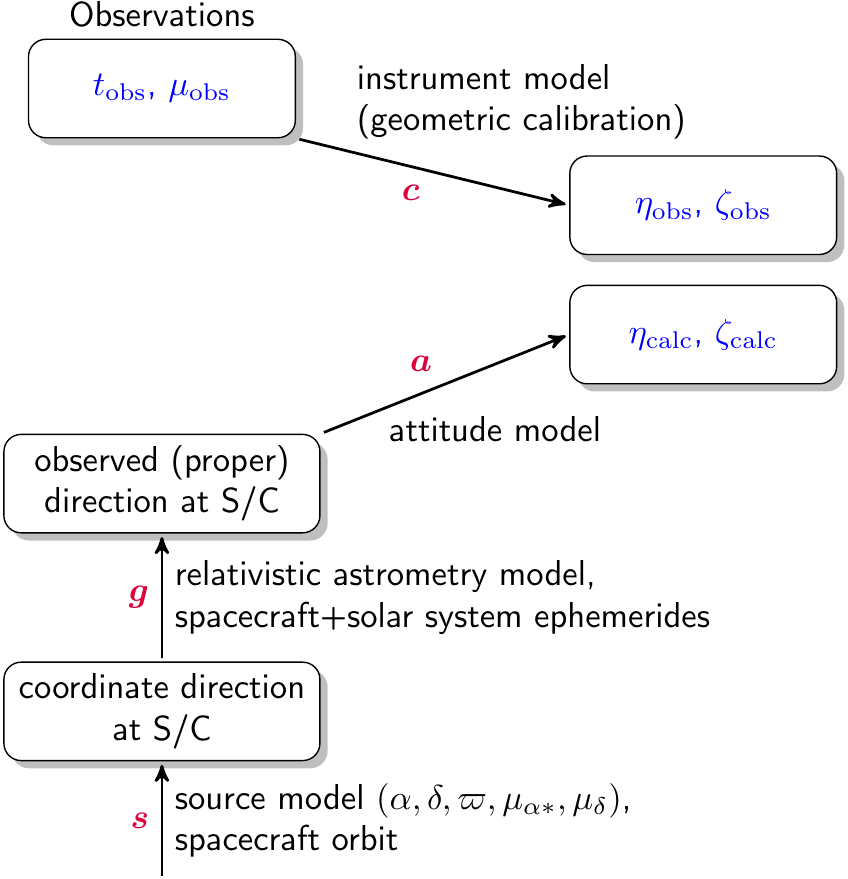}
    \end{minipage}
    \begin{minipage}[b]{0.48\textwidth}
        \includegraphics[width=\textwidth]{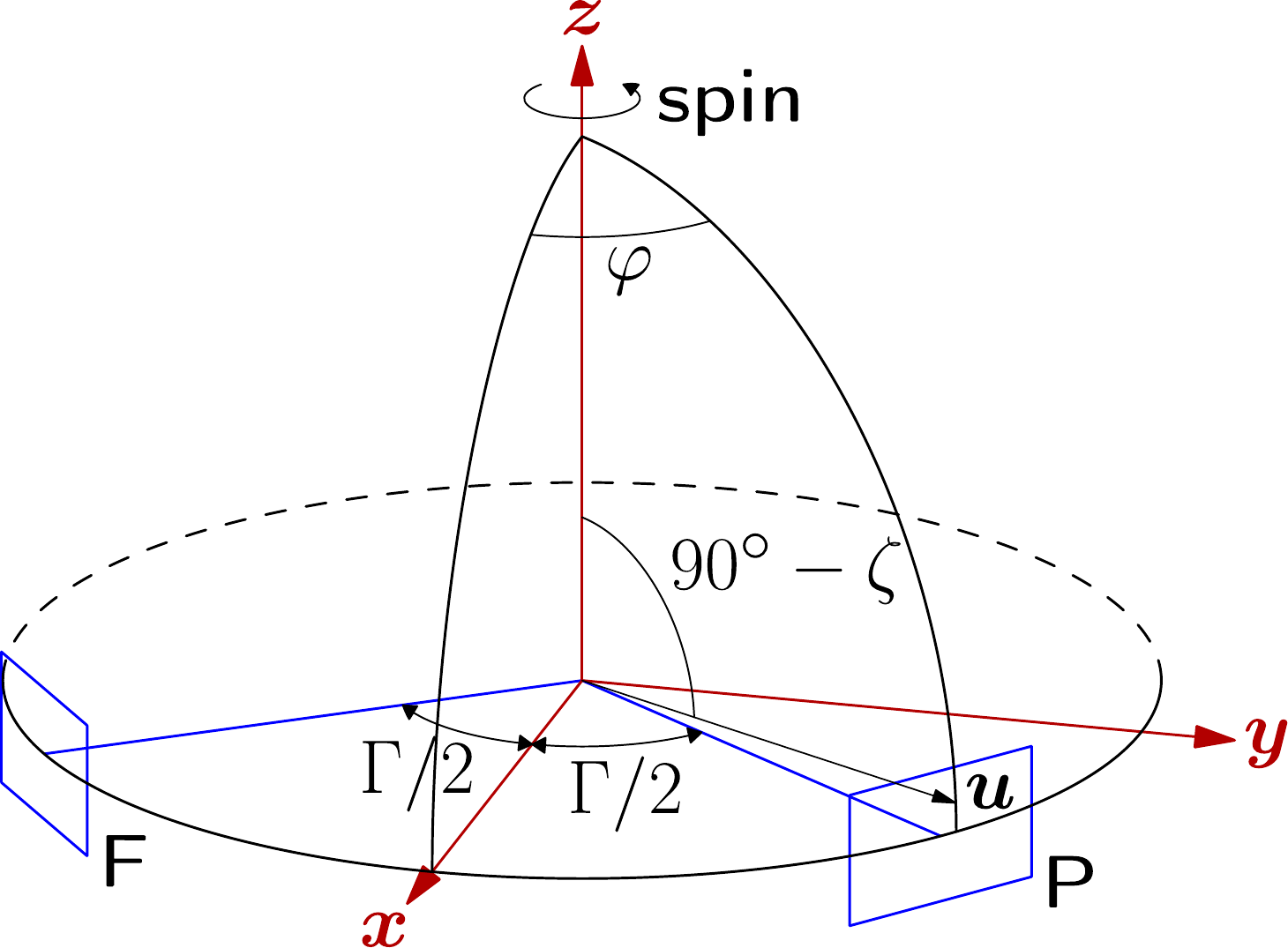}
    \end{minipage}
    \caption{The left panel shows schematically how the astrometric observables collected by {\gaia}, $t_\mathrm{obs}$ and
        $\mu_\mathrm{obs}$, are modelled in terms of the field angles, $\eta$ and $\zeta$. The latter describe the
        (proper) direction \vect{u} to a source as observed in the reference frame defined by the spacecraft axes
        \vect{x}, \vect{y}, \vect{z}, as shown in the right panel. For a source in the preceding field of view the
        angle $\eta$ is equal to $\varphi-\Gamma/2$ (with $-\pi\leq\varphi<\pi$) while in the following field of view
        $\eta=\varphi+\Gamma/2$. The angle $\zeta$ is defined as indicated in the figure. \textit{Credits: Right panel
    from \cite{Lindegren2012} (their Fig.~2), adapted with permission \copyright\ ESO.}}
    \label{fig:fa-agis}
\end{figure}

\begin{marginnote}
    \entry{Coordinate direction}{Unit vector corresponding to the difference between the spatial coordinates of two
    locations (space-time events, strictly speaking).}
    \entry{Barycentric coordinate direction}{The direction to a source as seen from the solar system barycentre, free
    from aberration and light bending effects.}
    \entry{Topocentric coordinate direction}{Source direction as seen from the observer, thus including the parallax
    shift, but free from aberration and light bending effects.}
    \entry{Proper direction}{The observable direction to a source as seen from the observer, including aberration
    and light bending effects.}
\end{marginnote}

Perhaps the easiest way to think about Equation~\ref{eq:minimization} is to consider the forward problem of
predicting where in the {\gaia} focal plane a source will be observed at a given moment in time. This involves four
steps which are schematically depicted in the left panel of Figure~\ref{fig:fa-agis}, which is a simplified version of
the schematic shown in Fig.~1 of \cite{Lindegren2012}:
\begin{enumerate}
    \item The source astrometric parameters \vect{s} determine the coordinate direction to the source at time $t$ as
        seen from the from the position of {\gaia} (i.e.\ the topocentric coordinate direction \vect{\bar{u}}), in an axis system
        co-moving with {\gaia} but aligned with the ICRS \citep[the so-called Centre of Mass Reference System or
        CoMRS,][]{Lindegren2012,Klioner2004}. The effects of proper motion, radial velocity, and parallax are accounted
        for. The calculation of the parallax shift requires knowledge of the orbit of {\gaia}.
    \item The source direction as seen from {\gaia} is affected by light bending by solar system objects and aberration
        due to {\gaia}'s motion. These effects are included in this step to calculate the proper direction \vect{u}
        to the source, still in the same axis system as above. The solar system ephemerides are required at this stage
        as well as the position and velocity of {\gaia}. A relativistic astrometry model is then applied to calculate
        the proper direction. The modelling is parametrized with a set of ``global'' parameters \vect{g}.
    \item The source direction as seen from {\gaia} is now translated to a reference frame defined by the spacecraft axes
        (Figure~\ref{fig:fa-agis}, right panel). This involves a pure rotation from the ICRS to the spacecraft frame,
        which describes the orientation, or attitude, of {\gaia}. The attitude must be modelled from the observations and
        is described with the parameters $\vect{a}$. After this step the proper direction \vect{u} is described in terms
        of the ``field angles'' $\eta$ and $\zeta$, which correspond to the along and across-scan directions
        (Figure~\ref{fig:fa-agis}, where $\eta_\mathrm{calc}$ and $\zeta_\mathrm{calc}$ are the predicted field angles). 
        \begin{marginnote}
            \entry{$\eta$}{Longitude-like angular coordinate of a source in a {\gaia} telescope field of view as defined in
            Figure~\ref{fig:fa-agis}.}
            \entry{$\zeta$}{Latitude-like angular coordinate of a source in a {\gaia} telescope field of view as defined in
            Figure~\ref{fig:fa-agis}.}
        \end{marginnote}
    \item The final step is to convert the field angles into the actual location of the source in the focal plane. This
        involves among others the optical projection from the telescope field of view to the focal plane, the precise
        locations and orientations of the detectors, and the properties of their pixel grids. The parameters \vect{c}
        describing this step are referred to as the ``geometric calibration'' parameters \citep{Lindegren2012}. In
        practice the geometric calibration is used to derive ``observed'' field angles $\eta_\mathrm{obs}$ and
        $\zeta_\mathrm{obs}$ from $t_\mathrm{obs}$ and $\mu_\mathrm{obs}$ (Figure~\ref{fig:fa-agis} and
        Section~\ref{sec:geocal}).
\end{enumerate}
When accounting for these steps Equation~\ref{eq:minimization} can be expanded by replacing the nuisance parameters
\vect{n} with the specific parameter sets \vect{a}, \vect{c}, \vect{g}. Before summarizing the approach to solving this
equation, some comments on the above elements are in order.

\subsubsection{Source model}
\label{sec:sourcemodel} 

The first step above involves the modelling of the position of the source in the Barycentric Celestial Reference System
\citep[e.g.,][]{Klioner2003}, i.e.\ relative to the solar system barycentre, which allows predicting its barycentric
coordinate direction. In the {\gaia} astrometric data processing all sources are treated as single objects moving with a
uniform space velocity relative to the solar system barycentre \citep[Eq.~3 in][]{Lindegren2012}. The source direction
\vect{\bar{u}} as seen from {\gaia} is calculated according to Eqs.~4--6 in \cite{Lindegren2012}. The topocentric
direction includes the parallax shift of the source as seen from {\gaia} which is given by the term
$\varpi\vect{b}_\mathrm{G}(t)/A_\mathrm{u}$ in Eq.~4 of \cite{Lindegren2012}, with $\vect{b}_\mathrm{G}(t)$ the
barycentric position of {\gaia} and $A_\mathrm{u}$ the astronomical unit. This implies the need to know the orbit of
{\gaia}. Solar system applications require the position to be known to better than 150~m such that main belt asteroid
position uncertainties are not affected by uncertainties in the position of {\gaia}. Bright star aberration corrections
require the velocity to be known to $2.5$~mm~s$^{-1}$ (Section~\ref{sec:grem}).

The above source model is strictly applicable only to single stars which are substantially smaller than 1 au in size.
The presence of stellar or sub-stellar companions requires introducing the orbital motion of the sources in the model,
while for very large (supergiant) stars the photocentre motion due to large convective cells in the atmosphere should be
accounted for \citep{Chiavassa2011, Chiavassa2018}. The latter complication is ignored in the {\gaia} data processing
and can lead to less reliable astrometry for physically large stars. The presence of non-single stars cannot be ignored,
but for determining the calibration parameters in the astrometric data processing (\vect{a}, \vect{c}, \vect{g}) only
sources for which there is no evidence that they have a companion are used. The processing of non-single stars with a
more complex source model is done in a separate, specialized process \citep[][Sect.~7]{gaiamission}.

\subsubsection{Relativistic astrometry model}
\label{sec:grem}

The topocentric direction \vect{\bar{u}} is transformed to the proper direction \vect{u} (cf.~Figure~\ref{fig:fa-agis})
using the prescriptions in \cite{Klioner2003}. To ensure that bright star astrometry is not affected by errors in the
relativistic modelling of the apparent source directions, the light-bending and aberration effects should be predictable
to $\sim1$~\muas\ accuracy. As illustrated in Table 1 in \cite{Klioner2003} this implies that the light bending effects
of the Sun and all the major planets should be accounted for, as well as the effect of a number of the moons and
asteroids in the solar system. This requires a precise knowledge of the ephemerides and masses of solar system bodies.
To compute the aberration effects at the $1$~\muas\ level the velocity of {\gaia} should be known to $2.5$~mm~s$^{-1}$.
The solar system and spacecraft ephemerides are taken as known (they are not solved for from the observations). For
\gdr{1} and \gdr{2} the INPOP10e solar system ephemeris was used \citep{Fienga2013}. The spacecraft orbit is obtained at
the required level of accuracy through a combination of spacecraft tracking at the {\gaia} Mission Operations Centre,
located at the European Space Operations Centre (ESOC) in Darmstadt, Germany, and optical observations of {\gaia} to
determine its position in the plane of the sky \citep{Altmann2014}. More detail can be found in \cite{gaiamission}.

The remaining aspects of the transformation from \vect{\bar{u}} to \vect{u} depend on a set of global parameters
\vect{g}, which for example include the parametrized post-Newtonian parameter $\gamma$ describing the strength of light
bending. The parameters \vect{g} are normally considered known (e.g., $\gamma=1$) in the {\gaia} data processing but can
in principle be solved for. This would allow tests of General Relativistic models, such as the determination of the
value of $\gamma$ from {\gaia} observations. In particular the scanning law of {\gaia} has been optimized such that
observations of bright stars near Jupiter are made a number of times during the mission. These data will allow an
attempt at measuring the quadrupole moment of light deflection by Jupiter. How {\gaia} astrometry can be used in these
and other tests of fundamental physics is discussed by \cite{Klioner2014}. Such tests will probably only be done in
later phases of {\gaia} data processing operations, when the other parameter sets in the astrometric solution are
sufficiently well known and understood.

For the \edr{3} astrometric solution the set of global parameters was extended with a spin related instrument distortion
model, aimed at suppressing parallax zero-point variations \citep{edr3astrometry}. This is discussed in
Section~\ref{sec:zonal}.

\subsubsection{Attitude model}
\label{sec:attcal} 

The attitude describes the orientation of the spacecraft frame (Figure~\ref{fig:fa-agis}) with respect to the CoMRS. The
set of parameters \vect{a} describes how the attitude varies as a function of time. Errors in the spacecraft attitude can
translate into additional noise or errors on the source astrometric parameters (Section~\ref{sec:systematics}). To
achieve the astrometric performance targeted for bright stars, the attitude reconstruction for {\gaia} should be
accurate to the $10$~\muas\ level. The details of the attitude modelling are given in \cite{Lindegren2012}.
\cite{Risquez2013} present detailed simulations of the {\gaia} attitude as it evolves under the influence of solar
radiation pressure torques, the noise in the micro-propulsion system, and the effects of micro-meteoroid hits and minute
structural changes in the spacecraft \citep[the micro-clanks described in][]{gaiamission}. They predict that the
attitude knowledge will be limited to the $20$~\muas\ uncertainty level, which would add an additional $7$~\muas\ rms to
the uncertainties of bright star parallaxes. As of \gdr{2} the attitude modelling has not yet reached this accuracy
level and it is one of the areas in which improvements are being made for future data releases.

\subsubsection{Geometric calibration}
\label{sec:geocal} 

The last step in the modelling of the astrometric observables is in practice done differently than implied in step 4
above \citep{Lindegren2012}. Rather than transforming the field angles to locations in the focal plane, the CCD pixels
are instead projected into the space of field angles through the geometric calibration model with parameters \vect{c}.
The source locations are measured as pixel coordinates (AL and AC, cf.~Figure~\ref{fig:gaiaobs}) in the data stream,
translated to the observation time $t_\mathrm{obs}$ and the continuous across-scan coordinate $\mu_\mathrm{obs}$.
Because of the TDI observation mode, the measured locations must be referred to some observation time halfway through
the time it takes to cross a CCD. The reference is taken to be a line of pixels halfway along the CCD, which is traced
out in field angle space as the ``observation line'' $[\eta(\mu), \zeta(\mu)]$.  Due to distortions in the optical
projection from telescope field of view to focal plane, and the details of the CCD locations and orientations, as well
as the pixel grids, the observation lines in general are not straight lines in $(\eta,\zeta)$, as depicted in
Figure~\ref{fig:gaiaobs} and in Fig.~4 of \cite{Lindegren2012}.

The geometric calibration model must accommodate effects on various spatial and time scales. \cite{Lindegren2012}
discuss a division into large-scale calibration effects, such as thermal variations in the optics, the detectors and the
mechanical support structures, and small-scale calibration effects, such as irregularities in the CCDs themselves. The
TDI observation mode leads to a very significant simplification of the geometric calibration model in that any details
in the CCD structure along scan (at the pixel level) are smeared out, such that only variations across-scan in the
observation lines need to be modelled (with $4500$ AL pixels this means a savings of a factor 4500 in calibration
parameters). The large-scale calibration effects are modelled over relatively short timescales with low order
polynomials in $\mu$. The small-scale effects are expected to be stable over longer timescales, but require detailed
spatial modelling at the level of a few AC pixels. This translates into differing demands on the amounts of observations
needed to calibrate a given effect. In particular it should be noted that the CCD gating, implemented to prevent the
saturation of bright sources, introduces difficulties in that each gate represents a separate ``instrument''. The
relative sparsity of bright sources makes the geometric calibration over short timescales more difficult for the
different gate settings \citep[a similar argument holds for the attitude calibration, as each gate sees a different
effective attitude,][]{BastianBiermann2005}.

\subsubsection{CCD pixel level data model}
\label{sec:pixmodel}

The time when a source was observed is derived by fitting the sample values of the image in the observation window
around a source (cf.\ Figure~\ref{fig:gaiaobs}) with a model that includes the total flux, local background level, and
the image location $\kappa$, a continuous representation of the AL pixel coordinates \citep{Lindegren2012}. The samples
are modelled with a point spread function (PSF), or line spread function (LSF) for the one-dimensional windows
transmitted for the majority of sources. The image location is translated to observation time through a known relation
between $t$ and $\kappa$. For a subset of sources the observation windows are two-dimensional which then also allows the
derivation of the AC location $\mu$ (using a PSF model in this case).

The PSF/LSF model includes optical as well as electronic effects on the shape of the source image, and should also
account for charge transfer inefficiency effects. The source colour, which one might have expected to be a part of the
source parameters, is accounted for at this level through the effect of colour on the details of the PSF/LSF shape.
Details on the PSF/LSF modelling as employed for \edr{3} can be found in \cite{Rowell2020}.

\subsubsection{Solving the minimization problem}
\label{sec:solvesacg}

With the above components of the observation modelling for {\gaia} in mind, the model problem in
Equation~\ref{eq:minimization} can now be written as:
\begin{equation}
    \min_{\vect{s},\vect{a}, \vect{c}, \vect{g}} \Vert\vect{f}^\mathrm{obs}(t_\mathrm{obs},\mu_\mathrm{obs}\mid\vect{c})
    - \vect{f}^\mathrm{calc}(t_\mathrm{obs} \mid\vect{s},\vect{a},\vect{g})\Vert_{\cal M}\,,
    \label{eq:minsacg}
\end{equation}
which is a simplified formulation of Eqs.~24--26 in \cite{Lindegren2012}. Solving this system of equations is done by
splitting the problem into a ``calibration'' part (solving for \vect{a}, \vect{c}, \vect{g}) and a ``source update''
part (solving for \vect{s}). The calibration part is solved from a subset of ``primary'' {\gaia} sources, for which
there is no evidence that they are non-single stars. Once the calibrations are solved the parameters of all {\gaia}
sources are updated, keeping \vect{a}, \vect{c}, and \vect{g} fixed. \cite{Lindegren2012} estimate that for $10^8$
primary sources $\sim5.5\times10^8$ parameters have to be solved from $\sim8\times10^{10}$ basic observables.
\cite{Bombrun2010} showed that for these numbers a direct solution of the minimization problem was many orders of
magnitude beyond computing capabilities in 2010. This is mainly due to the attitude and geometric calibration models
linking together sources observed in the same field of view or at the same time in the two fields of view, leading to
cross-terms between sources in the normal equations matrix over scales of $\sim 0.7^\circ$ and $\sim106.5^\circ$.

The solution chosen for the {\gaia} astrometric processing is the so-called Astrometric Global Iterative Solution
\citep[AGIS,][]{Lindegren2012}. The minimization problem in Equation~\ref{eq:minsacg} is tackled by iterating four
partial solutions until convergence is achieved. Three of the four parameter sets are fixed and one solves for the
remaining set. For example, assuming the source parameters are approximately known (from a starting catalogue or a
previous iteration) together with the geometric calibration parameters (e.g., derived from the nominal instrument
parameters), and taking the global parameters as given, one can solve for the attitude parameters. In subsequent
iterations one solves for \vect{c}, \vect{g}, and \vect{s}, until convergence is achieved. Upon convergence, the
parameters \vect{a}, \vect{c}, \vect{g} as determined from the observations of the primary sources are then fixed and
used to solve for the astrometric parameters of all sources. 

Obviously this summary skips over many practical complications that require considerable and continuing efforts to
solve. The details can be found in \cite{Lindegren2012}, with specific updates and improvements that were implemented
for \gdr{1} and \gdr{2} described in \cite{Lindegren2016} and \cite{Lindegren2018}.

\subsection{Systematic errors in {\gaia} astrometry}
\label{sec:systematics}

The scientific results from the analysis of the first two {\gaia} data releases very convincingly demonstrate the
exquisite quality of the astrometric data, where in particular \gdr{2} represents a major jump in the availability of
high accuracy parallaxes and proper motions for large numbers of stars all over the sky. Nevertheless, {\gaia}
astrometry is not and was not expected to be perfect and a number of works have demonstrated the presence of systematic
errors in the parallaxes and proper motions, in particular the catalogue validation papers accompanying the data
releases \citep{Arenou2017, Arenou2018}. The systematic errors are very small, the global zero-point offset for \gdr{2}
parallaxes being $30$--$80$~\muas.  Nevertheless the systematics cannot be ignored. At the bright end of the {\gaia}
survey the systematic errors should remain well below the random errors, which are at the tens of \muas\ level in
{\gaia} DR2 for stars brighter than magnitude $12$--$14$, to ensure the correct interpretation of, for example, subtle
features in a Hertzsprung-Russell diagram. At the faint end of the survey there is the potential to measure a precise
parallax for the Magellanic clouds by averaging over millions of stars, however this is currently limited by the
systematic errors, as the parallaxes of these galaxies are in the $15$--$20$~\muas\ range. The {\gaia} celestial
positions and proper motions also exhibit systematic errors, in the form of regional patterns on the sky \citep[such as
seen in the maps of the astrometric parameters of quasars, see Fig.~11 in][]{gcrf2}, and potentially in the form of a
global misalignment and spin of the {\gaia} celestial reference frame with respect to the ICRF. In the following
subsections the various types of systematic errors in {\gaia} astrometric data will be discussed, ending with a short
discussion on how to characterize the systematics.

\subsubsection{Global systematic errors: reference frame}
\label{sec:syserrframe}

The global astrometry concept ultimately rests on building a network of angles measured between sources on the
sky. The angle between two sources on the celestial sphere is invariant with respect to the orientation or spin (change
of orientation over time) of the sphere. Specifically for a Hipparcos/{\gaia}-type mission this means that an overall
change in the positions of sources due to a change of the orientation of the celestial sphere is completely degenerate
with a change in the spacecraft attitude \citep{Butkevich2017}. Hence, special measures have to be taken to align the
orientation of the resulting reference frame to the ICRF, and to ensure that it does not exhibit an average spin with
respect to the distant quasars. The principles of the construction of the celestial reference frame for \gdr{2} (the
{\gaia}-CRF2) are described in \cite{Lindegren2012,Lindegren2018} and its properties are described in \cite{gcrf2}. At
$G=19$ the alignment of {\gaia}-CRF2 to the ICRF is believed to be accurate to $20$~\muas, while
the frame is non-rotating at the $<20$~\muaspm\ level \citep{dr2paper,gcrf2}. The alignment of the {\gaia} reference frame
and ensuring it shows no average spin, are done with a specially selected set of quasars \citep{Lindegren2018, gcrf2}
which are typically faint ($G>19$) and have bluer colours than most stellar sources in the {\gaia} catalogue. Although the
astrometric data processing for {\gaia} is set up to ensure a consistent solution across all source brightness and colour
ranges, there are numerous factors that can cause inconsistencies between the reference frame at the faint and bright
ends of the {\gaia} survey, in particular the abrupt change in the instrument configuration at $G=13$, where 2D instead of
1D observation windows are used and where measures against CCD saturation start to take effect \citep{Lindegren2020a}. This
is shown in Fig.~4 of \cite{Lindegren2018} where the bright stars (at $G<13$) show an average spin at the $0.1$~\maspm\
level compared to the quasars. The \gdr{2} bright source reference frame was further investigated by
\cite{Lindegren2020a,Lindegren2020b} using VLBI observations of radio stars. The {\gaia} and VLBI astrometry of the radio
stars were used to jointly estimate the orientation and spin parameters of the \gdr{2} bright star reference frame,
where the assumption was made that the VLBI astrometry is on the (non-rotating) ICRF. The earlier estimates of the spin
were confirmed, while the value of VLBI observations of radio stars in characterizing the bright {\gaia} reference was
demonstrated. It is stressed by \cite{Lindegren2020a} that publications on VLBI astrometry should always provide the
full astrometric solution as well as the epoch for the solution, where especially the celestial positions are of great
historical value in reference frame investigations.

\subsubsection{Global systematic errors: parallax zero-point}
\label{sec:plxzp} 

A second type of systematic that affects all sources equally is that of the global parallax zero-point. As explained in
Section~\ref{sec:spaceastrometry} the Hipparcos/{\gaia} concept enables the measurement of absolute parallaxes which means
that in principle for a collection sources located at sufficiently large distances (such as quasars) the average
measured parallax should be zero. It is well known that this is not the case for \gdr{2} \citep{Lindegren2018}, the
quasar parallaxes being $-29$~\muas\ on average, which by definition is the parallax zero-point. In addition the
zero-point is known to exhibit variations as a function of (at least) source brightness, colour, and position on the
sky.  This was found both during the catalogue validation and in its first scientific use (see for example Figs.~12 and
13 in \cite{Arenou2018} and Fig.~16 in \cite{dr2dwarfsgcs} for examples of position-dependent parallax zero-point
variations), and in a number of subsequent investigations in which the parallax zero-point was estimated for different
samples of stars \citep[see for example][and references therein]{ChanBovy2020}. The causes for the zero-point variations
will be discussed in the next section, here we focus on the global parallax zero-point.

\begin{figure}[t]
    \centering
    \begin{minipage}[b]{0.48\textwidth}
        \includegraphics[width=\textwidth]{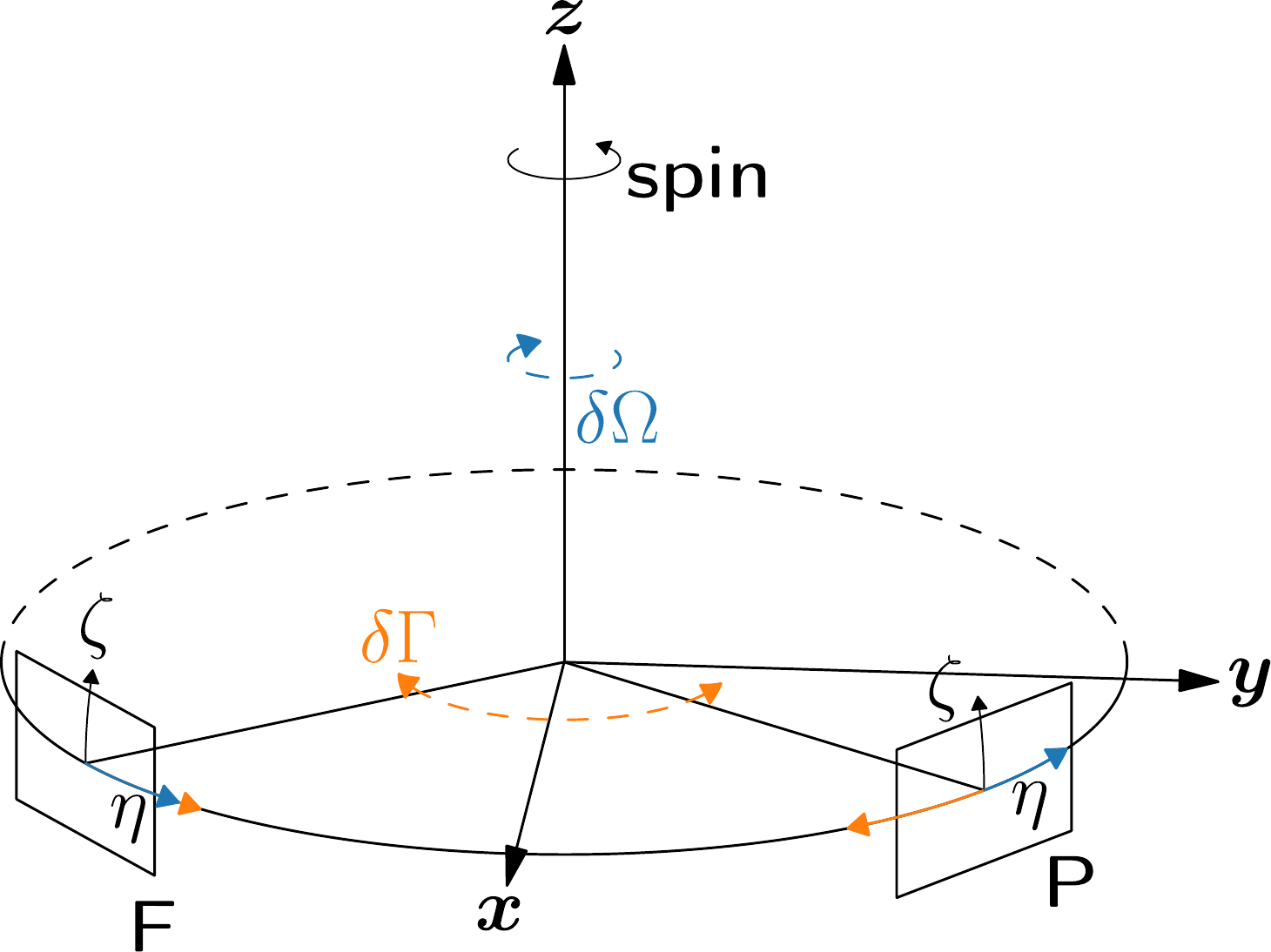}
    \end{minipage}
    \hfil
    \begin{minipage}[b]{0.48\textwidth}
        \includegraphics[width=\textwidth]{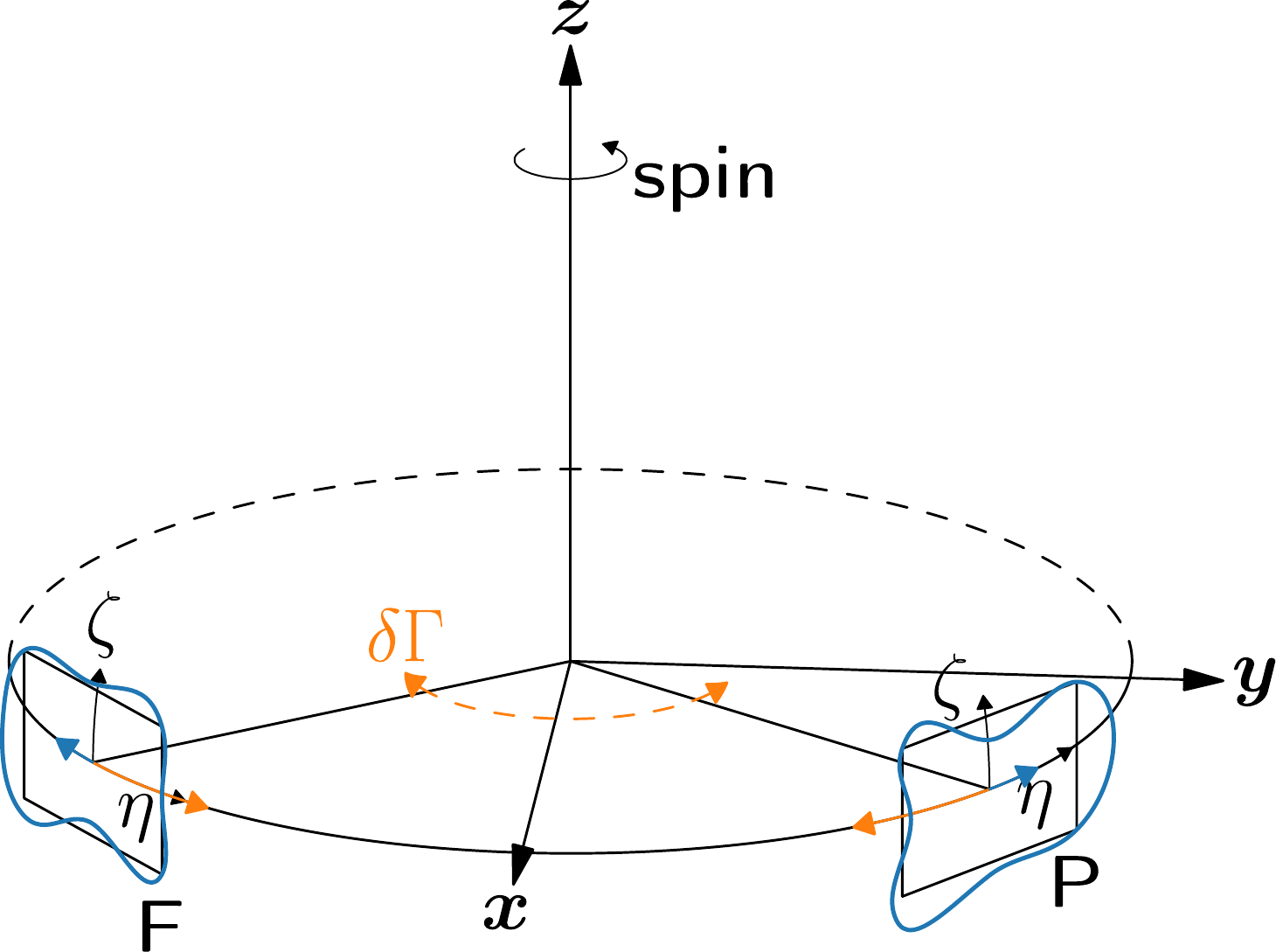}
    \end{minipage}
    \caption{Illustrations of the degeneracy between source astrometric parameter changes and combinations of changes in
        the basic angle, spacecraft attitude, and geometric calibration. The left panel shows how an increase in the
        basic angle combined with a negative rotation of the spacecraft around \vect{z} leads to changes in the apparent
        along-scan positions of a source (orange and blue vectors respectively). The right panel shows the same for a
        combination of an increase in the basic angle and changes in the focal plane projection on the sky (geometric
    calibration). \textit{Credits: Adapted with permission from figures in an unpublished technical note by S.~Klioner.}}
    \label{fig:bavdegeneracies}
\end{figure}

The astrometric parameters of a given source can be estimated thanks to the along and across-scan shifts in its
position, measured at a series of epochs, with respect to the position expected at some reference epoch, normally the
middle of the time interval covered by the mission. This is illustrated for the parallax in Figure~\ref{fig:alplxshift}.
For the rest of this discussion we focus on the AL shifts which carry most of the astrometric information. Shifts of
observed along-scan source positions can be caused by other effects, such as physical changes in the spacecraft or the
scientific instruments. Figure~\ref{fig:bavdegeneracies} illustrates three physical effects that can cause changes in
the measured AL positions of sources. The left panel shows how an increase in the basic angle $\delta\Gamma$ leads to an
AL shift of the source position to smaller values of $\eta$ in the preceding field of view and to larger values in the
following field of view. The same panel also shows the effect of a small change in the spacecraft attitude caused by a
small rotation $\delta\Omega$ around the $z$-axis, here in the sense opposite to the spacecraft spin. This leads to an
increase in $\eta$ in both fields of view. The right panel in Figure~\ref{fig:bavdegeneracies} shows how a change in the
details of the projection of the focal plane on the sky (due to evolving optics or thermoelastic changes in the focal
plane structure, for example) can lead to effects in $\eta$ very similar to a basic angle increase. Errors in the
spacecraft velocity components can also produce changes in the measured AL shifts, in this case due to apparent
aberration effects moving the sources toward the direction of motion.

Comparing Figure~\ref{fig:bavdegeneracies} to Figure~\ref{fig:alplxshift} qualitatively demonstrates how certain
physical changes in the spacecraft and instruments cause apparent shifts in source positions as measured by {\gaia}. These
apparent shifts could be confused for a real parallax effect and thus lead to systematic errors in the estimated
parallaxes. This is developed rigorously for the case of basic angle and attitude changes by \cite{Butkevich2017}. They
show how certain combinations of basic angle and attitude changes are indistinguishable from a global parallax offset
$\delta\varpi$ applied to all observed sources. One can develop the variations of the basic angle $\delta\Gamma$ as a
general harmonic series in the spin phase $\Omega$ of the spacecraft with respect to the direction to the solar system
barycentre (see Figure~\ref{fig:alplxshift}):
\begin{equation}
    \label{eq:bavharmonic}
    \delta\Gamma = \sum_{k\geq0} a_k\cos k\Omega + b_k\sin k\Omega\,,
\end{equation}
\cite{Butkevich2017} show that the $\cos\Omega$ component is completely degenerate with a global parallax offset
\citep[as well as specific attitude changes, see Eq.~17 in][]{Butkevich2017}. The prediction is that in the presence of
basic angle variations synchronous with the spacecraft spin period, the global parallax zero-point would amount to
$\delta\varpi\sim0.87a_1$.

This degeneracy between basic angle variations and the parallax zero-point was already known in Hipparcos times
\citep{Butkevich2017}. A major concern for both Hipparcos and {\gaia} was that the illumination of spacecraft by the sun
also varies with the spin period (due to lack of complete symmetry of the spacecraft structures about the spin axis),
which could thus lead to spin-synchronous basic angle variations and hence parallax zero-point offsets. To ensure the
targeted parallax accuracies for {\gaia} at the bright end, the basic angle was thus required to be stable to the few
$\muas$ level over the spacecraft spin period \citep{gaiamission}. This was not achieved in practice, with the actual
basic angle fluctuations being much larger at the $\sim1$~mas level\footnote{Note that mas-level basic angle
fluctuations amount to only tens of nm sized movements on the physical scale of {\gaia}.}, including a prominent 6-h period
component which is strongly coupled to the spin phase $\Omega$ of the spacecraft \citep{Lindegren2016}. If left
uncorrected these basic angle variations would thus lead to a $\sim0.87$~mas parallax zero-point for the {\gaia} data
releases. Because it was anticipated that maintaining the basic angle stable at a few \muas\ would be extremely
challenging, a metrology device capable of measuring the basic angle changes was developed and prototyped early on in
the {\gaia} project \citep{Meijer2009, Gielesen2013, Mora2014}. This so-called Basic Angle Monitor on board {\gaia} can
measure the basic angle changes to a few \muas\ precision over short timescales of 10--15 min \citep{gaiamission}.
This information on the basic angle fluctuations was used in the astrometric data processing for both \gdr{1} and DR2 to
correct for the effects on the parallax estimation \citep{Lindegren2016,Lindegren2018}. Despite the corrections based on
the Basic Angle Monitor an overall parallax zero-point nevertheless remains in the first two {\gaia} data releases. This
is partly because the monitor measures the basic angle variations as seen from a specific part of the {\gaia} focal
plane \citep[see Fig.~4 in][]{gaiamission}, which means that the metrology data is not fully representative of the
effective basic angle variations seen over the rest of the focal plane.

\cite{Butkevich2017} point out that the degeneracy between parallax and the 6-h basic angle variations is not perfect.
Three factors contribute to slightly breaking the degeneracy: the finite size of the telescope fields of view causes
slight variations in the parallax shift that can be distinguished from basic angle variation effects; the Sun as a
driver of the basic angle fluctuations leads to a basic angle varying with the inverse square distance between {\gaia} and
the Sun (which changes over the course of a year); for the same reason the basic angle varies not with the spin phase of
the spacecraft with respect to the direction to the solar system barycentre, but with the spin-phase referred to the
slightly different direction to the sun. Thus in principle there is a handle in the {\gaia} measurements to self-calibrate
the effect of the $\cos\Omega$ component of the basic angle fluctuations and remove the parallax zero-point offset. For
reasons discussed in the next section this is very non-trivial and is an area of active research and development for
future {\gaia} data releases. Finally, it should be stressed that basic angle fluctuations on timescales longer than the
6-h spin period can be calibrated as part of the astrometric data processing and are included in the geometric
calibration parameters \vect{c} from Section~\ref{sec:dataprocessing}.

\subsubsection{Zonal errors}
\label{sec:zonal} 

\cite{Butkevich2017} note that the basic data processing approach for {\gaia}, as captured in Equation~\ref{eq:minsacg},
contains many potential degeneracies, of which they discuss in detail the degeneracy between \vect{s} (which includes
the parallax), attitude \vect{a}, and the basic angle component of \vect{c}. In principle degeneracies between
parameters in the astrometric solution are not a problem as long they concern nuisance parameters only. For example a
degeneracy between subsets of \vect{a} and \vect{c} would lead to errors in the interpretation of the attitude and
geometric calibration but this would be harmless for the parameters \vect{s} of interest. However in practice there are
degeneracies between all four parameter groups, meaning in particular that changes in combinations of \vect{a},
\vect{c}, or \vect{g} may be misinterpreted as changes in \vect{s}, where it should be stressed that positions and
proper motions may also be affected. The effects on \vect{s} of these complex degeneracies are imprinted on the
celestial sphere after filtering through the {\gaia} scanning law and this could be an explanation for the regular
patterns seen in, for example, the median parallaxes and proper motions of stars in the LMC \citep{Arenou2018,
dr2dwarfsgcs}. The way this works can qualitatively be understood as follows\footnote{Based on a detailed elaboration in
unpublished DPAC technical notes by T.~Hilger and S.~Klioner.}. The attitude and geometric calibration of the spacecraft
both vary as a function of time. The attitude is continuously forced to vary as part of the scanning of the sky (and can
undergo short-timescale changes due to micro-meteoroid hits for example), while the geometric calibration varies as the
optics evolve and the {\gaia} instruments undergo thermoelastic changes, for example driven by solar illumination
changes and instrument aging. Now suppose that particular combinations of errors in \vect{a} and \vect{c} are
indistinguishable from certain changes to \vect{s}. If these errors in \vect{a} and \vect{c} evolve over time, the scan
law will transform these to changes in \vect{s} as a function of position on the celestial sphere, which could thus lead
to an imprint of specific systematic error patterns. This also implies that these patterns will occur all over the sky.
In practice they are easily visible only in areas such as the LMC due to the high density of stars with similar
parallaxes and proper motions. The size scale of these patterns can be expected to be of the order of the {\gaia} field
of view AC ($0.7^\circ$) but could be as small as the AC height of a CCD projected on the sky ($\sim0.1^\circ$).

A detailed understanding of how degeneracies between \vect{s}, \vect{a}, \vect{c}, and \vect{g} translate to zonal
errors may lead to specific strategies to handle the calibrations within AGIS in such a way as to suppress or mitigate
these errors. This has been the focus of a considerable research effort within the DPAC over the past decade. For the
\edr{3} astrometric solution a first version of a model that can deal with spin-related instrument distortions on short
timescales (less than a few {\gaia} spin periods) has been successfully implemented. This model, the parameters of which
are included in \vect{g}, is summarized in \cite{edr3astrometry}, who show that indeed the zonal parallax and proper
motion errors are significantly reduced compared to \gdr{2}. By capitalizing on the above mention effects that remove
the complete degeneracy between basic angle variation and the parallax zero-point, the model was also used to correct
the parallax zero-point to a value closer to zero for \edr{3}.

\subsubsection{Brightness and colour-dependent systematic errors}
\label{sec:mcdependencies} 

The systematic errors (as well as the uncertainties) in the {\gaia} astrometry show clear dependencies on source apparent
magnitude and colour \citep{Arenou2018, Lindegren2018}. The magnitude dependencies are driven by the instrument
configuration changes as a function of source brightness \citep{gaiamission}. For sources brighter than $G=13$
two-dimensional observation windows are used in combination with CCD gates to prevent image saturation. For sources
fainter than $G=13$ one-dimensional windows are used, where at $G=16$ a switch to shorter 1D windows occurs. By using
gates the integration time is reduced for bright stars (see the box on  \href{sec:gaiabox}{\gaia}). This means that a
slightly different effective spacecraft attitude is seen for these observations \citep{BastianBiermann2005}, but also
that the PSF and geometric calibrations are slightly different. For the 2D windows a full PSF model is employed in the
image location, while for the 1D windows a LSF is used. Hence each window/gate combination effectively introduces a new
``instrument'' to calibrate. The number of such ``instruments'' grows in particular at the bright end where the number
of calibration sources becomes smaller and more sparsely distributed on the sky. This makes the estimation of \vect{a}
and \vect{c} more vulnerable to systematic errors. In addition the high signal to noise ratio in the source images makes
the image location process more sensitive to PSF modelling deficiencies. The changes in the instrument configuration
related to gates and observation windows are clearly visible in the astrometric uncertainty distribution shown in
\cite{Lindegren2018} and also in the photometric uncertainties shown in \cite{Evans2018}. In both cases the most
prominent effect is that for sources brighter than $G=13$ there is a clear ``calibration floor'' in the uncertainties,
where systematic errors dominate.

The colour dependencies are introduced to a large extent by deficiencies in the PSF and LSF models. The precise shape of
a source image in the focal plane depends also on its colour and this must be taken into account both in the calibration
of the PSF/LSF models and in their application, when estimating the image location to obtain $t_\mathrm{obs}$ and
$\mu_\mathrm{obs}$ \citep{Rowell2020}. For \gdr{1} and DR2 the PSF/LSF models were assumed to be independent of colour
and time, leading to systematically biased image locations for sources that did not have a colour matching the effective
colour of the PSF/LSF model. This was handled for {\gaia} DR2 by adding terms in the astrometric solution that can model
the image location shifts as a function of time, colour, and magnitude \citep{Lindegren2018}, but this cannot completely
prevent the occurrence of colour-dependent systematics. The assumption of a PSF/LSF that is constant in time is a poor
approximation to the actual evolution of the {\gaia} image characteristics, driven by thermal relaxation of the payload as
well as the effects of contamination early in the mission \citep{gaiamission}. Thus the image location systematics will
also evolve with time which can lead to further zonal effects on the sky after filtering through the scan law.
There are numerous other effects that complicate the image location process and introduce further systematic errors.
These include the estimation of the local sky background, knowledge of the source colours (which depends on the
photometric data processing for {\gaia}), and charge transfer inefficiency effects on the PSF/LSF (which can be ignored for
the early {\gaia} data releases, but grow over time). See \cite{Rowell2020} for more details on the PSF/LSF modelling.

The challenge for upcoming {\gaia} data releases is to suppress as much as possible the systematic errors (global and
zonal), such that they will not start dominating the overall accuracy budget of the increasingly precise {\gaia} data
(thanks to a longer period of measurement). In particular in the areas of self-calibration of the basic angle
variations, the removal of zonal errors, and the improved modelling of the PSF/LSF, major efforts are being undertaken
within the DPAC. Significant improvements have already been achieved for \edr{3} as described in
\cite{edr3astrometry}.

\subsubsection{Characterization of systematic errors}
\label{sec:syserrchar} 

The unprecedented quality of the {\gaia} data releases as well as their all-sky coverage at high angular resolution make
the assessment of the systematic errors a very challenging task. The characterization of the systematic errors in the
{\gaia} astrometry is done at several levels during the data processing, both internally to the astrometric solution
\citep{Lindegren2016, Lindegren2018}, and at the level of the catalogue validation \citep{Arenou2017,Arenou2018}. Good
internal checks on systematic errors are the comparisons of astrometric solutions constructed for different subsets of
the observations (to get a handle on the variance in the zonal errors) and the examination of the parallaxes and proper
motions derived for quasars (to understand the uncertainty statistics, the global parallax zero-point, and the reference
frame quality). External checks are harder to come by as there is only very limited astrometric data that is of
comparable quality to {\gaia}. Nevertheless numerous studies have been undertaken to estimate the parallax zero-point for
different subsets of sources where one can have confidence in independent distance estimates \citep[see examples
in][]{Arenou2018}, or where a good understanding of the astrophysics of the source samples allows for joint estimations
of the parallax zero-point and intrinsic source properties. Examples include the use of Cepheid and RR Lyrae variables
\citep[e.g.,][]{Sesar2017, Riess2018} and Red Clump stars \citep[possibly in combination with asteroseismological data,
e.g.,][]{ChanBovy2020, Hall2019, Khan2019, Zinn2019}. These studies have confirmed the magnitude, colour and, sky
position dependencies of the parallax zero-point.  The complex and difficult-to-characterize variations in the
zero-point are the main motivation to leave the astrometry results from the {\gaia} data processing uncorrected upon
publication. 

Ultimately the insights into the systematic errors in the {\gaia} data releases will be driven by the continued
scientific exploration of the data, demanding consistency between the data and the astrophysical picture of the Milky
Way. A specific example is the study by \cite{Schoenrich2019}, which explicitly demands consistency between the
astrometric data and the spatial and velocity distribution of the stars in the subset of \gdr{2} for which radial
velocities were measured. They estimated a parallax zero-point consistent with other estimates for the bright end of
\gdr{2}, and identified an unexplained correlation between the parallax zero-point offset and the parallax uncertainty.

\section{GAIA SCIENCE HIGHLIGHTS}
\label{sec:highlights}

\subsection{Revolutionizing science from the solar system to the distant universe}
\label{sec:usecases}

The publication of \gdr{1} \citep{dr1paper} marked the start of a new era in the availability of fundamental
astronomical data. Parallaxes and proper motions were provided for 2 million sources, reaching sub-milliarcsecond
accuracies and providing a first taste of absolute microarcsecond astrometry in the optical for the proper motions of
the subset of Hipparcos stars ($\sim60$~\muaspm\ average uncertainties). In addition an unprecedented map of the sky was
provided, with positions at the milliarcsecond accuracy level for $1.1$ billion sources out to 20th magnitude. This
vastly improved astrometric reference was used to anchor other surveys. This led to the production of large and accurate
proper motion catalogues by combining \gdr{1} and older sky surveys
\citep[e.g.][]{Altmann2017,Deason2017,Tian2017,Zacharias2017}. However, \gdr{2} \citep{dr2paper} represents a more
profound transformation of astronomy, providing: a high precision sky map for $1.7$ billion sources at an angular
resolution significantly exceeding that of ground-based large-area optical sky surveys; microarcsecond to milliarcsecond
quality parallaxes and proper motions for $1.3$ billion sources out to the survey limit at $G\sim21$
\citep{Lindegren2018}; the first realization of the optical celestial reference frame constructed on the basis of
extragalactic sources only \citep{gcrf2}; an all sky homogeneous multi-colour photometric survey \citep{Evans2018,
Riello2018}; and a large radial velocity survey of some 7 million sources out to $G\sim12$ \citep{Katz2019,
Sartoretti2018}. In addition \gdr{2} provides light curves and properties for $550\,000$ variable stars
\citep{Holl2018,dr2variables}, astrophysical parameters (effective temperature, extinction, radius, and luminosity) for
some 160 million sources \citep{Andrae2018}, and a first catalogue of astrometry of minor bodies in the solar system
\citep{Spoto2018}.

As of September 2020 the papers describing \gdr{1} and DR2 collectively were cited $\sim6400$ times\footnote{As counted
on September 11 2020 using \url{https://ui.adsabs.harvard.edu/}.}, which gives a first impression of the enormous impact
of the {\gaia} data releases. The data is being used in many different ways, ranging from looking up the parallax,
proper motion, photometry, or radial velocity for a specific source, to analyses of large subsets of the {\gaia} data,
or even the full data set. The ready availability of fundamental astronomical data for $1.7$ billion sources has quickly
made \gdr{2} and indispensable part of astronomical research today, to the point of making it hard to imagine a time
when this data was not available. In the following subsections a highly selected overview of science results from
{\gaia} will be given, starting in the solar system and going at all the way to the distant quasars. Early science
highlights from \gdr{1} can be found in \cite{Brown2018} and the corresponding proceedings of the first IAU Symposium on
{\gaia} results held in 2017 \citep{RecioBlanco2018}. 

Two decades ago \cite{FreemanBlandHawthorn2002} introduced the field of Galactic Archaeology, the unravelling of the
formation history of the Milky Way by using the combination of accurate phase space and chemical composition data for
vast numbers of stars. They predicted that the field would be revolutionized by the {\gaia} mission data. We have now
arrived at this point, as demonstrated in the review by \cite{Helmi2020} on the insights provided by {\gaia} on the
early evolution of the Milky Way.

\subsubsection{Solar system}
\label{sec:solsys}

{\gaia} autonomously decides which sources to observe, in principle the only criteria being that a source is brighter
than the survey limit ($G\sim20.7$) and point-like \citep{deBruijne2015, gaiamission}. Minor bodies in the solar system,
such as main belt asteroids, near-Earth objects, and Kuiper-belt objects (referred to as ``solar system objects'' in the
following), move with median speeds of about 7~mas~s$^{-1}$ relative to stars in {\gaia}'s along-scan direction
\citep{TangaMignard2012}. During the detection stage the images of these objects will thus be smeared slightly (by a
fraction of a pixel) which in general leaves them sufficiently point-like that they will be measured by {\gaia}.
Extrapolating from the asteroid population known at the time, \cite{TangaMignard2012} estimated that about $350\,000$
solar system objects will be observed by {\gaia}, including a few moons of major planets. {\gaia}'s exquisite astrometry
combined with its spectrophotometric data will revolutionize studies of solar system objects through the combination of
high precision orbital and taxonomic classifications \citep{Delbo2012}.  \gdr{2} provides a foretaste in the form of
astrometry and $G$-band photometry for a sample of about $14\,000$ solar system objects. As shown in \cite{Spoto2018},
the astrometry is already at the level where uncertainties in the individual position measurements range from 1 to
10~mas. This represents an improvement by two orders of magnitude over the average astrometry available in the archives
\citep[][find average residuals of $\sim400$~mas for orbital fitting of the (relative) CCD astrometry collected by Minor
Planet Center]{Desmars2013} . For a subset of the solar system objects just the 22 months of data collected for \gdr{2}
enable estimates of the orbital parameters comparable in quality to what can be obtained from ground based data
collected over many decades \citep[Fig.~32 in][]{Spoto2018}.

However, the full exploitation of the {\gaia} epoch astrometry of asteroids is possible only by coupling it to the
available archive data. This is strictly required to measure subtle secular effects, important to understanding the
dynamical evolution of the asteroid belt, such as the Yarkovsky acceleration due to the recoil of the emission of
thermal photons. The large disparity in accuracy between space- and ground-based astrometry necessitated implementation
of a very sophisticated weighting scheme \citep{Spoto2019}, and a totally new approach to mitigating the effect of zonal
and regional errors \citep{Tanga2020} in pre-{\gaia} stellar catalogues. 

The main impact of \gdr{1} and \gdr{2} on solar system science has been the availability of a high accuracy all sky map
of star positions, parallaxes, and proper motions. This enables much more precise predictions of occultations of stars
by solar system objects \citep{Desmars2019a}, greatly enhancing the success rate and scientific value of occultation
campaigns. Orbits of solar system objects are better determined by recalibrating the existing astrometry against the
{\gaia} catalogue. \cite{Spoto2017} showed that orbit predictions from observations made over short time intervals can
be much improved already with \gdr{1}, increasing the reliability of hazard predictions for near-Earth objects. The
quality of the astrometry derived from past occultations of {\gaia} stars improves to a level that several ($\sim$10)
good occultation observations are by themselves sufficient for orbit determinations with accuracies comparable to those
based on decades of existing measurements.

\cite{Desmars2019b} fitted light curve data from past occultation campaigns to derive precise astrometric positions of
Pluto with respect to the \gdr{2} positions of the occulted stars. The resulting improved ephemeris for Pluto is precise
to the milliarcsecond level for the period 2000--2020 \citep[a precision not attainable with direct astrometry of Pluto
itself,][]{Desmars2019b}, which allows for better predictions of future occultations. The light curve modelling must
account for refraction effects in Pluto's atmosphere, from which one can infer properties of the atmosphere (temperature
and pressure profiles). \cite{Meza2019} used occultation data from 2002--2016 to infer an increase in Pluto's
atmospheric pressure over the period 1988--2016. This finding is based on a model that simulates volatile (N$_2$) cycles
over seasonal and astronomical timescales, and takes into account topography data from the New Horizons mission,
including the location of large ice reservoirs. The pressure increase is explained by the enhanced heating of nitrogen
ice at the onset of Pluto's northern spring. The pressure was predicted to peak within the next few years and then drop
as the solar insolation decreases (due to Pluto's increasing distance to the sun) and nitrogen condenses onto Pluto's
colder southern regions. \cite{Arimatsu2020} indeed present evidence for a drop in pressure based on an occultation
observation from 2019.

The prediction of Pluto's shadow trajectory on Earth for the July 19 2016 occultation was improved thanks to the
pre-release of the \gdr{1} position of one star. \gdr{2} astrometry was pre-released to support the observations of the
occultation of a star by Triton on October 5 2017 \citep{MarquesOliveira2018}, leading to successful observations of the
central flash in the light curve. The latter enables studies of the atmospheric temperature profile of Triton down to
low altitudes \citep{MarquesOliveira2019}.

Another exciting application concerns the observations of occultations by the Centaur 10199 Chariklo, the only known
minor planet with rings. By combining the observed occultation of 5 stars in \gdr{1}, \citet{Leiva2017} were able to
derive plausible shape models for Chariklo, along with precise volume and density estimates. The accuracy of the
occultations also provides a hint of topographic features of a size comparable to those of the icy moons of Saturn.
Based on the total derived mass of the object, the rings appear to fall in the 3:1 mean motion resonance with respect to
the rotation period of Chariklo.

The above offers a dramatic illustration of the power of an accurate sky map to faint magnitudes, which enables the
monitoring of the atmospheric evolution of small bodies at billions of kilometers from Earth. A possibly even more
impressive illustration is the support that was offered to the extended New Horizons mission which was to visit the
Kuiper Belt object Arrokoth after its visit to Pluto. Pre-release \gdr{2} astrometry was provided to the New Horizons
team in order to refine the orbit determination for Arrokoth, through a more accurate astrometric calibration of HST
imaging of this object \citep[for which parallaxes and proper motions were required,][]{Buie2020, Porter2018}. The
improved orbit was essential to the navigation of the New Horizons spacecraft. In addition the pre-release astrometry
was used to plan the observation of four stellar occultations by Arrokoth. Analysis of the occultation data showed that
Arrokoth is a contact binary with a 10\% albedo (compare Fig.~10 of \citet{Buie2020} to Fig.~1 of \citet{Stern2019}),
and allowed to further refine its astrometry and orbit. The results were used in the detailed planning of the Arrokoth
flyby. \cite{Buie2020} and \cite{Porter2018} emphasize that the availability of \gdr{2} astrometry enables much more
precise predictions of stellar occultations by Kuiper Belt objects, well in advance of the events so that large-scale
observation campaigns can be organized. This opens up many opportunities for obtaining shape measurement of Kuiper Belt
Objects at resolutions of 1~km or better \citep{Buie2020}, and for getting upper limits on the presence of tenuous KBO
atmospheres.

Other space missions that currently profit from stellar occultations for accurate planning and target characterisation
are LUCY (NASA), which will explore the asteroid Trojans of Jupiter \citep{Noll2020}, and DESTINY+ (JAXA) planned to
visit 3200 Phaethon \citep{Arai2020}. The occultation campaign for this asteroid of only $\sim$6~km in size, very
challenging due to extremely high requirements put on the predictions, was the first one to be successful for a member of
the near-Earth object category.

\subsubsection{Exoplanets and proto-planetary disks}
\label{sec:exoplanets}

{\gaia} will eventually detect planets astrometrically through an analysis of the epoch astrometry in cases where the
simple single star source model does not fit the data \citep{gaiamission}.  \cite{Perryman2014} predicted that some
$21\,000$ exoplanets with masses down to $\sim1$--$15$~$M_\mathrm{J}$ should be discovered out to $\sim500$~pc during
the nominal five-year {\gaia} mission, and up to $70\,000$ if the mission lifetime is doubled to ten years. However this
requires the collection of astrometric data over a sufficient time interval and very well understood astrometric
uncertainties at the level of individual position measurements. Hence exoplanet catalogs are only expected for later
{\gaia} releases. Nevertheless \gdr{2} significantly impacted the field of exoplanet studies through the better
characterization of exoplanet host stars, and studies of the discrepancies between long time baseline proper motions
derived from combining Hipparcos and {\gaia} positions, and the proper motions measured by Hipparcos or {\gaia} alone.

The precise estimation of exoplanet properties such as radius, mass, and density, require precise knowledge of the
properties of the host star, such as its radius and evolutionary state. \gdr{2} lists radii and luminosities for a
subset of the stars in the catalogue \citep{Andrae2018}, while independently \cite{FultonPetigura2018} and
\cite{Berger2018} used the \gdr{2} parallaxes to derive precise ($5$--$8$\%) estimates of radii of stars in the
\textit{Kepler} field. This allowed, for example, to confirm and better characterize the presence of a gap around
$2$~$R_\oplus$ in the size distribution of small close-in planets \citep{Berger2018, FultonPetigura2018}.
\cite{Berger2020a} presented a refined catalogue of parameters for $186\,301$ \textit{Kepler} stars with median radius
accuracies of $4$\% and including more stellar parameters, such as ages and densities. This allowed \cite{Berger2020b}
to make a more refined study of the planet radius gap as a function of stellar mass and age, finding hints in the ratio
of super-earths to sub-neptunes as a function of host age, that core powered mass loss may be the driving mechanism for
the radius gap. They caution that the sample is as yet too small to rule out the photo-evaporation mechanism.  In
another application of the {\gaia} data, advantage is taken of the high resolution sky map to properly interpret Kepler
and TESS light curves, which may be affected by blending of several sources into the large PSFs and detector pixels used
by these missions \citep[e.g.,][]{Cloutier2020}.

Combining {\gaia} and Hipparcos data allows for the derivation of proper motions over a $\sim24$~yr baseline. The
long-term proper motions can be compared to the Hipparcos and \gdr{2} proper motion measurements which cover baselines
of $\sim3.5$~yr and 22 months, respectively. Discrepancies between the three measurements can point to the presence of
unseen companions orbiting the target stars, including sub-stellar objects and exoplanets \citep{Mignard2009,
Michalik2014}. \cite{SnellenBrown2018} used the proper motion discrepancy to constrain the orbital period and mass of
the young planet $\beta$~Pictoris~b. They took advantage of the edge-on orbit of the planet which tightly constrains the
expected direction of the stellar reflex motion. A planet mass of $11\pm2$~$M_\mathrm{J}$ was derived which provides
support for so-called hot-start models for gas giant formation. The mass of $\beta$~Pic~b was confirmed in studies by
\cite{Dupuy2019} and \cite{Nielsen2020}, in which radial velocities and high angular resolution imaging data were added
to the analysis of the proper motion discrepancy. \cite{Brandt2018} applied the proper motion discrepancy technique to
the full Hipparcos Catalogue, and derived a catalogue of Hipparcos-\gdr{2} accelerations aimed at facilitating a search
for sub-stellar and dark companions. \cite{Brandt2019} employed this catalogue to constrain the masses of directly
imaged companions, including the most precise mass estimate for Gl 758B. \cite{Kervella2019} presented an independent
catalogue of Hipparcos-\gdr{2} proper motion discrepancies, and use their results to confirm that Proxima Centauri is
bound to $\alpha$~Cen, and set upper limits of $0.1$--$0.3$~$M_\mathrm{J}$ on the mass of potential planets orbiting
Proxima within 1 to 10 au. They stress the power of {\gaia} and the Hipparcos-{\gaia} combination for the detection of
long orbital periods, where the detection of tangential motion anomalies at accuracies of $1$~m~s$^{-1}$~pc$^{-1}$ is
already possible with Hipparcos and \gdr{2}. Future {\gaia} data releases will increase the sensitivity due to the
increased time base line and the increased precision of the {\gaia} proper motions.

Finally, for the study of the early stages of planetary systems formation, the parallaxes provided by \gdr{2} allow for
distances to proto-planetary disk systems to be established to better then one percent precision. Hence a precise
linear scale can be added to the spectacular images of disk substructure presented for example by \cite{Andrews2018}.

\subsubsection{Observational Hertzsprung-Russell diagrams}
\label{sec:hrds}

To illustrate the quality of \gdr{2}, \cite{dr2hrds} presented a set of splendid observational Hertzsprung-Russell
diagrams. These reveal a wealth of detailed substructure, such as the binary sequence, a detailed view of the giant
branch \citep[Figs.~8 and 10 in][]{dr2hrds}, and the newly discovered split of the main sequence for halo stars into a
blue and a red sequence \citep[Figs.~20 and 21 in][]{dr2hrds}. This is a finger print of the formation history of the
halo of the Milky Way which is discussed extensively in the review by \cite{Helmi2020}. The features in the HR-diagrams
are visible thanks to the high precision parallaxes and photometry combined with the dense sampling of
absolute-magnitude-colour space. \cite{JaoFeiden2020} present a detailed analysis of the distribution of stars along the
main sequence at $\bpminrp>1.7$ and unveil many subtle features in the form of local density enhancements or gaps. The
gap near $M_G\approx10$, discovered by \cite{Jao2018}, is confirmed and attributed to variability in stellar radii and
luminosities over a narrow range in mass, $\sim0.34$--$0.36$~\msun, caused by non-equilibrium $^3$He fusion in the
stellar core and intermittent mixing between core and envelope convection zones. The other features remain as yet
unexplained but \cite{JaoFeiden2020} speculate that subtleties in stellar atmospheric physics could be the cause.  This
is reminiscent of the features in the white dwarf observational HR diagram (Figure~\ref{fig:wdhrd}) discussed in the
next section. 

\begin{figure}[t]
    \centering
    \includegraphics[width=\linewidth]{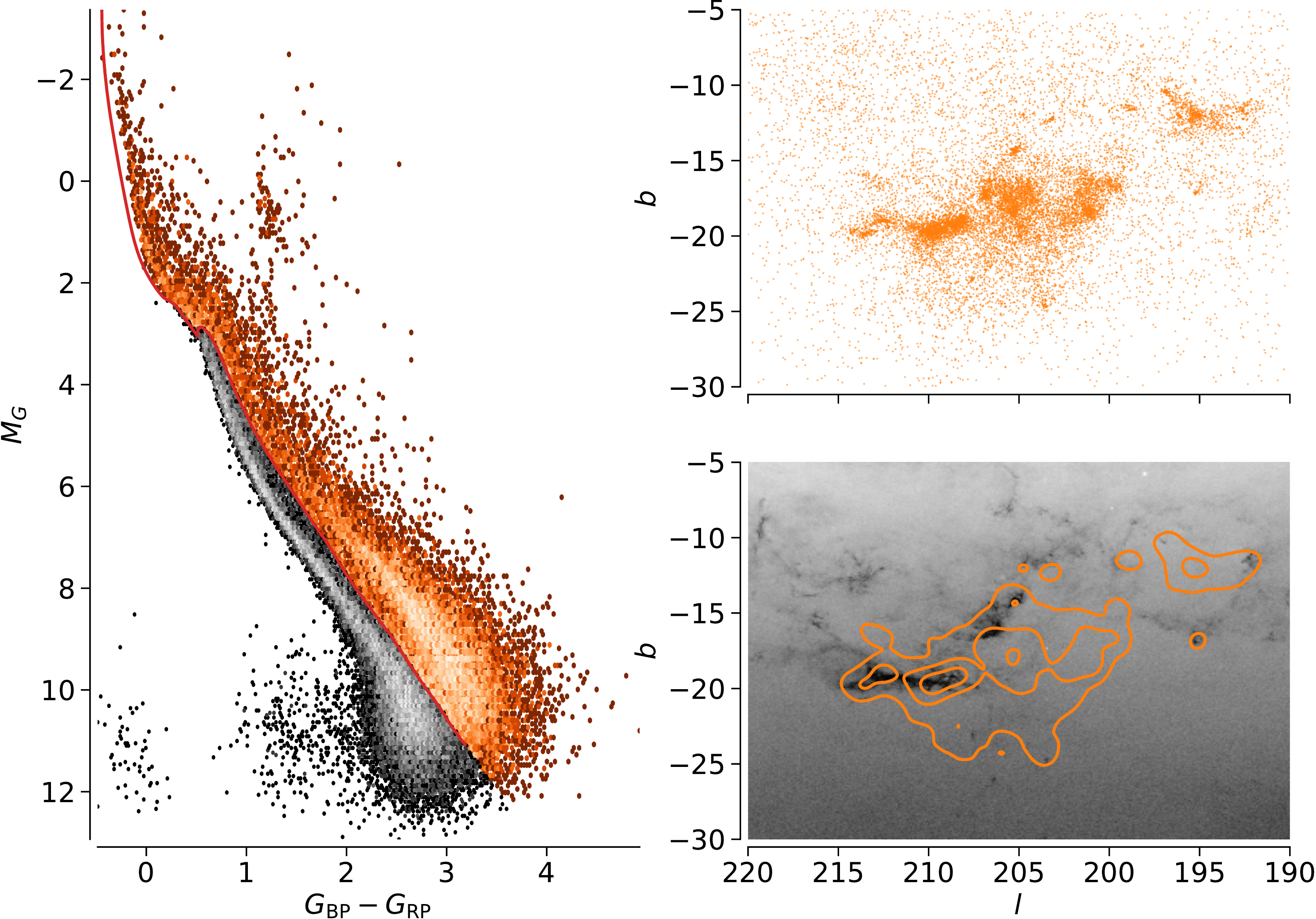}
    \caption{Selection of young stars in the \gdr{2} observational HR diagram for the Orion region. The left panel shows
        the HR diagram for the $40\,167$ stars selected as explained in the text. The line is a $14$~Myr
        ($\log_{10}(\text{age})=7.15$) solar metallicity isochrone from the MIST library \citep[version
        1.2,][]{Choi2016, Dotter2016} and is used to separate the young population above the isochrone from the rest of
        the stars in the diagram. The top right panel shows the distribution of the young population ($16\,692$ stars)
        which is clearly very clustered and is closely associated with the dust features in the Orion region, as shown in
        the bottom right panel. The dust features are outlined as dark features in the map of number counts of stars
        located beyond $500$~pc.}
    \label{fig:orion}
\end{figure}

\gdr{2} provides precise parallaxes for stars in the solar neighbourhood down to faint magnitudes. This allows for a new
way to pick out the young stellar population by selecting low mass stars still located on the pre-main sequence, which
are now easily recognizable by their position in the observational HR diagram, well above the main sequence in
luminosity. A first application of this technique was presented by \cite{VillaVelez2018} who showed for the
Scorpio-Centaurus-Lupus sky region how a simple selection on parallax combined with isolating the young stars in the
\gdr{2} observational HR diagram beautifully reveals the sub-groups of the Sco~OB2 association. \cite{Kounkel2018}
applied the HR diagram selection to a combined \gdr{2}-APOGEE-2 sample to characterize the subgroups in the Orion star
forming region. An illustration of the observational HR diagram selection of young stars, based solely on \gdr{2}, is
shown in Figure~\ref{fig:orion} for the Orion region. The stars were selected from \gdr{2} according to:
$190^\circ\leq\ell\leq220^\circ$, $-30^\circ\leq b\leq-5^\circ$, $2.0\leq\varpi\leq3.5$~mas,
$\varpi/\sigma_\varpi\geq5$, and a proper motion filter selecting all stars within $5$~\maspm\ from
$(\mura,\mudec)=(-0.5,1)$~\maspm. In addition data quality filters were applied, requiring the renormalized unit weight
error
\citep[RUWE,][]{ruwe}\footnote{\url{https://www.cosmos.esa.int/web/gaia/dr2-known-issues\#AstrometryConsiderations}} to
be less then $1.4$ and applying Eq.~C.2 from \cite{Lindegren2018}. This results in a list of $40\,167$ stars from which
the young population can be selected as illustrated in the left panel of Figure~\ref{fig:orion}. The rough selection of
stars above the $14$~Myr isochrone obviously includes a few evolved stars, but this does not affect the clearly
clustered nature of the distribution of the young stars on the sky (top left panel). The bottom right panel of
Figure~\ref{fig:orion} shows that, as one would expect, the selected young population is closely associated with the
dust features in Orion. \cite{Zari2018} applied this combination of a simple kinematic and HR diagram selection to the
full \gdr{2} sample within 500~pc from the Sun and presented 3D maps of the spatial and age distribution of low mass
pre-main sequence stars, as well as the spatial distribution of the population of young blue (OBA) stars. The maps
outline the well known star forming regions around the sun, and according to \cite{Zari2018} cast doubt on the existence
of the Gould Belt as an entity. They argue that the impression of a ring-like distribution of young stars is created by
chance features in the 3D distribution \citep[a conclusion reached earlier by][on the basis of Hipparcos
data]{BouyAlves2015}.

\subsubsection{White dwarfs}
\label{sec:wd}

\gdr{2} for the first time allowed for the selection of field white dwarfs (WDs) directly from their location in the
observational HR diagram, which previously was only possible for WDs in clusters \citep{GentileFusillo2019}. This allows
for the construction of large all-sky samples of WDs, mostly free from colour and proper motion selection biases.
\cite{GentileFusillo2019} presented a catalogue of almost half a million WD candidates, where the selection was guided
by a spectroscopically confirmed sample from the Sloan Digital Sky Survey \citep[see][]{GentileFusillo2015,
Hollands2017}. The sample contains some $260\,000$ high probability WD candidates.  \cite{JimenezEsteban2018} combined
\gdr{2} and Virtual Observatory tools to select a sample of $73\,221$ WDs, showing that the {\gaia} WD inventory is
almost complete within 100~pc. \cite{PelisoliVos2019} presented a catalogue of candidate extremely low-mass WDs.

\begin{figure}[t]
    \includegraphics[width=0.6\textwidth]{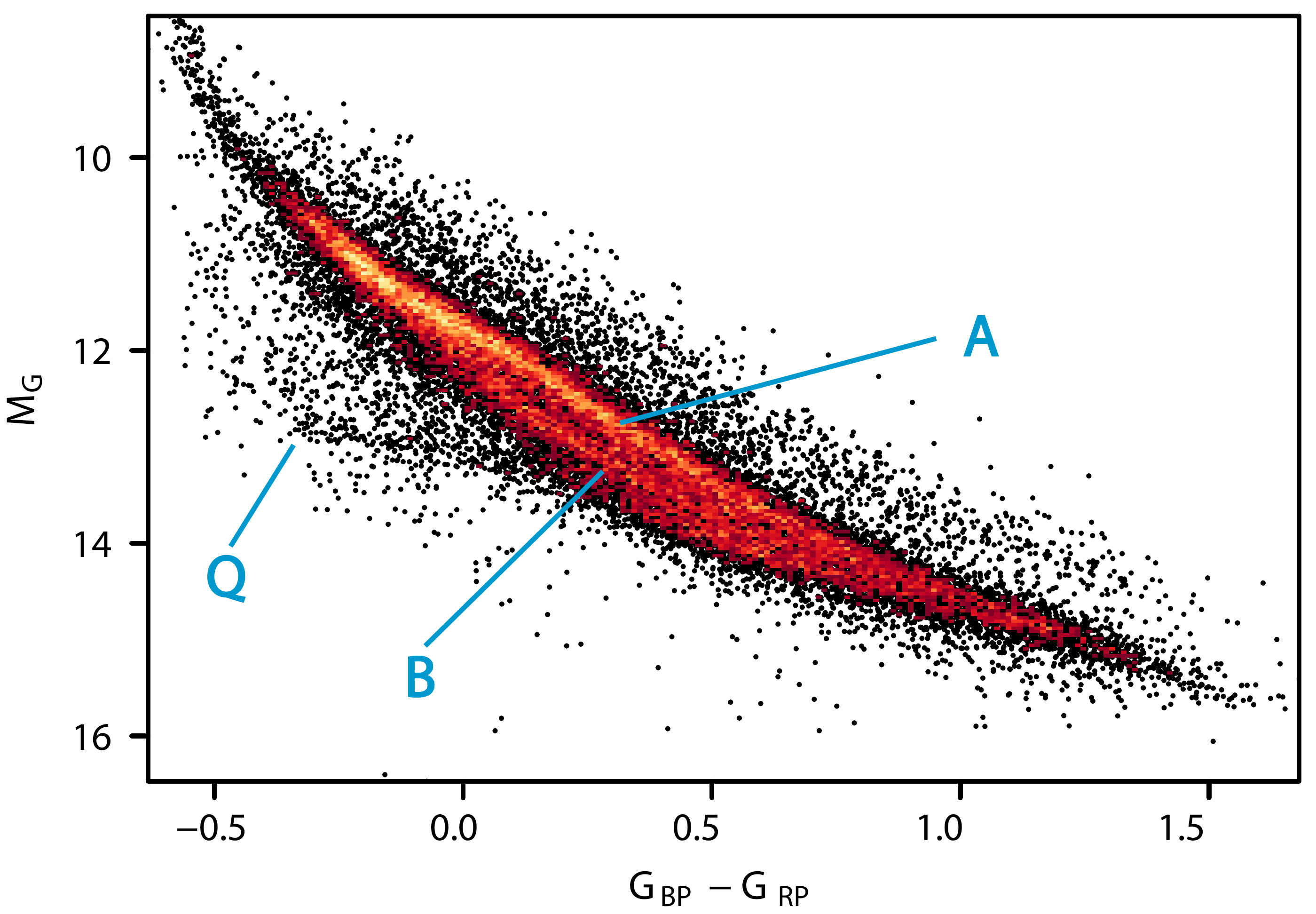}    
    \caption{The \gdr{2} observational Hertzsprung-Russell diagram for $26\,264$ white dwarfs with parallaxes better
    than 5\%. The indicated branches of the white dwarf sequence are discussed in the text. \textit{Credits: Figure
    adapted with permission from \cite{dr2hrds}; \copyright\ ESO.}}
    \label{fig:wdhrd}
\end{figure}

A first exploration of the WDs in \gdr{2} was presented in \cite{dr2hrds}. The observational HR diagram for WDs with
parallaxes better than 5\% was examined (Figure~\ref{fig:wdhrd}), which showed unprecedented substructure, including a
bifurcation of the WD sequence (features labeled ``A'' and ``B'' in Figure~\ref{fig:wdhrd}), as well as a branch
``Q'' running more or less horizontally across. Sequence A coincides with the cooling sequence for $0.6$~\msun\ pure H
atmosphere WDs \citep[][Fig.~14]{dr2hrds}. Sequence B shows the maximum separation from A around $\bpminrp=0.1$ and at
that point coincides with the $0.8$~\msun\ pure H WD cooling sequence, which was unexpected. In colour-colour space the
bifurcation of the WD sequence is also seen but in that case seems to correspond to a split between $0.6$~\msun\ pure H
and pure He atmospheres \citep[the same puzzling result is shown in][]{JimenezEsteban2018}. With neither mass nor
atmosphere effects seemingly able to explain the A and B sequences, several other explanations were put forward soon
after the appearance of \gdr{2}. \cite{Kilic2018} proposed that a combination of both effects is needed and attribute
the B sequence to relatively massive WDs that formed through mergers. \cite{ElBadry2018} showed that the bifurcation is
still present in a sample of WDs all confirmed spectroscopically to be of the same atmospheric composition (DA) and
propose that the shape of the field star initial-final mass relation is the cause of the bifurcation, with the slope of
this relation being flatter for $3.5\lesssim m_\mathrm{ini}/\msun\lesssim5.5$. The slope changes in the initial-final
mass relation would then produce a secondary peak in the WD mass distribution at $\sim0.8$~\msun, leading to sequence B.
They explain the lack of bifurcations in  observed cluster WD sequences by the need for a mixture of ages to produce the
secondary mass peak. \cite{Bergeron2019} in an extensive study further investigated whether a higher than average WD
mass or a different atmospheric composition is responsible for sequence B. They showed that the interpretation of the
location of WDs in the HR or colour-colour diagram in terms of their mass is quite sensitive to the assumed atmospheric
composition \citep[an issue pointed out also in][]{GentileFusillo2019}. In particular the masses in sequence B would
shift from $0.8$ to $0.6$~\msun\ when assuming a He atmosphere with a small admixture of H instead of a pure H
atmosphere, which applies over the temperature range $\sim12\,000$ to $\sim5000$~K where the split between sequences A
and B is clearly visible. This leads to a picture of the \gdr{2} observational WD HR diagram being populated mostly by
``normal'' white dwarfs of $\sim0.6$~\msun, but with a varying H/He mixture in the atmosphere as a function of
temperature. At both high ($>12\,000$~K) and low temperatures ($<5000$~K) the different atmosphere compositions are then
indistinguishable in colour-magnitude or colour-colour space.

The detailed structure revealed in the \gdr{2} observational HR diagram for WDs also constrains the interior structure
of these stars. \cite{Tremblay2019} showed that the Q branch in the WD HR diagram (Figure~\ref{fig:wdhrd}) corresponds
to white dwarfs that are undergoing crystallization of their interiors. The corresponding release of latent heat leads
to a delay of the cooling, which takes WDs from the bright blue end of the HR diagram to the faint red end, by about a
billion years.  This causes a pile-up of WDs along the Q branch. This is the first observational confirmation of a
prediction made over 50 years ago by \cite{VanHorn1968}. \cite{Bergeron2019} showed in addition that the location of the
crystallization sequence is quite sensitive to the interior composition of WDs, and showed empirically for the first
time that, as expected, WD interiors consist of a mixture of carbon and oxygen. By analyzing the WD luminosity function,
\cite{Tremblay2019} find strong evidence that in addition to the latent heat from crystallization WD cooling is further
delayed by the energy released from $^{16}$O sedimentation in the crystallizing cores. \cite{Cheng2019} focused their
analysis on the high mass end of the Q branch and concluded that a small fraction of the $\sim1.1$~\msun\ WDs experience
an additional cooling delay of $6$--$8$~Gyr, which could possibly be explained by $^{22}$Ne sedimentation in their
interiors. \cite{Bauer2020} propose that the formation of solid clusters of $^{22}$Ne in the liquid C/O plasma enables
the much more rapid sedimentation required to explain the long additional delay.

\subsubsection{Open clusters}
\label{sec:ocs}

The field of open cluster studies received a tremendous boost from \gdr{2}, with the parallaxes, proper motions, and
homogeneous all-sky two-colour photometry enabling the accurate characterization of known clusters as well as the
discovery of many new ones. Many works focus on individual clusters and these will mostly not be discussed here. This
overview is focused on the large-scale cluster studies that have been conducted with \gdr{2} data.

The massive \gdr{2} data set invites the application of machine learning methods and automated statistical analyses to
the discovery and characterization of the many thousands of open clusters contained in the data. \cite{Cantat2018}
applied the unsupervised UPMASK method \citep{Upmask2014, CantatTgas2018} to a list of known and claimed clusters and
were able to determine membership and parameters for 1229 clusters. In addition 60 new clusters were discovered
serendipitously in the fields examined. \cite{CastroGinard2018} developed a cluster identification method which uses
DBSCAN \citep{Ester1996} to find initial clusterings of sources in proper motion and parallax space, and then confirmed
whether a cluster is found using an artificial neural network. Having found 23 new clusters, \cite{CastroGinard2019}
then adapted this method for use on a supercomputer and applied it to the full \gdr{2} subset at $|b|\leq20^\circ$ and
$G\leq17$. They discovered 582 new open clusters in the Galactic disk, after accounting for overlaps with the \gdr{2}
cluster lists of \cite{Cantat2018,Cantat2019}, \cite{CastroGinard2019}, \cite{Sim2019}, and \cite{LiuPang2019}. Despite
the bonanza of new clusters, the results from the search for open clusters in the Perseus and Galactic anti-centre
directions \citep{Cantat2019, CastroGinard2019} show that there is still a significant incompleteness in the inventory
of open clusters within $\sim2$~kpc from the sun. A related issue is that many of the previously catalogued open
clusters turn out to be spurious, as shown in the works by \cite{Kos2018} and \cite{Cantat2018}. \cite{Kos2018}
demonstrated that four sparse stellar aggregates with NGC numbers, classified as open clusters in 1888, are in
fact chance projections on the sky. Most of the clusters not confirmed by \cite{Cantat2018} are located at low Galactic
latitudes and toward the inner disk. \cite{CantatAnders2020} made a particularly extensive effort to confirm open
clusters claimed to exist in various literature sources. Their Fig.~3 shows that most of the clusters claimed to exist
at large heights above the Galactic plane are spurious. This is in particular true toward the inner Galaxy, with
NGC~6791 the only confirmed cluster above 500~pc inside the solar circle. The work of extending and cleaning up the
Milky Way open cluster inventory should continue as an accurate cluster inventory is essential, for example in
understanding cluster disruption timescales as a function of their properties and the Galactic environment. Toward the
fainter end of the {\gaia} survey ($G>18$) the combination with complementary data from photometric and spectroscopic
surveys will be crucial.

\begin{figure}[t]
    \centering
    \includegraphics[width=\linewidth]{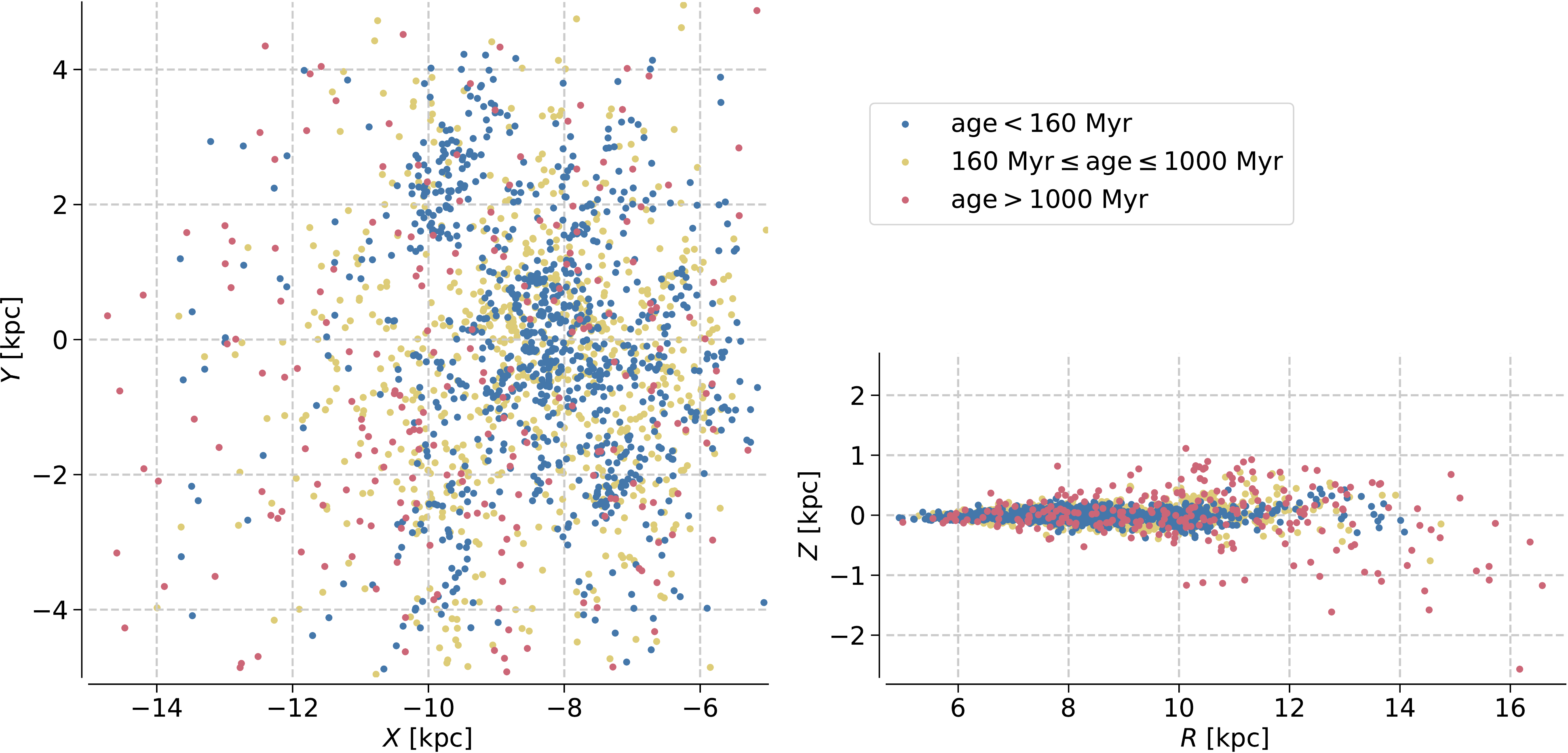}
    \caption{Projections of the 3D positions of $1867$ open clusters listed in \cite{Cantat2020} in Galactocentric
        coordinates. The dots indicate the open cluster positions, colour coded by age as indicated in the legend. The
        left panel shows the locations of the $1799$ clusters closest to the Sun projected onto the Galactic plane, with
        the Sun at $(-8.178,0)$.  The young clusters along $X\sim-10$ roughly outline the Perseus spiral arm with the
        gap identified in \cite{Cantat2020} located at $(-10,1)$. The right panel shows the projected cluster positions
        in distance from the Galactic centre $R$ and vertical height $Z$. The most distant clusters Berkeley 29 and
        Saurer 1* (at $(R,Z)\sim(20.5,1.8)$) are not included in the right panel. \textit{Credits: Open cluster data
    from \cite{Cantat2020}.}}%
    \label{fig:cantatocs}
\end{figure}

While the {\gaia} data readily allow for estimates of the distances and motions of open clusters, their characterization in
terms of age and chemical composition, essential when using clusters to probe the evolution of the Milky Way, requires
more painstaking efforts. In view of the large number of clusters with homogeneous data from \gdr{2} automated methods
for cluster characterization were thus developed and applied. \cite{Bossini2019} applied a Bayesian cluster parameter
estimation method to a sample of 269 open clusters from \cite{Cantat2018}, focusing on low-extinction objects. They used
only \gdr{2} data which ensures a homogeneous determination of age, extinction and distance modulus for these clusters.
The resulting ages show large differences with existing cluster catalogues, which may be due to the more reliable
membership lists from {\gaia} combined with the precise and homogeneous photometry. \cite{Cantat2020} extended the existing
\gdr{2} open cluster inventory to a total of 2017 clusters, containing $\sim230\,000$ stars brighter than $G=18$. They
subsequently developed a neural network method for cluster parameter estimation (distance, age, reddening), which was
trained mostly on the parameters from \cite{Bossini2019}. Reliable parameters were obtained for 1867 clusters. The
resulting cluster 3D positions were used to trace the structure of the Milky Way disk out to $\sim4$~kpc
(Figure~\ref{fig:cantatocs}). The young clusters ($\text{age}\lesssim150$~Myr) roughly trace the expected spiral arm
pattern, outlining particularly well the Perseus and local arms. An apparent gap is seen in the Perseus arm, which
\cite{Cantat2020} argue is physical. The reach of this \gdr{2} open cluster sample is not quite enough to trace the
outer disk or constrain the geometry of the warp.  \cite{Anders2020} examined the distribution of ages as determined by
\cite{Cantat2020} for 834 open clusters withing 2~kpc. Accounting for the completeness of the cluster sample as a
function of age, they find that the age distribution is significantly shifted toward younger ages, peaking around
$150$~Myr, as compared to recent determinations based on the MSWC catalogue \citep{Kharchenko2013, Piskunov2018,
Krumholz2019}. The open cluster age function as determined from \gdr{2} thus seems compatible with the tidal disruption
model proposed by \cite{Lamers2005} with a typical destruction timescale of $1.5$~Gyr for $10^4$\msun\ clusters.
\cite{Anders2020} estimate a current cluster formation rate of $0.55^{+0.19}_{-0.15}$~Myr$^{-1}$~kpc$^{-2}$, which for
the typical cluster mass in their sample would imply that only $16^{+11}_{-8}$\% of stars in the solar vicinity form in
bound clusters.

For individual evolved open clusters the exquisite astrometry enables a much better characterization of the expected
tidal tails that should be present around them. Two independent studies of the Hyades cluster by
\cite{MeingastAlves2019} and \cite{Roeser2019} reveal the tidal tails extending over about 200 pc in space, populated by
hundreds of Hyades stars identified in the tidal tails \citep[including 5 white dwarfs,][]{Roeser2019}. In both studies
the tidal tail structure clearly shows an S-shape when projected onto the Galactic plane, which corresponds well to
predictions from $N$-body models of open clusters orbiting in the Gravitational potential of the Milky Way disk.  Both
\cite{Fuernkranz2019} and \cite{Tang2019} mapped the tidal tails around the Coma Berenices cluster and in the process
discovered a neighbouring moving group separated by only a few \kms\ in velocity space but without a cluster core.
\cite{Meingast2019} discovered a large stellar stream close to the sun, also without a discernible core, and likely the
remnant of a tidally disrupted cluster or association. The age of this system was estimated to be around 1~Gyr but this
was revised downwards to $\sim120$~Myr by \cite{Curtis2019}. The younger age was confirmed in a follow-up study by
\cite{Ratzenboeck2020}. Part of the confusion can be traced to the fact that searches done for (new) nearby stellar
systems exclusively in \gdr{2} are vulnerable to missing the high mass end, as those stars tend to be bright and may be
missing from \gdr{2} \citep{dr2paper}. It thus remains important to complement \gdr{2} data with Hipparcos data at the
bright end.

The above studies hint at a large network of criss-crossing, relatively young (few hundred Myr), stellar clusters,
associations, and streams in the solar neighbourhood. This network was mapped with \gdr{1} data by \cite{Oh2017}, while
\cite{KounkelCovey2019} presented what appears to be complex filamentary structure in the young star distribution in the
solar neighbourhood, which they proposed is primordial. \cite{Meingast2020} cast doubt on this finding on account of the
very large velocity dispersion in the proposed filaments, and present robust evidence for an overlapping network of
extended young clusters. \cite{Meingast2020} tentatively attribute this to a combination of primordial filamentary
structure and tidal disruption. The picture is likely to grow more complex as the {\gaia} astrometry improves.
Disentangling all the information into a coherent picture of the formation and disruption of stellar aggregates and
their diffusion into the field star population cannot be done with spatio-kinematic studies only.  Complementary
photometric and spectroscopic data are required in order establish which stellar aggregates are co-natal, through age
and chemical composition information. The interpretation must be guided by advanced simulations of star clusters
dissolving into the Milk Way disk, such as those presented by \cite{Kamdar2019}.

\subsubsection{The phase spiral}
\label{sec:spiral}

\begin{figure}[t]
    \centering
    \includegraphics[width=0.8\linewidth]{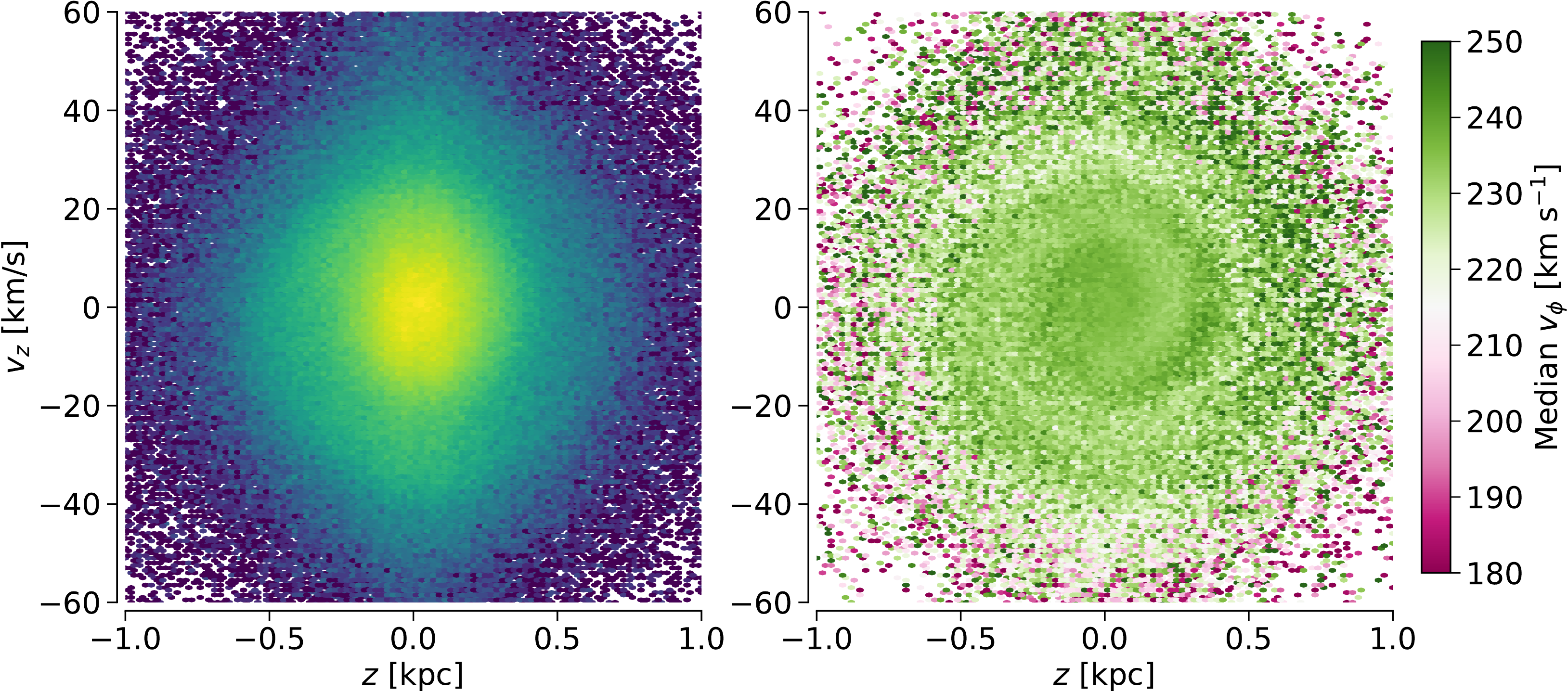}
    \caption{The distribution of stars near the solar circle in the vertical position-velocity plane. The data are shown
        for stars selected on their cylindrical Galactic coordinate according to $|R-R_\odot|\leq100$~pc. The colour coding
        in the left panel represents the star number counts in bins of $\Delta z=0.02$~kpc by $\Delta v_z=1$~\kms. In
        the right panel the colour coding shows the median values of $v_\phi$ for the same bins. \textit{Credits: Figure
            adapted by permission from Springer Nature: Nature 561:360 (A dynamically young and perturbed Milky Way disk,
            T.~Antoja et al., left and right panels of their Fig.~1)
    \copyright\ 2018.}}%
    \label{fig:spiral}
\end{figure}

\cite{dr2kinematics} analyzed a sample of $6.4$ million F-G-K stars, for which precise parallaxes
($\sigma_\varpi/\varpi<20$\%) and full 6D phase space information are available from \gdr{2}, and mapped the Milky Way
disk kinematics between 5 and 13~kpc from the Galactic centre and up to 2~kpc above and below the disk mid-plane. The
results show a very rich and complex velocity field in the Galactic disk, strongly confirming the growing evidence from
the past decade that the Milky Way disk is not in equilibrium. The most spectacular evidence for a disturbed disk was
presented by \cite{Antoja2018} in the form of the ``phase spiral'' visible in the distribution of stars in the
plane of vertical velocity $v_z$ vs.\ vertical position $z$ with respect to the Galactic plane. Figure~\ref{fig:spiral}
shows the phase spiral which is visible directly in the number counts in the panel on the left, but is even more
clearly visible in the panel on the right when the bins in $z$-$v_z$ are colour coded by the median azimuthal velocity
$v_\phi$ of the corresponding stars. The sample used for this figure consist of $764\,370$ stars selected to have
precise parallaxes, good quality astrometry (renormalized unit weight error less than $1.4$), reliable radial velocities
\citep[removing the potentially spurious radial velocities listed in][]{Boubert2019}, and located in galactocentric
cylindrical radius at $|R-R_\odot|\leq100$~pc. The barycentric Cartesian velocities were transformed to galactocentric
cylindrical coordinates as described in \cite{Antoja2018}, using different values for $R_\odot$
\citep[$8178$~pc,][]{Gravity2019} and $z_\odot$ \citep[$20.8$~pc,][]{BennettBovy2019}. By combining \gdr{2} with
spectroscopic data from the second data release of the GALAH survey \citep{Buder2018}, \cite{BlandHawthorn2019} showed
that the phase spiral is visible at different galactocentric radii and azimuths and across different stellar
populations (age, metallicity), thus demonstrating that \cite{Antoja2018} discovered a disk-wide phenomenon.

\cite{Antoja2018} pointed out the similarity of the phase spiral to the process of phase mixing in two dimensional
systems \citep[cf.\ Fig.~1 in][]{Tremaine1999}, implying that the disk of the Milky Way was disturbed relatively
recently on a cosmological timescale. They presented a toy model of a disturbed self-gravitating disk which reproduced
qualitatively the phase spiral. \cite{BinneySchoenrich2018} and \cite{DarlingWidrow2019} presented more sophisticated
models, showing how the effects of a massive intruder passing through the Galactic plane or bending waves excited in the
disk manifest as spirals in the $z$-$v_z$ plane. A massive intruder would in fact excite waves in the disk
\citep[e.g.,][]{Laporte2018} and \cite{Antoja2018} pointed to the previous pericentric passage of the Sagittarius dwarf
galaxy as the event that caused the observed phase spiral. Detailed numerical simulations presented in
\cite{BlandHawthorn2019} strengthened the case for this explanation, with the excitation of the spiral timed at
$\sim0.5$~Gyr ago. Although \cite{Laporte2019} showed that the phase spiral was visible in simulations that were run
prior to the appearance of \gdr{2} \citep{Laporte2018}, none of the pre-\gdr{2} works on disturbances of the Milky Way
disk predicted the specific features unveiled by \cite{Antoja2018}.

Further observational evidence that the Sagittarius dwarf is related to the phase spiral was uncovered by
\cite{RuizLara2020} through a stellar population synthesis analysis of the \gdr{2} observational HR diagram of stars
within 2~kpc from the sun. The star formation history they inferred shows peaks at $5.7$, $1.9$, and $1$~Gyr ago,
coinciding with the past pericentre passages of the Sagittarius dwarf galaxy. This suggests that the Sagittarius dwarf
may have played an important role in the build up of the stellar mass of the Milky Way's disk, where it is a curious
coincidence that the somewhat extended peak in the star formation history around $5.7$~Gyr ago covers the interval in
time some $4.6$~Gyr ago during which the Sun and the solar system formed. 

\cite{BlandHawthornTepperGarcia2020} elaborate the model of \cite{BinneySchoenrich2018} to further understand how the
phase spiral is excited and how it evolves after an impulsive disturbance. They time the emergence of the phase spiral
at about 380~Myr after the last passage of the Sagittarius dwarf through the Galactic disk. Fig.~7 in
\cite{BlandHawthornTepperGarcia2020} shows how the impact induces an $m=2$ bending mode in the disk. The vertical
displacement evolves in time and azimuth around the disk which is compatible with the conclusion reached by
\cite{Poggio2020} that the warp in the disk is precessing, likely as the consequence of a recent or ongoing encounter
with a satellite galaxy. It seems clear that the Sagittarius dwarf galaxy and the phase spiral are linked but there is
work to be done to tie the whole story together neatly. The observation of the phase spiral together with advances in
the mapping of the Sagittarius stream \citep{Antoja2020, Ramos2020} illustrates that, as implied by
\cite{BlandHawthornTepperGarcia2020}, with \gdr{2} the era of ``Galactic seismology'' has started
\citep[the term Galactoseismology was introduced by][]{Widrow2012}.

\subsubsection{Stellar streams}
\label{sec:streams}

Prior to the launch of {\gaia}, \cite{Ivezic2012} reviewed the impact of the large photometric and spectroscopic surveys
that were in operation during the first decade of the 21st century. These surveys clearly established the complex and
dynamical nature of the Milky Way. In particular the Sloan Digital Sky Survey opened up the possibility of peering deep
into the halo and through matched filter techniques pick out many substructures, including streams\footnote{The stellar
streams discussed here concern those recognizable as such in configuration space. Streams identified as overdensities in
the space of integrals of motion are not considered.}, thought to be remnants of dwarf galaxies or globular clusters
disrupted by tidal forces as they orbit in the gravitational potential of the Milky Way. The release of \gdr{2} clears
our view of streams in the halo by filtering out the foreground stars on the basis of their large parallaxes and/or
proper motions, and at the same time providing a $\sim1$~\maspm\ precise proper motion survey of the stars in streams.

\gdr{2} led to a substantial expansion of the inventory of streams criss-crossing the halo, in particular thanks to the
application of the \texttt{STREAMFINDER} algorithm \citep{Malhan2018a} to the \gdr{2} astrometric and photometric data
for sources beyond $1$~kpc and at high latitude \citep[$|b|>20^\circ$,][]{Malhan2018b, Ibata2019a}. Many known streams
were identified as well as candidate streams which may be artifacts of the {\gaia} scanning law and await confirmation.
Five new streams were identified beyond 5~kpc and eight new structures in the inner halo at heliocentric distances
between 1 and 10~kpc. Of the latter the ``Fimbulthul'' stream is noteworthy as it was subsequently identified as the
tidal stream of the $\omega$~Centauri globular cluster \citep{Ibata2019b}, lending strong support to the suspicion that
$\omega$~Cen is the remnant core of an accreted dwarf galaxy.

Many studies using \gdr{2} concentrated on individual streams, the results for two streams are highlighted here. The
study of the GD-1 stream by \cite{PriceWhelanBonaca2018} is a beautiful illustration of the power of {\gaia} at
filtering out foreground stars. By considering only stars at $\varpi<1$~mas in the sky region around GD-1,
\cite{PriceWhelanBonaca2018} were able to narrow down the GD-1 candidates through a proper motion selection, and then
applied a matched filter to Pan-STARRS photometry to isolate the stars in \gdr{2} that are members of the stream.
\gdr{2} transformed the view of GD-1. The mapping of GD-1 was extended by $20^\circ$ across the sky such that it now
extends over $80^\circ$. In addition plausible locations for the stream progenitor and gaps within the stream were
identified, as well as stars that are clearly part of the stream but located off the main stream track. Modelling by
\cite{Bonaca2019}, aimed at explaining the stream gaps as well as the off-track stars, suggested that GD-1 may have been
perturbed by a massive object, the models preferring a dense perturber with a mass in the range $10^5$--$10^8$~\msun. No
known globular cluster could be identified as the possible perturber, leaving as a tantalizing possibility that the
perturber was a dark matter subhalo, predicted to exist in $\Lambda$CDM. In a follow-up study \cite{Bonaca2020} acquired
high resolution spectroscopy for the off-track stars and confirmed that they are kinematically and chemically consistent
with the main stream. The radial velocity offset between the stream and the spur then allows us to constrain the orbit of
the perturber. The possible perturber orbits overlap with the present day debris of the Sagittarius stream, suggesting
that the perturber may have originated from the Sagittarius dwarf galaxy. An interaction with the Sagittarius dwarf was
also suggested by \cite{deBoer2020}, who further extended the mapping of GD-1. \cite{Ibata2020} re-examined the GD-1
stream, adding radial velocity information and investigating sample contamination, and concluded that the apparent gaps
in the stream can also be explained as the projected morphology caused by epicyclic motion of the stream stars in a
smooth Galactic potential. In addition \cite{Ibata2020} point out that the sky region around GD-1 shows artifacts in the
{\gaia} completeness function (see their Fig.~10) that are clearly related to the scanning law and could cause or
enhance the appearance of gaps. The picture of GD-1 thus appears quite complex. Further insights will come from a better
understanding of {\gaia} selection function effects, a cleaner GD-1 sample, and more complete modelling in which GD-1 is
orbiting in the halo in the presence of larger systems such as the Sagittarius dwarf galaxy.

The advantages of an all sky astrometric and photometric survey are very well illustrated by the study of the Orphan
stream by \cite{Koposov2019}. No longer constrained by the footprint of surveys such as SDSS, they used and all-sky RR
Lyrae sample extracted from \gdr{2} to guide the subsequent selection of members of the Orphan stream. The stream now
spans $\sim210$ degrees across the sky, translating to some 150~kpc in physical length. This led to the detection of the
changing orbital pole position along the stream, which corresponds to twists in its shape. In addition the proper
motions show a strong across-track component. These observations point the Orphan stream having been perturbed by a
massive object. Modelling by \cite{Erkal2019} showed that a perturbation of the Orphan stream by the Large Magellanic
Cloud very nicely explains the observational data. The model for the Orphan stream in the combined potential of the
Milky Way and the LMC can be used to estimate the mass of the latter as $1.38^{+0.27}_{-0.24}\times10^{11}$~\msun. In
addition the Milky Way mass within $50$~kpc is estimated to be $3.80^{+0.14}_{-0.11}\times10^{11}$~\msun\ and the
prediction is made that the reflex motion of the Milky Way in response to the LMC should produce a bulk upwards motion
of the stellar halo. These two studies of the Orphan stream demonstrate the need for a much more complex modelling of
streams orbiting in the Milky Way potential. The effects of massive perturbers such as the LMC on both the streams and
the Milky Way must be accounted for in order to properly interpret the observations.

\subsubsection{From globular clusters to the Andromeda nebula}
\label{sec:dwarfsgcs}

The studies of tidal streams above demonstrate the capability of {\gaia} to probe the Milky Way far out into the halo.
Indeed many globular clusters, dwarf galaxies, and even the M31 and M33 galaxies are within reach of {\gaia} astrometry. A
first demonstration of the reach of \gdr{2} was provided by \cite{dr2dwarfsgcs}, who presented the proper motions and
orbits of 75 globular clusters, nine dwarf galaxies, one ultra-faint system and the Small and Large Magellanic Clouds.
Several works followed up on and extended the results presented.

The orbits of dwarf galaxies provide constraints on the Milky Way's mass. \cite{dr2dwarfsgcs} assumed that the Leo I
dwarf galaxy is bound and derived a Milky Way mass of $0.9\times10^{12}$~\msun. \cite{Fritz2018} examined the proper
motion data for 39 (candidate) dwarf galaxies out to 420~kpc and find that the observed satellite kinematics are best
explained by their high mass halo alternative of $1.6\times10^{12}$~\msun. \cite{Callingham2019} used the \textsc{EAGLE}
cosmological hydrodynamics simulations to predict the distribution function for satellite galaxies. They infer the Milky
Way's halo mass by calculating the likelihood that the observed satellite population is drawn from this distribution
function and find a value of $1.2\times10^{12}$~\msun. \cite{EadieJuric2019} use the \gdr{2} globular cluster kinematics
to estimate the total mass of the Milky May and arrive at $0.7\times10^{12}$~\msun. The latter study relies on the 150
globular cluster proper motions derived from \gdr{2} by \cite{Vasiliev2019}, who derived the mean proper motions after
accounting for systematic errors. The Milky Way mass derived by \cite{Vasiliev2019} is $1.2\times10^{12}$~\msun\ with
large uncertainties. The division into low ($\sim10^{12}$\msun) and high ($\sim1.6\times10^{12}$~\msun) estimates for the
Milky Way mass \citep{BlandHawthornGerhard2016} is still present but will likely be resolved once \edr{3} is available.

Besides the orbits and the distribution function of globular clusters \citep{dr2dwarfsgcs, Vasiliev2019} these systems
can also be studied in detail individually with {\gaia}. \cite{dr2dwarfsgcs} demonstrated that the signature of rotation
can clearly be detected with \gdr{2} in at least five globular clusters. This was extended by \cite{Bianchini2018} to
the detection of rotation at the $3\sigma$ level for 11 globular clusters. For eight clusters it is in addition possible
to constrain the inclination angle of the rotation axis with respect to the line of sight. The rotation of globular
cluster can be measured thanks to the availability of proper motions for large number of stars in each cluster. In the
Magellanic Clouds many millions of stars are observed by {\gaia} and consequently \cite{dr2dwarfsgcs} were able to present
exquisite kinematic maps for the LMC and SMC, deriving a proper motion based rotation curve for the LMC that is
competitive with line-of-sight velocity curves. \cite{Vasiliev2018} used half a million red giants in the LMC to derive
a two-dimensional kinematic map, consisting of the mean proper motion and its dispersion out to 7~kpc from the LMC
centre. This kinematic map was used to measure the rotation curve, the velocity dispersion profile, and the orientation
of the LMC. In a sense {\gaia} provides the tangential motion equivalent of the line-of-sight kinematic maps derived from
integral field unit spectrograph data for distant galaxies.

Moving further out into the local group \cite{VanDerMarel2019} selected stars in M31 and M33 from \gdr{2}, taking care
to remove contaminant sources and bad measurements. The resulting sample of stars allows the measurement of the mean
proper motions of M31 and M33, but more impressively, an averaging of the proper motions over sectors in azimuth around
the two galaxies reveals the signature of their rotation. These are consistent with the literature but at the moment are
not very constraining. The rapid improvement in proper motion precision for future {\gaia} data releases (see
Section~\ref{sec:nearfuture}) should permit rotation measurements that are competitive. The mean proper motions measured
with \gdr{2} were combined with existing measurements as inputs to numerical orbit integrations, the results of which
imply that M33 must be on its first infall with respect to M31. The \gdr{2} proper motion for M31 also implies that it
is on a less radial orbit toward the Milky Way than found from HST proper motions. Hence a future collision with the
Milky Way will happen somewhat later and at a larger pericentre. \cite{TepperGarcia2020} used the \gdr{2} proper motions
measurements in a combined $N$-body and hydrodynamical simulation of the M31 and M33 orbital history that takes into
account the effects of dynamical friction and of mass loss due to tidal stripping. They conclude that the \gdr{2} proper
motions do allow for a second-infall scenario for M33, with the first encounter $6.5$~Gyr ago bringing these galaxies to
within $\sim50$~kpc of each other.

\subsubsection{QSO and AGN studies}
\label{sec:qsoagn}

This overview of \gdr{2} science results started in the solar system where {\gaia} observes any minor body that is
bright enough and sufficiently point-like in appearance to its telescopes. At the other end of the universe {\gaia}
automatically observes galaxies that have bright cores as well as distant quasars which are point-like. {\gaia} in
principle could observe tens of millions of galaxies located at redshifts up to $\sim0.3$ \citep{Robin2012}, but many of
these will not be selected for observation on board \citep{deBruijne2015}. Of order 1 million galaxies are expected to
be observed by {\gaia} and these can be morphologically characterized at the $0.2''$ level, which represents a unique all
sky survey \citep{KroneMartins2013}. The characterization of galaxies is part of the data products for future {\gaia}
data releases. 

\gdr{2} contains $550\,000$ identified quasars which materialize the celestial reference frame, but also form an
interesting sample scientifically. \cite{gcrf2} presented and examined the properties of the {\gaia}-CRF2 and it was
noted that a small fraction of the ICRF3 sources in \gdr{2} show significant offsets between their radio and optical
positions, even after the alignment of the optical and radio frames. It is well-known that there can be physical reasons
for such offsets, the optical centre of emission for a QSO or AGN does not necessarily coincide with the radio centre.
This was exploited by \cite{Plavin2019} who analyzed the radio-optical position offsets for a sample of $\sim4000$ AGN.
For the subset with the most precisely determined radio jet position angles they found that the position offsets for
$\sim70$\% of the sources are along the radio jet. Due to synchrotron opacity effects, the core of the radio emission is
expected to be slightly offset ($<1$~mas) from the AGN nucleus downstream along the jet, while the core of the optical
emission will be determined by the relative brightness of the nuclear region and jet in the optical. The observed
downstream offsets of the optical positions are explained by the presence of optical jets $20$--$50$ pc long. The
upstream offsets are explained as the accretion disk dominating the optical source position, with possibly a minor
contribution from radio offsets due to unaccounted for structure in the radio jet or frequency-dependent synchrotron
opacity. The \cite{Plavin2019} sample is large enough to allow for studies of the radio-optical offsets as a function of
redshift and colour and the results indicate that the distribution of offsets is consistent with the unified scheme for
AGN classification.

The high spatial resolution of the {\gaia} sky map combined with its high completeness to $G\sim20$ makes it a great
tool to search for gravitational lens systems harboring multiply imaged quasars. Several groups capitalized on this,
discovering new quasar lenses in imaging surveys such as DES and Pan-STARRS, guided by the \gdr{1} catalogue
\citep[e.g.,][]{Agnello2018, Ostrovski2018}. The improvements in the completeness and angular resolution in \gdr{2}
prompted automated searches for clusters of sources in the sky map that could be the images of a lensed quasar.
\cite{Delchambre2019} conducted a blind search for lenses by extracting some 2 million source clusters from \gdr{2}, and
then using simulations to train a machine learning model to identify promising lensed quasar candidates, of which 15
were confirmed as newly identified quadruply lensed quasars.  Starting from a more targeted search, guided by existing
quasar catalogues, \cite{KroneMartins2018} presented two new quadruply imaged quasars which were entirely discovered in
\gdr{2}. \cite{Ducourant2018} present a detailed analysis of the \gdr{2} positions of the HE0435-1223 system and show
that the precision of the {\gaia} astrometry allows for significantly better constraints on the parameters of the
lensing model compared to HST astrometry, and that parameter correlations are significantly reduced. Future
{\gaia} data releases will feature gravitational lens catalogues thus promising a bright future for studies of strong
gravitational lensing.

\subsubsection{Creative uses of the {\gaia} data}
\label{sec:hacking}

Massive and richly annotated data sets such as the {\gaia} data releases lend themselves to investigations that go
beyond the traditional or obvious uses of the data. A few examples of creative uses of \gdr{1} and DR2 are highlighted
here. The potential of the \gdr{1} sky map for the discovery of new stellar systems was realized by \cite{Koposov2017},
who produced an all-sky map of potential source over-densities. This led to the discovery of two new star clusters, one
of which, Gaia-1, had always been hiding in plain sight right next to Sirius. This discovery was possible thanks to
anti-blooming measures in the {\gaia} CCDs \citep{Crowley2016} which enable high spatial resolution to be maintained
even very near the brightest stars in the sky. The reality of this cluster was confirmed in a spectroscopic follow-up
study by \cite{Simpson2017} who concluded that Gaia-1 is a rather massive ($\sim10^4$~\msun) cluster with a metallicity
of $[\text{Fe}/\text{H}]=-0.13\pm0.13$ and an age of about $3$~Gyr. The orbit of the cluster as calculated by
\cite{Simpson2017} shows that it can travel as far as $\sim1$~kpc from the Galactic plane, perhaps making this a thick
disk cluster. \cite{Koch2018} measured an iron abundance of $-0.62\pm0.1$ and a moderately enhanced $\alpha$-element
abundance, thus providing further evidence that Gaia-1 is part of the thick disk population. The precise nature of
Gaia-1 remains unclear. \cite{Mucciarelli2017} find a solar abundance for the cluster, and \cite{Carraro2018} disputes
the association of Gaia-1 with the thick disk by pointing out that the height of the cluster above the warped disk is
less than $0.5$~kpc.

To track down variables stars for which no light curves were published in \gdr{1} \citep[except for the modest sample
presented in][]{Clementini2016}, \cite{Belokurov2017} used the fact that the photometric uncertainties quoted in
\gdr{1} reflect the scatter in the individual observations made for each source. This leads to overestimates of the
uncertainty on the mean $G$ band value for variable sources, making these stars stand out in a diagram of the
uncertainty in $G$ vs.\ $G$. By calibrating against samples of known variable stars \cite{Belokurov2017} were able to
identify candidate RR Lyrae stars in a field covering the Magellanic clouds. These candidate RR Lyrae beautifully
outline the LMC and SMC and in particular reveal the bridge of old stars between the two Milky Way companions. A
combination of \gdr{1} with GALEX data revealed the existence of a bridge of younger stars, offset from the bridge of
old stars and coincident with the known HI bridge between the LMC and SMC.

In a similar vein, \cite{Belokurov2020} examined the astrometric data quality indicators in \gdr{2} in order to identify
potential unresolved binaries across the observational HR diagram for sources within 400~pc from the sun. The sources
they selected for study generally have good quality astrometric solutions as indicated by the value of the RUWE, which
peaks around unity. However there is a tail of sources with RUWE values up to 4 and these could be unresolved binaries
as pointed out in \cite{ruwe}. \cite{Belokurov2020} showed that the positions in the HR diagram where binaries are
expected (such as the location $0.75$ magnitudes brighter than the main sequence) indeed manifest elevated median values
of the RUWE. A translation between the RUWE value and the binary component separation on the sky was proposed and
verified against known spectroscopic binaries. This allowed \cite{Belokurov2020} to apply their binary detection method
to various science cases. In one application they show that for stars hosting high-mass Jupiters the peak in the RUWE
distribution shifts away from unity for hosts where the planet is orbiting close to the star. This suggests that the
existence of hot Jupiters can be explained as the consequence of inward migration driven by the effects of another
companion of the host star, perhaps through the Kozai-Lidov mechanism. This finding from \gdr{2} astrometric data
quality indicators may be confirmed when in future {\gaia} data releases binary and exoplanet catalogues based on the
epoch astrometry will appear.

\cite{Voggel2020} show how astrometric and photometric data quality indicators in \gdr{2} can be used to search for
globular clusters and tidally stripped galaxy nuclei around galaxies out to $\sim25$~Mpc. Applying this to the case of
Centaurus A they were able to identify 632 candidate luminous globular clusters in the halo of this galaxy, of which 5
have been confirmed as bona fide clusters through follow-up spectroscopy.

\subsection{Some lessons learned}
\label{sec:lessons}

I list here a few lessons learned from the scientific exploitation of the \gdr{1} and in particular \gdr{2} catalogues
which have implications for future directions in survey astronomy.
\begin{itemize}
    \item The first {\gaia} data release already made clear that indeed this mission will deliver on the promise to
        revolutionize studies of the Milky Way and play a key role in unravelling the formation history of our home
        galaxy. This was evident from the study by \cite{Helmi2017}, and indeed as the title of that paper states,
        the {\gaia} data releases represent a ``box full of chocolates'', containing a flavour for every astronomical
        taste as amply demonstrated by the breadth of topics discussed above. The main and lasting impact of {\gaia}
        therefore is the availability to the astronomical community of fundamental data for sources all across the sky,
        in the form of accurate astrometry, photometry, and radial velocities. This makes {\gaia} an anchor for many
        other surveys, either as an astrometric, photometric, or radial velocity reference or as a source for
        well characterized observing targets. As a consequence, \gdr{2} has quickly become an indispensable part of the
        astronomical ecosystem.
    \item Many of the studies discussed above rely on the combination of \gdr{2} and other photometric and/or
        spectroscopic surveys. The latter benefit from the addition of 3D spatial and kinematic data provided by {\gaia},
        while the {\gaia} astrometry and radial velocities are really only useful in combination with the astrophysical
        characterization of sources. In designing future survey strategies from which to decide on instruments to build,
        this powerful combination of astrometry, photometry, and spectroscopy should be kept in mind, where for example
        priority could be given to the collection of the fundamental data most limiting to the astronomy applications at
        a given time.
    \item The discovery and subsequent follow-up work on the phase spiral (Section~\ref{sec:spiral}), as well as
        the Galactic archaeology work on the Galactic halo reviewed by \cite{Helmi2020}, demonstrate that dense sampling
        of Galactic phase space is essential to uncover subtle but important features in the phase space of our galaxy.
        The dense and precise sampling of phase space also fully enables the field of Galactic seismology. This argues
        for massive surveys, where in the disk it should be possible to penetrate the obscuring dust. The dense phase
        space sampling combined with precise parallaxes and photometry also allows for discovering and studying subtle
        features in the observational HR diagram (Sections~\ref{sec:hrds} and \ref{sec:wd}).
    \item Although \gdr{2} represents an intermediate {\gaia} data release, based on a fraction of the amount of
        data that will eventually be collected, its legacy value is already immense in that it would represent the
        standard for the coming decade in terms of the astrometric and photometric precision. Future releases
        superseding \gdr{2} will have even longer lasting legacies and it should thus be ensured that the
        {\gaia} data archive remains technically up to date and easily interoperable with other surveys. There is also a
        very strong case for making available lower level data products from {\gaia} to enable a future reprocessing of
        (parts of) the data \citep{Brown2010, Brown2012}.
    \item Conspicuously absent from the major results obtained with {\gaia} data is the determination of basic
        structural parameters of the Milky Way, such as the radial profile of disk stars \citep[for the pre-{\gaia}
        review of Galactic structural parameters see][]{BlandHawthornGerhard2016}. The main reason is the lack of
        a well described selection function for the {\gaia} data releases. A major task for the coming years will be to
        derive the selection function in order to ensure that the exquisite positional and kinematic data provided by
        {\gaia} can be used for population studies. The first very encouraging attempts have been made
        \citep{Boubert2020a, Boubert2020b}, demonstrating the need to engage the DPAC to ensure that as yet unpublished
        information relevant to the {\gaia} selection function is made public.
    \item A lesson very much taken to heart by the community is that the majority of the {\gaia} parallaxes will remain
        at relative accuracy levels which make it dangerous to directly invert their values in order to estimate
        distances \citep[see the extensive discussion in][]{Luri2018}. Nevertheless, all the {\gaia} parallax
        information should be used to constrain or calibrate alternative distance indicators which can then extend the
        reach of {\gaia}, for instance to ensure that the full relative accuracy of proper motions can be exploited.
        Standard candles are a natural choice but ideally (photometric or spectroscopic) distance indicators across the
        HR diagram should calibrated and progress is being made in this direction. The use of parallaxes in combination
        with spectroscopy and/or photometry to make accurate Bayesian distance estimates was demonstrated by
        \cite{SandersDas2018} and \cite{Anders2019}. The first steps toward accurate luminosity calibrations across the
        HR diagram on the basis of {\gaia} parallax information were taken by \cite{Leistedt2017, Anderson2018}, and
        \cite{Cranmer2019}.
\end{itemize}

\section{PANORAMA FOR THE COMING DECADE}
\label{sec:nearfuture}

\subsection{Upcoming {\gaia} data releases}
\label{sec:newreleases}

The astronomical community can look forward to more {\gaia} data releases over the coming decade. The next release,
\edr{3} will appear in December 2020 and provides updated astrometry and photometry based on 34 months of
observations \citep{edr3paper}. The parallaxes will improve by 20\% in precision but for the proper motions the
improvement will be a factor of 2, thanks to the longer time baseline of the observations (see below). The full \gdr{3}
will appear in 2022 and will contain: updated and new radial velocities (for some 30 million stars out to
$\grvs\sim14$); astrophysical parameters for sources based on the BP/RP prism and RVS spectra; the prism and RVS spectra
for a subset of sources; an extended catalogue of variable stars; the first catalogue of binary stars, including
eclipsing, spectroscopic, and astrometric binaries; astrometry for some $100\,000$ solar system objects and reflectance
spectra for a subset of $\sim5000$ asteroids; QSO host and galaxy morphological characterization; and the light curves
for all sources in a $5.5^\circ$ radius field around M31. This illustrates that the science topics such as discussed in
Section~\ref{sec:usecases} will profit from an increasingly rich set of {\gaia} data providing the astrophysical
characteristics of sources.

\gdr{4} will be based on all data collected during the nominal mission lifetime and a part of the extended mission
($5.5$ years of observations), and will feature parallaxes that are more precise by a factor $1.7$ with respect to
\gdr{2}, while for proper motions the gain is a factor of $5.2$. This release will feature an exoplanet catalogue,
predicted to contain some $21\,000$ Jupiter mass planets \citep{Perryman2014}. However the main change that \gdr{4} will
introduce is the release of all the epoch data from {\gaia}, meaning for each source the astrometric, photometric,
radial velocity, and BP/RP/RVS spectra times series.  This represents a vast volume of data and creates significant new
possibilities, where for example the community can conduct independent searches for companions to {\gaia} sources using
the epoch astrometry.

\begin{figure}[t]
    \centering
    \includegraphics[width=0.8\linewidth]{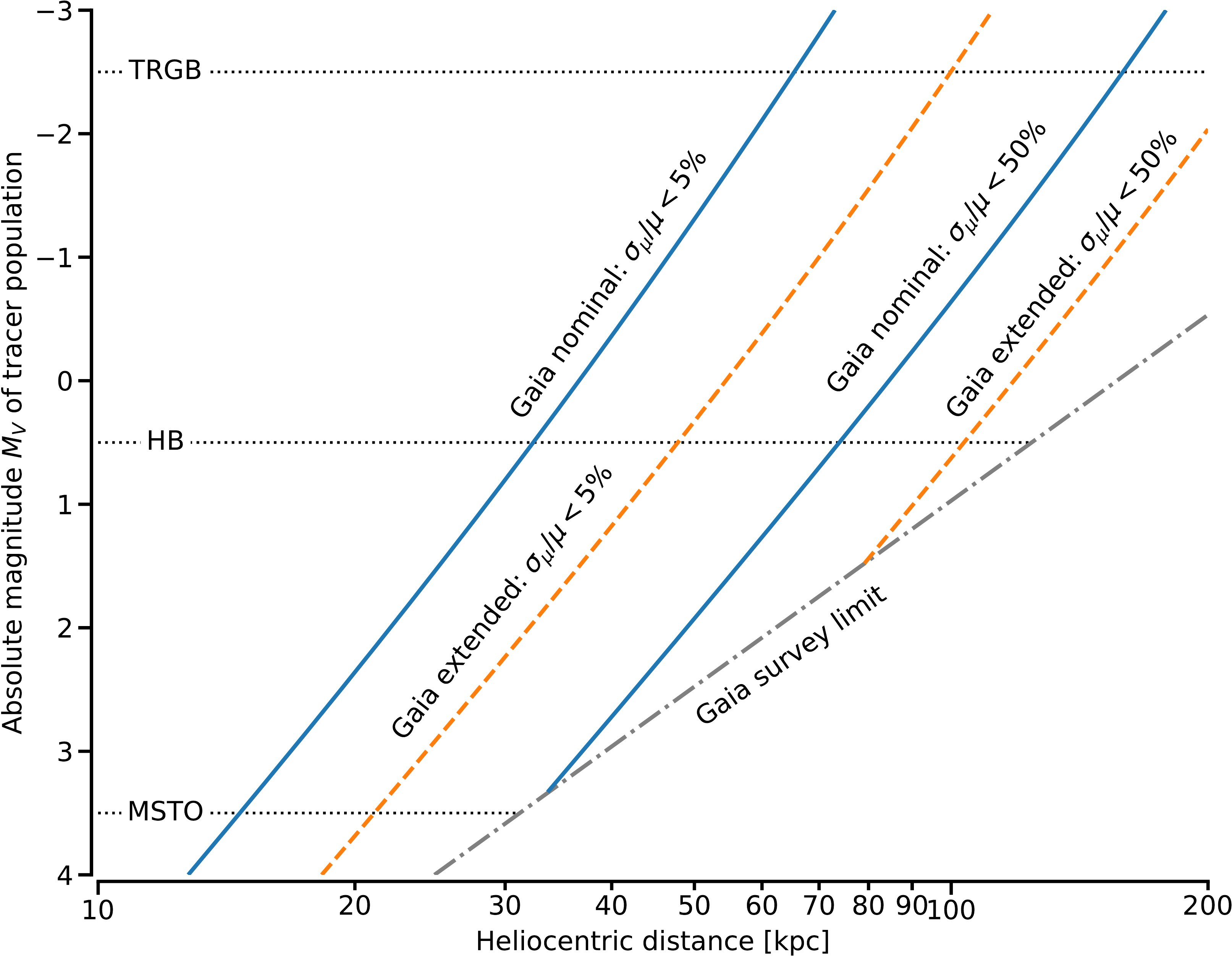}
    \caption{Predicted relative proper motion uncertainty $\sigma_\mu/\mu$ for three tracer populations in the Absolute
        magnitude vs.\ distance plane. The tracers are the main sequence turnoff (MSTO), horizontal branch (HB), and tip
        of the red giant branch (TRGB) stars. The lines show contours of constant $\sigma_\mu/\mu$ for the nominal 5-yr
        {\gaia} mission (solid lines) and the extended 10-yr mission (dashed lines). The dot-dashed line indicates the
        {\gaia}
        survey limit ($G=20.7$). The diagram illustrates the gain in distance out to which a certain relative proper
        motion uncertainty can be obtained when doubling the {\gaia} mission lifetime. Note the logarithmic horizontal
    scale. \textit{Credits: Figure adapted with permission from \cite{Mateu2017}, their Fig.~15.}}%
    \label{fig:extension}
\end{figure}

On July 16 2019 the nominal five year {\gaia} mission ended and the project transitioned to the extended mission phase.
It is expected that {\gaia} can continue to operate until early 2025, the limiting factor to its lifetime being the
propellant for the micro-propulsion system \citep[see][Sect.~3.2.1]{gaiamission}. Once the fine attitude control is lost
the images collected by {\gaia} will no longer be of the quality needed for astrometry. The {\gaia} mission has formally
been extended to the end of 2020, with further extension pending approval. It is expected however that a full ten year
mission can be realized. This doubling of the mission lifetime means the precision for all {\gaia} data products,
including parallaxes will improve by 40\% (reflecting the $\sqrt{t}$ behaviour of the signal to noise). For proper motions
the improvement also includes the longer time baseline, the precision improving proportional to $t^{1.5}$, which
implies a factor of $2.8$ better proper motions (with respect to the 5 year mission) at a given magnitude or an increase
by a factor $22$ of the volume over which a given proper motion precision can be achieved. This is visualized for a
range of bright stellar tracer populations in Figure~\ref{fig:extension}. For more complex motions in binaries and
exoplanet systems the gains are even more impressive. Measuring perturbations of the assumed motion on the sky for a
single star is equivalent to measuring second and higher order derivatives of the celestial positions with respect to
time (the proper motion being the first derivative). Each derivative of time introduces another factor of $t$
improvement, hence acceleration sensitivity increases by a factor of $5.6$. Unambiguously determining the orbital period,
mass, and distance of the perturbing body is only possible when the fourth derivative of the position can be measured
(shown by L.~Lindegren in the proposal submitted to ESA for the extended {\gaia} mission).  In this case the sensitivity
increases by a factor of $22.6$. Doubling the mission lifetime also means that the sensitivity to long orbital period
companions increases, leading for example to an increase in exoplanet detections to $\sim70\,000$ \citep{Perryman2014}.
For solar system objects the 10 year mission lifetime would allow measuring astrometry for main belt asteroids over two
orbital periods, and for Trojans a full orbit would be covered. There is thus much to look forward to in the {\gaia}
data release based on the full extended mission!

\subsection{Synergies}
\label{sec:synergies}

As laid out in \cite{BlandHawthornFreeman2000} and \cite{FreemanBlandHawthorn2002}, unravelling the formation
history of the Milky Way can only be achieved by combining distance and kinematic information with knowledge of the
detailed chemical compositions of the stars. Hence, while {\gaia} is central to the field of Galactic archaeology, its
full power can only be unleashed by combining its data with other sky surveys. As summarized in \cite{Helmi2020}, the
community has made a concerted effort since the approval of the {\gaia} mission in 2000 to propose, build, and operate
large ground based spectroscopic surveys. These collect radial velocity data at the faint end of the {\gaia} survey
($G>15$), and the crucial chemical composition information needed to untangle the formation history of the Milky Way.
The combination of {\gaia} and photometric sky surveys opens up the way to distances and astrophysical parameters for all
stars surveyed by {\gaia}, as demonstrated for example by \cite{Anders2019}. Photometric surveys and {\gaia} are also a
powerful combination when it comes to mapping the 3D distribution of interstellar dust as demonstrated by the works of
\cite{Lallement2019, Green2019}, and \cite{Hottier2020}. The Vera Rubin Observatory's Legacy Survey of Space and Time
\citep[LSST, ][]{LSST2019} is around the corner, and like other surveys will profit from the astrometric and photometric
reference system provided by {\gaia}. The synergies between LSST and {\gaia} were already anticipated by
\cite{Ivezic2012}. As illustrated in their Fig.~21, LSST will in some sense be the deep extension of {\gaia} from
$r\sim20$ to $r\sim24$, with astrometric uncertainties that are anticipated to be comparable to {\gaia} performance at
19th magnitude and then smoothly extending the uncertainty vs.\ magnitude curve by roughly 5 mag.  The stacked LSST
images will reach depths of $r\sim27$ meaning that LSST can peer deep into the outer halo of the Galaxy, resolving main
sequence turn-off stars out to 300~kpc \citep{LSST2019}. This is further motivation to build on the {\gaia} parallaxes
to calibrate photometric distance indicators in order to turn the LSST imaging as accurately as possible into a deep
spatial map.

\section{FUTURE DIRECTIONS FOR MICROARCSECOND ASTROMETRY}
\label{sec:future}

In considering future directions for astrometry a number of options can be considered depending on the science case:
increasing the precision and/or accuracy achieved for a given source brightness; extending a given precision level to
fainter limits; extending a given precision to a different wavelength domain. A trade-off has to be made between the
various options which also involves a decision between narrow- and wide-angle astrometry, with higher precision
typically easier to achieve at narrow angles. An additional important consideration is the maintenance of the dense and
highly accurate optical reference frame that {\gaia} provides. In the following I only briefly comment on ground based
advances coming up and then focus on the future of space astrometry.

\subsection{Ground-based microarcsecond astrometry}
\label{sec:groundmuas}

The efforts to achieve microarcsecond astrometry across the radio frequency domain, through technology and calibration
method improvements, are discussed in \cite{RiojaDodson2019, RiojaDodson2020}. Microarcsecond astrometry is considered
``routine'' at $8$ to $22$~GHz frequencies with good progress being made at higher frequencies. The prospect of the SKA
is driving the developments to improve astrometric precision at low $\lesssim8$~GHz frequencies, where systematic errors
due to the effects of the ionosphere present the main challenge.

In the optical/IR domain on the ground GRAVITY is currently the only instrument capable of astrometry at the tens of
microarcsecond level over narrow angles. \cite{IrelandWoillez2018} argue that in principle the current generation of
single optical/IR telescopes and long baseline interferometers can achieve astrometric precisions at the single \muas\
level, provided instruments are developed with sufficient control over and knowledge of the optical path taken by the
astronomical light, as well as a sufficient knowledge of the astrometric (i.e.\ interferometer) baseline. It will be
important to ensure that such instruments can access as much of the sky as possible for targets as faint as possible.
The generation of giant optical/IR telescopes currently in the planning and construction phases will feature aperture
sizes of up to $39$~m in the case of the European Extremely Large Telescope (ELT). In principle this should allow for
astrometry at the $\sim50$~\muas\ level, provided image locations can be determined at the level of $0.5$\% of a
resolution element and that various sources of systematic error can be overcome. The MICADO instrument under development
for the ELT is aimed at reaching the $50$~\muas\ astrometric accuracy level over a 1 arcsec field of view
\citep{Rodeghiero2019, Trippe2010}. This would for example allow the measurement of single-epoch tangential motions to
$\sim20$~\kms at $100$~kpc, and with a few years of observations would produce precisions at the $5$~\kms\ level,
sufficient to measure the internal kinematics in dwarf galaxies. 

\subsection{Space astrometry prospects}
\label{sec:spaceprojects}

In the coming decade a number of space mission with astrometric capabilities will be launched
\citep[cf.][]{Vallenari2018}. Of these the Small-JASMINE mission
\citep{SmallJasmine2020}\footnote{\url{http://www.jasmine-galaxy.org/index.html}} is a dedicated space astrometry
mission which will survey an area of $0.7^\circ$ radius around the Galactic centre as well as a region along the
Galactic plane ($-2^\circ<l<0.7^\circ$ and $0^\circ<b<0.3^\circ$) in the $H_\mathrm{w}$ band at $1.1$--$1.7$~$\mu$m. The
astrometric accuracies targeted are $25$~\muas\ and $25$~\muaspm\ for parallaxes and proper motions, respectively, at
$H_\mathrm{w}<12.5$. At $H_\mathrm{w}=15$ the accuracies are expected to be $\sim125$~\muas.  Both upcoming NASA space
observatories, the James Webb Space Telescope and the Nancy Grace Roman Space Telescope, can be used for astrometry. For
the latter an extensive science case was presented along with the expected astrometric performances
\citep{Sanderson2019}. Relative proper motions can be measured to $25$~\muaspm\ for the proposed High Latitude Survey,
while astrometric precisions of $3$--$10$~\muas\ are anticipated for relative astrometry of exoplanets or using the
spatial scanning technique. Absolute astrometry to $0.1$~mas can be achieved with the Roman Space Telescope by anchoring
the measurements to the {\gaia} reference frame. The {\gaia}-CRF would be the limiting factor and \cite{Sanderson2019}
emphasize the significant benefits of an extended {\gaia} mission in terms of the reference frame accuracy at the time
the Roman Space Telescope is operational.

On a longer timescale two white papers on space astrometry missions have been submitted to ESA as part its Voyage 2050
long-term planning process. The ``Faint Objects in Motion'' paper \citep{Malbet2019} makes the case for narrow-angle
astrometry from space, targeting end-of-mission accuracies from $10$~\muas\ at $R\sim14$--$22$ for dark matter and
compact object studies, down to $0.15$~\muas\ at $R\sim1$--$18$ for exoplanet studies. The mission concept relies on a
single $0.8$~m telescope for which the PSF is about $0.2$ arcsec in size. This implies that image location systematics
must be under control to 1 part in $200\,000$ in order to achieve the desired differential astrometric accuracy. This
includes contributions from random noise sources, instabilities in the optical aberrations and focal plane geometry, and
variations of detector quantum efficiency between pixels. The development of sophisticated metrology at the telescope
and focal plane level is required. Much of that work has already been done in the context of the SIM PlanetQuest
programme \citep{Unwin2008}, and as part of the preparations for the NEAT mission proposal \citep{Malbet2012}. 

The other Voyage 2050 white paper concerns the GaiaNIR mission proposal \citep{GaiaNIR2019}. This mission would operate
on the same principles as {\gaia} except for conducting its observations over a wide optical to near-infrared band
($400$--$1800$~nm). The science case is laid out in \cite{GaiaNIR2019} and is as broad as that for {\gaia}. Out to 20th
magnitude some 8 billion sources would be observed, most of the gain being in the Galactic plane, where the dust limits
the reach of {\gaia}. The resulting survey would on its own have an astrometric performance comparable to {\gaia} and
allow for synergies with a host of other instruments in space and on the ground, in particular the large spectroscopic
surveys in the infrared such as SDSS-V and MOONS. The dense sampling of a much larger volume over the Galactic disk would
address one of the lessons learned from Section~\ref{sec:lessons}. A design study of the GaiaNIR concept was undertaken
by ESA with the support of experts from the Gaia community and the resulting report is publicly
available\footnote{\url{https://sci.esa.int/s/8a65kZA}}. The major technology challenge for a revolving scanning mission
in the infrared will be the development of corresponding TDI capable detectors. An alternative would be to include an
optical element which would be capable of stabilizing images in the focal plane as the spacecraft rotates.

\subsection{Reference frame maintenance}
\label{sec:crfmaintenance}

An essential element for the future of astrometry and the science it enables is the maintenance of the celestial
reference frame. This frame is ultimately defined by the positions of the sources in an astrometric catalogue. The use
of these positions at any given epoch $t_\mathrm{o}$ (to serve as a reference for astrometric observations) requires
propagating them from the catalogue epoch $t_\mathrm{c}$ to $t_\mathrm{o}$. The reference frame as materialized by
quasars will have a zero spin but with some uncertainty which propagates from $t_\mathrm{c}$ to $t_\mathrm{o}$. An
additional uncertainty is introduced by the errors on the determination of the acceleration of the solar system
barycentre relative to the quasar rest frame. The acceleration is manifest as a proper motion pattern in the quasars,
and uncertainties therein also propagate into future positions.

\begin{figure}[t]
    \centering
    \includegraphics[width=0.8\linewidth]{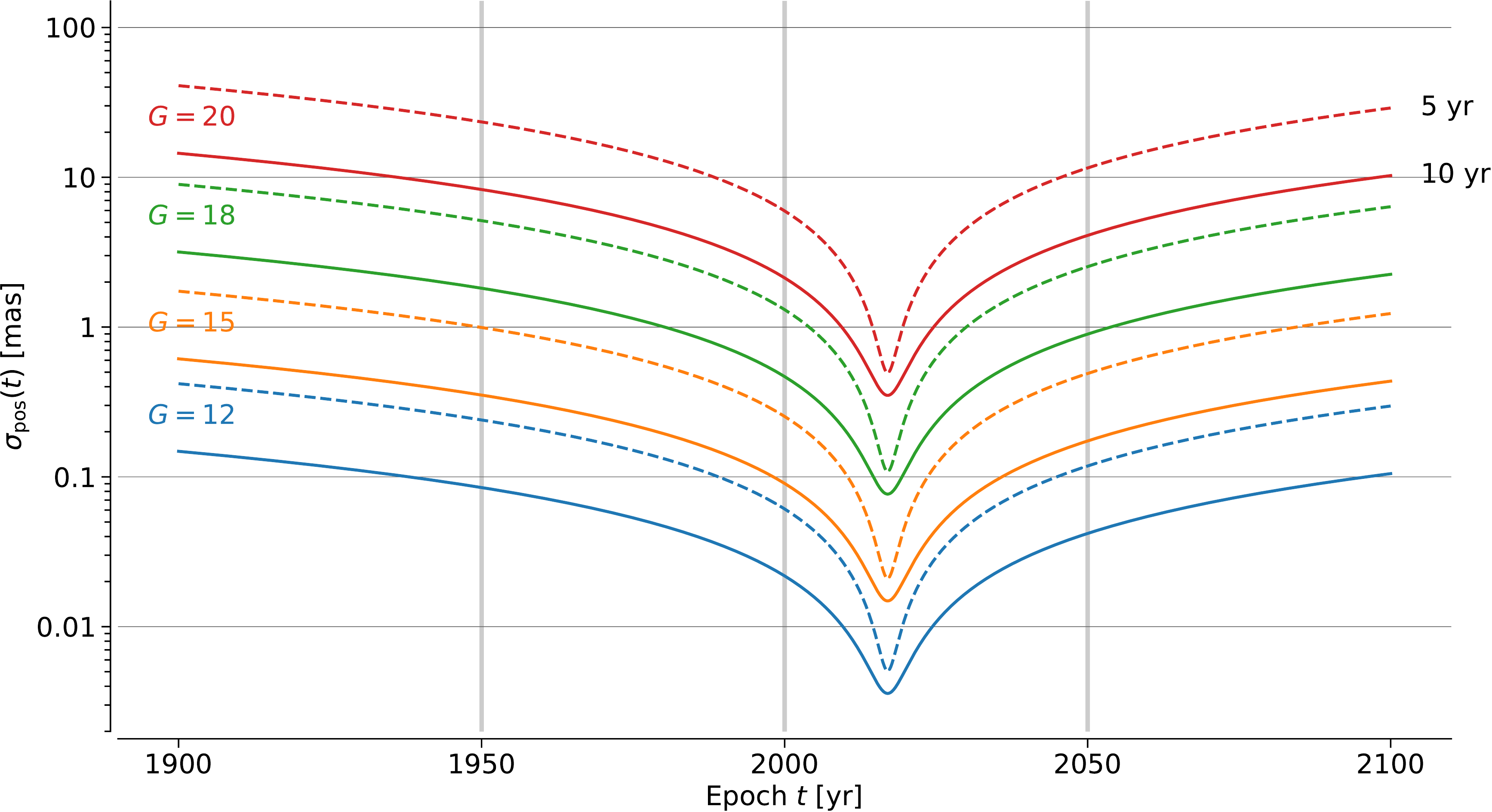}
    \caption{The evolution of the {\gaia} source position uncertainties over time. The curves show how the uncertainty
        on the celestial position of a source \citep[to be precise, the semi-major axis of the position uncertainty
        ellipse as defined by Eq.~B.1 in][]{Lindegren2018} degrades over time as one moves away from the mid-epoch of
        the mission.  This is caused mainly by the uncertainty in the proper motions. The curves shows the uncertainty
        degradation for 4 different magnitudes and for the 5-yr (dashed lines) and 10-yr (solid lines) mission life
        times. \textit{Credits: Figure adapted from \cite{Hobbs2016}, right panel of their Fig.~4 which was created by
    F.~Mignard.}}
    \label{fig:crfevolution}
\end{figure}

In practice however, the {\gaia}-CRF will most often be used by anchoring new astrometric observations to the {\gaia}
positions of stars. The stars provide a much denser reference frame over a larger range in brightness, making a larger
area of the sky accessible to (relative) astrometry. The degradation of the stellar reference frame is dominated by the
uncertainties at $t_\mathrm{c}$ in the proper motions of the stars \citep[the rigorous propagation of astrometric
parameters also depends on the parallax and radial velocity, e.g.,][]{Butkevich2014}, leading to a degradation in time
of the positional uncertainties, roughly proportional to $t_\mathrm{o}-t_\mathrm{c}$. This is illustrated for the case
of the {\gaia} in Figure~\ref{fig:crfevolution}. At the faint end of the {\gaia} survey the typical positional
uncertainty of $\sim0.5$~mas at the end of the 5-yr mission grows to about $3$~mas by 2025 and $10$~mas by 2050. To put
these numbers in context, \cite{Sanderson2019} state that {\gaia} catalogue position errors of a few mas translate to
$\sim0.02$ pixel errors in the Roman Space Telescope Wide Field Instrument, which significantly impacts the ability to
perform high precision absolute astrometry (i.e.\ to better then $0.01$ WFI pixel). Figure~\ref{fig:crfevolution} also
shows the positional uncertainties over time in case the {\gaia} mission life time is extended to 10 years. The factor
$2.8$ improvement in the proper motions leads to a stellar reference frame accurate to $1$~mas in 2025 and $4$~mas in
2050.  This would thus be of tremendous benefit for high precision absolute astrometry with other space or ground based
instruments. For example the MICADO team investigated the use of {\gaia} stars to aid in calibrating out the effects of
optical distortions on the astrometry \citep{Rodeghiero2018}, an effort that will again benefit from the best possible
accuracy for a network of faint star positions. Figure~\ref{fig:crfevolution} is deliberately extended into the past to
illustrate the fact that old observations can also be re-reduced on the {\gaia} reference frame, which remains better
than anything available in the past even over more than a century. As mentioned in Section~\ref{sec:solsys} this is used
for example to improve the orbits of asteroids from position measurements over long time spans.

Improving the reference frame by extending the mission life time of {\gaia} points to a very important benefit of the
GaiaNIR proposal. It would enable the combination of two independent astrometric surveys to derive proper motion
measurements over a time baseline of 20 years which leads to a factor of 20 improvement in the proper motion precision
\citep{GaiaNIR2019}. The stellar reference frame would both improve in precision and degrade much more slowly
\cite[positional uncertainties remaining below $\sim1$--$5$~mas out to 2100 as shown in Fig.~3 of][]{GaiaNIR2019}. The
improved proper motions are scientifically interesting by themselves and would also lead to improved parallaxes for the
stars in common between GaiaNIR and {\gaia}.

Apart from the science enabled by the dense and accurate stellar reference frame, maintenance and improvement of the
celestial reference frame as defined by the quasars is also of great scientific interest. It is well known that quasars
show variability in their optical positions which can be attributed to the internal source structure variations
\citep{Taris2011}, which can lead to apparent non-zero parallaxes and proper motions. However, even if the positions
would be perfectly stable, quasar proper motion patterns would still appear in high precision astrometric surveys
because the solar system barycentre is not at rest with respect to cosmological microwave background.
\cite{Bachchan2016} review these proper motion patterns and how they might be measured from the {\gaia} data. The most
prominent proper motion pattern is due to the acceleration of the barycentre and its expected value of
$\sim4.3$~\muaspm\ can be measured to high significance with {\gaia}. This can be done by solving for the acceleration
parameters as part of the overall reference frame determination (where for \gdr{2} only orientation and spin were
considered). The instantaneous velocity of the solar system with respect to he CMB will cause a quasar proper motion
pattern of order $1$--$2$~\muaspm, which is redshift-dependent and can in principle be used to measure the Hubble
constant. \cite{Bachchan2016} show that this is at the limit of the possibilities with {\gaia}, but in combination with
a mission like GaiaNIR this would become a secure measurement. Further examples of proper motion patterns are those due
to gravitational waves with periods longer than the {\gaia} mission length, and changes in the angular separation
between distant sources due to a possibly non-isotropic expansion of the universe. In the case of gravitational waves,
upper limits can be measured for their energy flux at frequencies below $1.5$~nHz, where {\gaia} alone is expected to
improve on existing limits by 1 to 2 orders of magnitude \citep{Klioner2014, MignardKlioner2012}. A combined
{\gaia}-GaiaNIR measurement would provide another factor 400 improvement \citep{GaiaNIR2019}. The anisotropy of
expansion causes proper motion effects at the $\sim0.02$--$2$~\muaspm\ level, which can only be measured with a
{\gaia}-GaiaNIR combination.

\subsection{Splitting the microarcsecond}
\label{sec:nas}

The SIM PlanetQuest Programme as well as the NEAT proposal and the white paper by \cite{Malbet2019} all aim for
sub-\muas\ accuracy for narrow-angle astrometry. In the context of the definition of science themes for the L2 and L3
missions in ESA's Cosmic Vision Programme, the white paper ``Space-Time Structure Explorer: Sub-microarcsecond
astrometry for the 2030s'' explored the science cases and the challenges of an all-sky absolute astrometry mission
aiming for the sub-\muas\ to nano-arcsecond regime \citep{Brown2013}. Fig.~1 in \cite{Brown2014} illustrates that such a
mission would allow the direct measurement of percent level parallaxes and a few \kms\ tangential motions for bright
tracer populations out to $0.1$--$1$~Mpc. Such a mission concept faces several major engineering and data processing
challenges. Assuming an optical/IR mission, the efficiency of current telescope/detector technology is already close to
100\%. The basic scaling relation for the precision (Equation~\ref{eq:precisionscaling}) then shows that major gains in
accuracy can only be obtained through increases in the aperture or baseline $B$. Continuing this line of reasoning, one
quickly ends up with an argument for an interferometric mission with collecting areas of a few m$^2$ and baselines of
order $100$--$1000$~m \citep[see also][]{Lindegren2007}. This implies the development of precision formation flying,
which is already being developed and tested in the context of the PROBA-3 and e-LISA missions, as well as elaborating an
operations concept that would allow for absolute astrometry. Further engineering challenges include the extreme
requirements on the thermo-mechanical stability of the spacecraft structures, on the attitude control, and on the
knowledge of the barycentric velocity of the spacecraft. The challenges mentioned above for the narrow angle astrometry
missions at the level of detector and optical distortion effects also apply here.

Data processing to achieve sub-{\muas} precisions will be complicated by (among others):
\begin{itemize}
    \item Relativistic modelling of astrometric measurements will have to be pushed to the nano-arcsec level in order to
        correctly interpret the raw data. Going beyond $0.1$~\muas\ precision in relativistic modelling would require
        research as well as improvements in our knowledge of the solar system (asteroid masses for example). 
    \item System calibration at this level will have to be much better (by an order of magnitude) than the
        astrometric accuracy aimed for and will thus be extremely challenging. The design of the instruments and mission
        concept will have to incorporate the data processing demands from the start. That is, any mission study and
        development phases should treat engineering and data processing on an equal footing. This is especially
        important where for example it is clear that certain stability requirements cannot be met through engineering
        but may be managed through a combination of metrology and sophisticated calibration concepts at data processing
        level.
    \item Approaching nano-arcsecond levels it is not obvious that simple models of the time dependence of source
        coordinates will be sufficient. In addition research into sources of astrometric jitter (such as star spots or
        microlensing in crowded regions) and their effect on the interpretation of image locations in the data stream is
        required.
    \item Another conceptual problem when approaching the nano-arcsecond regime is that for parallax measurements small
        ($<1$~au) sources which are sufficiently bright (hot) are required in large numbers (for global astrometry). It
        is not clear that at cosmological distances this requirement is fulfilled \citep{Lindegren2007}.
\end{itemize}

\section{CONCLUSIONS}
\label{sec:conclusions}

\begin{summary}[SUMMARY POINTS]
\begin{enumerate}
    \item {\gaia} is revolutionizing astronomy through the vast set of fundamental astronomical data that touches on
        every science topic. In parallel the celestial reference frame provided by {\gaia} enables the accurate
        astrometric and photometric calibration of past, current, and future sky surveys. Astronomical research and
        observations will profit from this legacy for many decades to come.
    \item \gdr{2} definitively demonstrated the power of an all-sky, high spatial resolution, high astrometric
        and photometric accuracy survey, enabling a vastly improved open cluster census, the tracing of stellar streams
        all across the sky, and fully mapping nearby satellite galaxies as well as the outer regions of globular
        clusters.
    \item \gdr{2} for the first time provides a dense sampling of Galactic phase space at high astrometric, photometric,
        and radial velocity precisions, which is essential for uncovering the subtle features in phase space as well as
        in the observational HR-diagram that are leading to new insights into the Milky Way and stars.
    \item \gdr{2} provides exquisite astrometry which nevertheless suffers from systematic errors. Armed with a detailed
        understanding of {\gaia} measurements and the data processing, efforts are underway to ensure that systematic
        errors remain well below the increasing precision levels of future data releases.
    \item Continuing down the path of microarcsecond optical/infrared absolute astrometric surveys is only feasible from
        space, and is required for the maintenance of the celestial reference frame which will anchor the ground
        based photometric and spectroscopic surveys as well as the extremely large aperture facilities of the future.
    \item The 21st century astronomer has an array of microarcsecond astrometry facilities at their disposal, with VLBI,
        GRAVITY, and HST providing vital capabilities in dust obscured and crowded sky regions where {\gaia}'s
        capabilities are limited.
\end{enumerate}
\end{summary}

Future efforts remain necessary both to maximize the scientific exploitation of data collected by the {\gaia} mission
but also to extend the microarcsecond astrometric survey capabilities into the future and possibly go one step beyond to
the nanoarcsecond regime.

\begin{issues}[FUTURE ISSUES]
\begin{enumerate}
    \item Achieving the highest possible accuracies at the bright end of the {\gaia} survey will remain challenging.
        Complementary VLBI astrometry of optically bright radio stars will remain essential as validation of the {\gaia}
        astrometry, and for accurately characterizing the bright {\gaia} reference frame. It is important that VLBI
        astrometry publications provide full astrometric solutions, including celestial positions and their observation
        epochs.
    \item The true power of {\gaia} lies in the combination of its data with photometric and spectroscopic surveys. The
        planning for future surveys should take into account the need to acquire the optimal mix of fundamental
        astronomical data at compatible accuracy levels.
    \item Determining the {\gaia} survey selection function is essential to unlocking some of the key science cases and
        will require a major research effort from the community, which should be supported by the release of as yet
        unpublished information relevant to the {\gaia} selection function. Future large survey plans should include
        from the very start the task to build and provide a survey selection function.
    \item Taking full advantage of the precision of {\gaia} proper motions out to large distances requires using the
        {\gaia} parallaxes to advance the precision of alternative distance indicators, especially photometric distance
        indicators across the HR diagram.
    \item As a next space astrometry mission the GaiaNIR option is the most feasible and will offer the widest set of
        science cases, the advantage of a dense sampling of the Milky Way disk complementing upcoming infrared
        spectroscopic surveys, while enabling a reset of optical reference frame accuracies to milliarcsecond levels
        into the next century and providing an unprecedented infrared reference frame.
    \item Nevertheless {\gaia} astrometry is likely to remain the standard for decades to come, including its treasure
        trove of astrometric, spectrophotometric, and spectroscopic time series. It is thus essential to invest effort
        into the long-term preservation of the {\gaia} data archive to keep it easily interoperable with other surveys
        and to enable the community to mine the epoch data and reprocess intermediate level {\gaia} data products in the
        light of new complementary data.
    \item Entering the nanoarcsecond absolute astrometry domain requires rethinking the Hipparcos/{\gaia} concept at
        several levels: the engineering challenges of wide-angle astrometry with a space interferometric system; the
        modelling of source motions; solving the issue of a lack of compact sources for parallax measurements over
        cosmological distances; and the modelling of light propagation in the presence of solar system light bending
        effects at the sub-nanoarcsecond level.
\end{enumerate}
\end{issues}

\section*{DISCLOSURE STATEMENT}
The author is not aware of any affiliations, memberships, funding, or financial holdings that might be perceived as
affecting the objectivity of this review. 

\section*{ACKNOWLEDGMENTS}
It is a great pleasure to thank the many DPAC and ESA colleagues as well as the Airbus Space engineer team, for
providing me over the years with their wisdom and insights that contributed in one way or another to this review taking
shape. It is their fantastic efforts over the past 25 years that have made {\gaia} a reality and incredible success. In
particular I wish to thank Paolo Tanga, Timo Prusti, and Jos de Bruijne for valuable inputs and comments on early
drafts.
This work has made use of data from the European Space Agency (ESA) mission {\gaia}
(\url{https://www.cosmos.esa.int/gaia}), processed by the {\gaia} Data Processing and Analysis Consortium (DPAC,
\url{https://www.cosmos.esa.int/web/gaia/dpac/consortium}). Funding for the DPAC has been provided by national
institutions, in particular the institutions participating in the {\gaia} Multilateral Agreement.
Funding from the Netherlands Organisation for Scientific Research (NWO) through grant NWO-M-614.061.414 and the
Netherlands Research School for Astronomy (NOVA) is gratefully acknowledged.
This work made use of the following software: Astropy, a community-developed core Python package for Astronomy
\citep[\url{http://www.astropy.org}]{astropy:2013, astropy:2018}, IPython
\citep[\url{https://ipython.org/}]{ipython:2007}, Jupyter (\url{https://jupyter.org/}), Matplotlib
\citep[\url{https://matplotlib.org}]{matplotlib:2007}, SciPy \citep[\url{https://www.scipy.org}]{scipy:2020}, NumPy
\citep[\url{https://numpy.org}]{numpy:2020}, scikit-learn
\citep[\url{https://scikit-learn.org/stable/}]{scikit-learn:2011} and TOPCAT
\citep[\url{http://www.starlink.ac.uk/topcat/}]{topcat:2005}.
This work has made use of NASA’s Astrophysics Data System.

%

\bibliographystyle{ar-style2}
\bibliography{brown}

\end{document}